\documentclass[twocolumn]{aastex63}

\newcommand{\totcan}{653~}

\usepackage{afterpage}

\usepackage{booktabs} % For \toprule, \midrule and \bottomrule
 
\usepackage{siunitx} % Formats the units and values
\usepackage{amsmath,amstext,color}
\usepackage{gensymb}
\usepackage{natbib}
\usepackage{hyperref}
\usepackage{longtable}
\usepackage{threeparttablex}
\usepackage{booktabs}

\usepackage{color}
\usepackage{graphicx}
\usepackage{amsmath}
\usepackage{amssymb}

\usepackage[]{appendix}

\begin{document}
\title{A Systematic Exploration of Kilonova Candidates from Neutron Star Mergers During the Third Gravitational Wave Observing Run}
\correspondingauthor{Jillian Rastinejad}
\email{jillianrastinejad2024@u.northwestern.edu}

\shorttitle{O3 Kilonova Candidates}
\shortauthors{Rastinejad et al.}
\newcommand{\NU}{\affiliation{Center for Interdisciplinary Exploration and Research in Astrophysics (CIERA) and Department of Physics and Astronomy, Northwestern University, Evanston, IL 60208, USA}}
\newcommand{\UA}{\affiliation{Steward Observatory, The University of Arizona, 933 North Cherry Avenue, Tucson, AZ 85721-0065, USA}}
\newcommand{\Keck}{\affiliation{W. M. Keck Observatory, 65-1120 Mamalahoa Highway, Kamuela, HI 96743, USA}}
\newcommand{\UCDavis}{\affiliation{Department of Physics, University of California, 1 Shields Avenue, Davis, CA 95616-5270, USA}}
\newcommand{\Padova}{\affiliation{Department of Physics and Astronomy Galileo Galilei, University of Padova, Vicolo dell'Osservatorio, 3, I-35122 Padova, Italy}}
\newcommand{\INAF}{\affiliation{INAF Osservatorio Astronomico di Padova, Vicolo dell'Osservatorio 5, I-35122 Padova, Italy}}
\newcommand{\LPL}{\affiliation{Lunar and Planetary Lab, Department of Planetary Sciences, University of Arizona, Tucson, AZ 85721, USA}}
\newcommand{\Stockholm}{\affiliation{Department of Astronomy, The Oskar Klein Center, Stockholm University, AlbaNova, 10691 Stockholm, Sweden}}
\newcommand{\UPenn}{\affiliation{Department of Physics and Astronomy, University of Pennsylvania, 209 South 33rd Street, Philadelphia, PA 19104, USA}}
\author[0000-0002-9267-6213]{J. C.~Rastinejad}
\NU

\author[0000-0001-8340-3486]{K.~Paterson}
\NU

\author[0000-0002-7374-935X]{W.~Fong}
\NU

\author[0000-0003-4102-380X]{D.~J. Sand}
\UA

\author[0000-0001-9589-3793]{M.~J. Lundquist}
\Keck

\author[0000-0002-0832-2974]{G.~Hosseinzadeh}
\UA

%alphabtical
\author{E.~Christensen}
\LPL

\author{P.~N. Daly}
\UA

\author[0000-0002-2575-2618]{A.~R. Gibbs}
\LPL

\author[0000-0002-3841-380X]{S. Hall}
\NU\UPenn

\author{F.~Shelly}
\LPL

\author[0000-0002-2898-6532]{S.~Yang}
\Stockholm

\begin{abstract}
We present a comprehensive analysis of \totcan optical candidate counterparts reported during the third gravitational wave (GW) observing run. Our sample concentrates on candidates from the 15~events (published in GWTC-2, GWTC-3 or not retracted on GraceDB) that had a $>$1\% chance of including a neutron star in order to assess their viability as true kilonovae. In particular, we leverage tools available in real time, including pre-merger detections and cross-matching with catalogs (i.e. point source, variable star, quasar and host galaxy redshift datasets), to eliminate 65\% of candidates in our sample. We further employ spectroscopic classifications, late-time detections and light curve behavior analyses, and conclude that 66 candidates remain viable kilonovae. These candidates lack sufficient information to determine their classifications, and the majority would require luminosities greater than that of AT\,2017gfo.  Pre-merger detections in public photometric survey data and comparison of catalogued host galaxy redshifts with the GW event distances are critical to incorporate into vetting procedures, as these tools eliminated $>$20\% and $>$30\% of candidates, respectively. We expect that such tools which leverage archival information will significantly reduce the strain on spectroscopic and photometric follow-up resources in future observing runs. Finally, we discuss the critical role prompt updates from GW astronomers to the EM community play in reducing the number of candidates requiring vetting. 
\end{abstract}


% SAGUARO forever!!

\section{Introduction}

Since the first detection of a compact object merger by gravitational waves (GWs) in 2015 \citep{LVC16}, the large number of detected mergers of black holes (BH) and/or neutron stars (NS) has contributed to the rapidly-growing field of multi-messenger astronomy. Each subsequent GW observing run has brought increased detector sensitivity and a larger survey volume to detect the mergers of binary black holes (BBHs), binary neutron stars (BNSs) and neutron star black holes (NSBHs) \citep{GWTC-1,GWTC2,GWTC-3,Abbott+21_NSBH}. BNS and some NSBH mergers are expected to produce kilonovae, optical-near-IR thermal transients powered by the radioactive decay of heavy $r$-process elements \citep{lipaczynski98,metzger+10,barneskasen13,tanaka+13,kawaguchi+16}. Given their relatively low peak luminosities ($\sim 10^{41}-10^{42}$ erg s$^{-1}$) and fast-fading nature (observable on $\sim$week timescales), discerning kilonovae from the wide array of optical transients is a long-standing challenge in this field.

The discovery of the first GW-detected BNS merger, GW170817 \citep{Abbott+17a}, and its kilonova AT\,2017gfo \citep{Arcavi+17,Coulter+17,Lipunov+17,Tanvir+17,Soares-Santos+17,Valenti+17}, was the first proof-of-concept for multimessenger astronomy between gravitational and electromagnetic (EM) waves. Positively identifying new kilonova counterparts to GW events will help to constrain their intrinsic and extrinsic diversity \citep{metzgerfernandez14,Shibata+19,Kawaguchi+20b,Gompertz+18,Ascenzi+19,Rossi+20,Rastinejad+21}. Further, by matching kilonova observations to models, one may infer their ejecta masses and compositions, therein elucidating the contribution of NS mergers $r$-process enrichment in the Universe. Given the high angular resolution of typical optical instruments, the discoveries of kilonova counterparts to GW events provide sub-arcsecond localizations, and thus crucial identifications to host galaxies and stellar populations. This in turn can enable constraints on the Hubble constant (through identification of the host galaxy; \citealt{LVC_GW170817_H0}), and lend insight into the environments which give rise to BNS/NSBH mergers. Finally, one can indirectly constrain the maximum mass of neutron stars (as, with a greater sample of mergers with optical counterparts, we can probe the upper end of component masses that produce kilonovae; \citealt{Fryer+15,Nicholl+21}).

The third and most recent LIGO-Virgo Collaboration (LVC) observing run (O3) took place from April 2019 to March 2020. The improved sensitivity of detectors produced a higher rate of detected compact object mergers at greater distances. This resulted in 125 published O3 events between the second and third Gravitational-Wave Transient Catalogs (GWTC-2 and GWTC-3), augmenting the previous collection of GW-detected mergers by a factor of $\sim$10 \citep{GWTC2,GWTC-3}. The O3 literature includes five mergers for which the mass distribution of at least one component falls within the upper limit of a NS, accounting for uncertainties ($\lesssim 3 M_{\odot}$). In addition, tens of merger events involving a NS were announced via the Gamma-ray Coordinates Network circulars (GCNs) in O3 that did not pass the traditional thresholds for inclusion in the published samples. Despite these numerous opportunities and subsequent efforts by the community, no credible EM counterpart to a GW event has been identified since AT\,2017gfo  \citep{Andreoni+19,Coughlin+19,Dobie+19,Goldstein+19,Gomez+19,Hosseinzadeh+19,Lundquist+19,Andreoni+20,Antier+20a,Antier+20b,Ackley+20,Garcia+20,Gompertz+20b,Kasliwal+20,Morgan+20,Pozanenko+20,Thakur+20,Vieira+20,Watson+20,Anand+21,Alexander+21,Becerra+21,Bhakta+21,Chang+21,de_Wet+21,Dichiara+21,Dobie+21,Kilpatrick+21,Oates+21,Ohgami+21,Paterson+21,Tucker+21,deJaeger+22}. A significant challenge for EM follow-up is the need to search large localization areas, which spanned $\sim$10--10,000~deg$^2$ for events in O3. 

Previously in \citet{Paterson+21}, the Searches After Gravitational-waves Using ARizona Observatories (SAGUARO) collaboration presented an analysis of optical candidate counterparts to 17 O3 events. Similar to other surveys, SAGUARO's observations of the large localizations returned thousands of candidate counterparts, $\sim$tens of which remained viable candidates after initial vetting \citep{Lundquist+19,Paterson+21}. In \citet{Paterson+21}, we also examined optical follow-up by the community, finding that only 65\% of reported candidates were ever re-observed. Among this follow-up, we found a high potential for redundant spectroscopic or photometric observations of candidates. In addition, we eliminated 12 previously ``open'' candidates as kilonovae by examining their photometric light curves and host galaxy redshifts. 

With the LIGO-Virgo-KAGRA's (LVK) fourth observing run (O4) on the horizon, we are still confronted with the challenge of the correct identification of optical counterparts amidst large localization areas. This is evidenced by the many remaining viable O3 kilonova candidates and the limited nature of spectroscopic and photometric follow-up resources. Thus, we are motivated to leverage the full arsenal of tools available at the time of follow-up (e.g., contextual catalog matching, existing survey observations) to conduct a uniform analysis of all kilonova candidates of any O3 merger involving an NS. We aim to examine what fraction of kilonova candidates could have been eliminated without targeted follow-up, and what fraction remain viable after thorough vetting.

In this work, we analyze \totcan O3 kilonova candidates across 15~GW events gathered from the GCNs and the Transient Name Server (TNS). We aim to (i) identify the most promising methods to eliminate candidates as kilonovae in real time (i.e., shortly after candidates have been identified and before they are reported in GCNs) and (ii) determine if, after exploiting all tools at our disposal, any of the candidates are still physically viable as kilonovae. In Section~\ref{sec:sample_coll} we describe our selection of 15 GW events and the corresponding \totcan candidate counterparts. In Section~\ref{sec:3} we apply tools to eliminate candidates that will be available in real time for O4. In Section~\ref{sec:4} we utilize all tools, regardless of if they are available in real time, to eliminate any remaining candidates. We examine the results of our analysis in Section~\ref{sec:discussion} and make recommendations for future GW observing runs. Finally, we present our conclusions in Section~\ref{sec:conclusion}. All magnitudes are reported in the AB system and are corrected for Milky Way dust extinction based on \citet{SchlaflyFinkbeiner11}. Throughout, we assume a standard cosmology of $H_{0}$ = 69.6~km~s$^{-1}$~Mpc$^{-1}$, $\Omega_{M}$ = 0.286, $\Omega_{vac}$ = 0.714 \citep{Bennett+14}

\section{Collection of Kilonova Candidates}
\label{sec:sample_coll}

\subsection{Event Selection}

Toward our aim of identifying any remaining, plausible kilonovae, we examine the candidate counterparts to GW events involving at least one NS. Though potentially only a small fraction of NSBH mergers produce EM emission \citep{Foucart2013,Shibata+19,Broekgaarden+21}, we search for candidates from all NSBH events, regardless of their mass ratio, spin or other properties. Our sample of GW events includes (i) events published in the LVC literature whose final mass distribution includes at least one component with a $>5$\% probability of being  $< 3 M_{\odot}$ (following the conservative upper limit on a NS mass assumed by \citealt{GWTC2} based on \citealt{RhoadesRuffini74,KalogeraBaym96}) and (ii) non-retracted events announced in the GCNs whose most recent classification likelihood of being a BNS or NSBH merger is $>1\%$. 

We first gather events from the GWTC-2 and GWTC-3 catalogs or other published LVC works \citep{LVC_GW190425,GWTC2, LVC_GW190814,Abbott+21_NSBH,GWTC-3}. Five GW events that were initially announced in the GCNs meet the first critera above (GW190425, GW190426\_152155, GW190814, GW200105\_162426 and GW200115\_042309; \citealt{LVC_GW190425,GWTC2, LVC_GW190814,Abbott+21_NSBH}). These events have False Alarm Rates (FAR) of $< 2.0$ yr$^{-1}$, which corresponds to an expected contamination fraction of $<10$\% \citep{GWTC2,Abbott+21_NSBH}. We also note that the GWTC-3 catalog \citep{GWTC-3} published three events that meet criteria (i), GW191113\_071753, GW191219\_163120, and GW200210\_092254, but they were not announced in real time via GCNs and all have estimated median distances beyond which typical nightly surveys depths would be sensitive to an AT\,2017gfo-like kilonova ($D_{\rm L} \gtrsim 550$~Mpc; \citealt{GWTC-3}); thus we do not include them in our sample.

Following criteria (ii) above, we augment our sample with 10 additional non-retracted events (S190510g, S190718y, S190901ap, S190910d, S190910h, S190923y, S190930t, S191205ah, S191213g and S200213t) that were reported in the GCNs but did not meet the threshold for inclusion in the GWTC-2 or GWTC-3 catalogs due to low FAR values found by offline analyses \citep{GWTC2,GWTC-3}. As most of these events were the subject of targeted optical searches reported to the GCNs and the aim of this work is to investigate how to improve future follow-up, we add them to our sample. We use the most recent localizations and properties reported on GraceDB\footnote{\url{https://gracedb.ligo.org}} for these unretracted, low-significance events.

Though an optical counterpart to a BBH merger has been claimed (\citealt{Graham+20}; but see also \citealt{Ashton+20}), we do not include BBH mergers in our analysis as their probability of producing detectable EM counterparts is exceedingly low compared to mergers involving an NS \citep{Perna+18}. We note that mergers in which one object falls in the ``Mass Gap'' ($3 < M / M_{\odot} < 5$; \citealt{lvc_2016_LRR}) also have the potential to create EM emission and thus received some follow-up during O3. Three events were initially announced in the GCNs as having a $>$95\% chance of being a Mass Gap event (GW190924\_021846, GW190930\_133541 and GW200316\_215756). However, final analyses of these events \citep{GWTC2,GWTC-3} find all three are most likely BBHs, and thus are not included in our sample. Our final sample of 15 GW events and their properties are listed in Table~\ref{tab:event_cand}. 

\begin{deluxetable*}{lcclcc}
\savetablenum{1}
\tabletypesize{\small}
\centering
\tablecolumns{6}
\tabcolsep0.15in
\tablecaption{Total Sample of Gravitational Wave Events and Electromagnetic Counterpart Candidates
\label{tab:event_cand}}
\tablehead {
\colhead {Event}		&
\colhead {Classification$^{*}$}		&
\colhead {Distance}	&
\colhead {GCN}		& 
\colhead {TNS}	&
\colhead {Combined Unique Total} \\
\cline{4-6}
&
&
\colhead {(Mpc)}	&
\multicolumn{3}{c}{(Number of Candidates)} 
}
\startdata
\textbf{GW190425} & BNS, NSBH & 157$^{+70}_{-70}$ & 16 & 33 & 34 \\
\textbf{GW190426$^{\dagger}$} & NSBH & 377$^{+180}_{-160}$ & 22 & 20 & 28 \\
S190510g & BNS$^{\ddagger}$ & 227$^{+92}_{-92}$ & 15 & 19 & 21 \\
S190718y & BNS$^{\ddagger}$ & 227$^{+165}_{-165}$ & 3 & 15 & 15 \\
\textbf{GW190814} & MassGap, NSBH & 241$^{+40}_{-50}$ & 31 & 95 & 96 \\
S190901ap & BNS$^{\ddagger}$ & 241$^{+79}_{-79}$ & 12 & 94 & 95 \\
S190910d & NSBH$^{\ddagger}$ & 632$^{+186}_{-186}$ & 0 & 7 & 7 \\
S190910h & BNS$^{\ddagger}$ & 230$^{+88}_{-88}$ & 16 & 59 & 59 \\
S190923y & NSBH$^{\ddagger}$ & 483$^{+133}_{-133}$ & 1 & 1 & 2 \\
S190930t & NSBH$^{\ddagger}$ & 108$^{+38}_{-38}$ & 10 & 181 & 183 \\
S191205ah & NSBH$^{\ddagger}$ & 385$^{+164}_{-164}$ & 8 & 27 & 27 \\
S191213g & BNS$^{\ddagger}$ & 201$^{+81}_{-81}$ & 12 & 23 & 23 \\
\textbf{GW200105$^{\dagger}$} & NSBH & 280$^{+110}_{-110}$ & 7 & 38 & 38 \\
\textbf{GW200115$^{\dagger}$} & NSBH & 300$^{+150}_{-100}$ & 10 & 13 & 13 \\
S200213t & BNS$^{\ddagger}$ & 201$^{+80}_{-80}$ & 5 & 12 & 12 \\
\hline
Total & & & 168 & 638 & 653 \\
\enddata
\tablecomments{
Bolded event names indicate the event met the significance threshold for inclusion in GWTC-2 or GWTC-3 \citep{GWTC2,GWTC-3}. Unbolded events were announced in the GCNs but not included in the GWTC-2 or GWTC-3 catalogs, though they have not been retracted. \\
$^*$Most probable non-terrestrial classifications. \\
$^{\dagger}$We abbreviate these event titles from their full names: GW190426\_152155, GW200105\_162426, and GW200115\_042309. \\
$^{\ddagger}$Most recent classification, according to GraceDB.}
\end{deluxetable*}

\begin{figure*}
\centering
\includegraphics[width=0.98\textwidth]{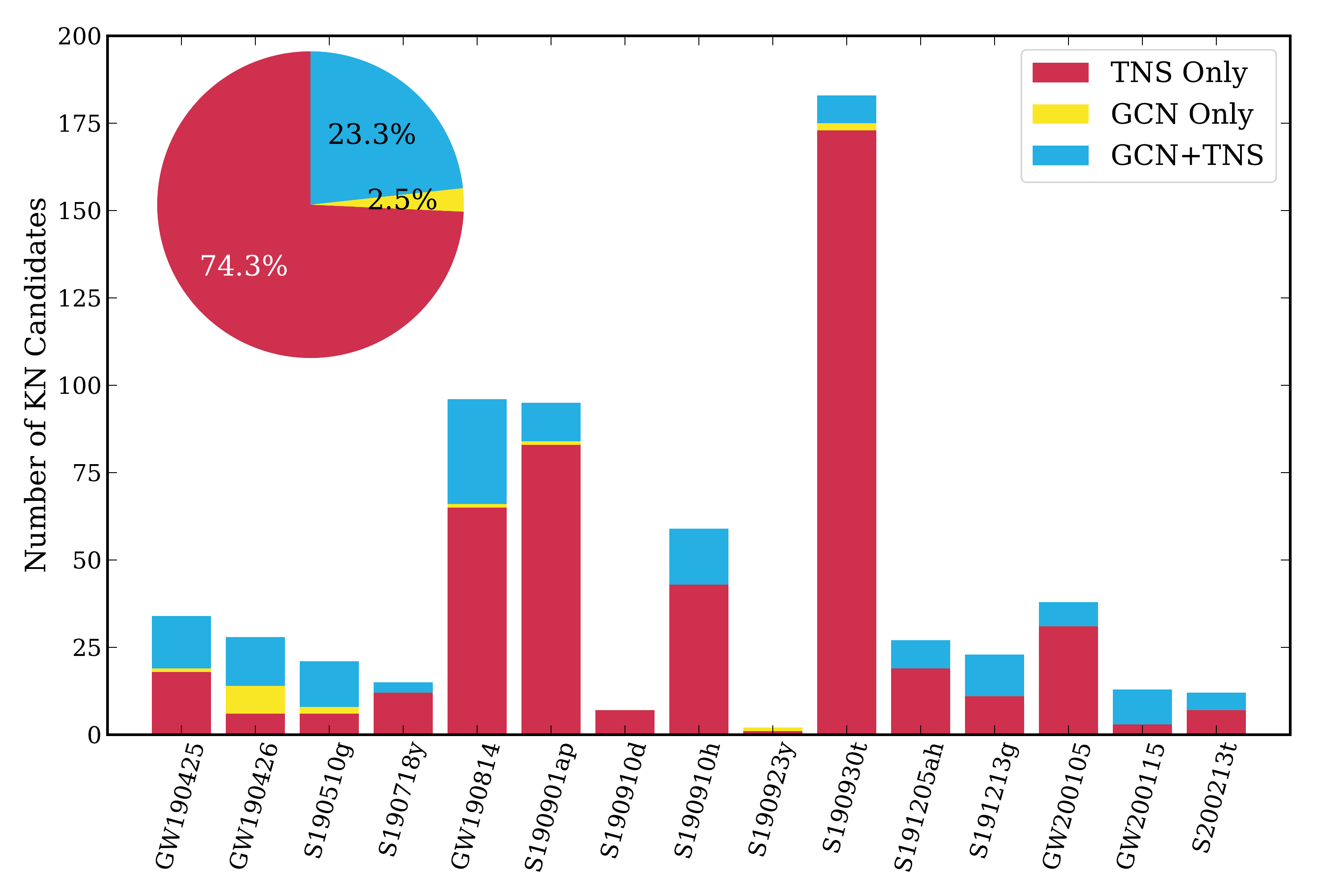}
\caption{The number of plausible kilonova candidates sorted by GW event after our initial criteria but before we eliminate candidates using ``Tools Available in Real-time'' (Section~\ref{sec:3}). Each bar is color-coded by the source of the candidate (GCN, TNS or both; Section~\ref{sec:sample_coll}). The most likely GW event classification is labeled at the top of each bar. We list two classifications for GW190425 and GW190814, as these events are well-studied and the literature is divided on their origin. \textbf{Inset}: A pie chart shows the distribution of the sources of all candidates meeting our initial criteria. Approximately one-quarter of candidates that we consider in our sample are reported in the GCNs, and almost all are reported to the TNS.}
\label{fig:cand_event_RO}
\end{figure*}

\subsection{GCN \& TNS Candidates}
\label{sec:gcn_tns_cand}

We gather candidate counterparts to each GW event from the GCNs and TNS\footnote{\url{https://www.wis-tns.org}}. GCNs are real-time notices to the community of EM follow-up to gamma-ray bursts (GRBs) or GW events. GCNs primarily contain candidates reported immediately following the event, while TNS is a more comprehensive database of newly discovered transients that is independent of GW events.

Our first step in optical candidate selection is to define initial criteria based on time, location and luminosity to use in our GCN and TNS searches. Our initial criteria for inclusion as follows: candidates (i) with $0 < \delta t < 5$ days (where $\delta t$ is the time between the event merger time and the discovery time of the candidate), and (ii) that are within the 90\% contour on the final localization map (for events published in the O3 catalogs or other LVK papers; \citealt{LVC_GW190425,GWTC2, LVC_GW190814,Abbott+21_NSBH}) or the most recent LALInference map available in GraceDB. We apply a third criteria, which filters out candidates that would be more luminous than the brightest kilonova model predictions (e.g., \citealt{Barbieri+20,Fong+21}) and short GRB kilonova candidates (see analyses of \citealt{Fong+21,Rastinejad+21}) at the GW-inferred event distance. We calculate the luminosity, $\nu L_{\nu}$, at the $1\sigma$ lower bound on the GW-inferred event distance (providing a conservative estimate) using the discovery filter pivot wavelength (or the $r$-band pivot wavelength if no filter is reported) and include only candidates fitting the criteria $\nu L_{\nu} < 10^{43}$ erg s$^{-1}$, $\approx 10$ times the luminosity of AT\,2017gfo and $\approx 5$ times the peak luminosities of the brightest kilonova models \citep{Barbieri+20,Fong+21}.

We apply these initial criteria to our GCN searches for the 15 events, collecting the name, RA, Dec, discovery magnitude and time of discovery of each reported candidate. Since our goal is to investigate real-time tools that might eliminate the need for follow-up, we include all GCN candidates that pass the initial criteria, regardless of if they were subsequently eliminated by further GCNs or the literature. However, we do keep track of which candidates were later eliminated. In total, we find 168 candidates across 15 events reported to the GCNs, 96 (57.1\%) of which were subsequently eliminated as reported in the GCNs. One event had no candidates reported (S190910d), while GW190814 had the most candidates reported that meet our initial criteria (31). 

We next apply the initial criteria to all transients reported to TNS in 2019 and 2020, regardless of their TNS classification. We gather 638 candidates from TNS across the 15 events, the majority ($\gtrsim$90\%) of which are not classified as a particular transient type as of 12/2021. The event with the second-largest 90\% localization region, S190930t, had the greatest number of candidates from TNS (181).

To remove any duplicate candidates, we cross-match between the GCN and TNS samples using the unique TNS name (if reported in GCNs) and by eliminating matches within $2''$. Together, our sample includes 652 unique candidates. However, one transient is a candidate to both S190910d and S190910h, and thus we ``double-count'' it by independently considering it for each GW event. Thus, after our preliminary cuts, we consider a total of \totcan candidates corresponding to the 15 GW events. 

\subsection{Total Sample}

We present the number of candidates collected in the GCNs, TNS, and in our total sample in Table~\ref{tab:event_cand}. In Figure~\ref{fig:cand_event_RO} we show a histogram of the number of candidates per event. With the exception of GW190814, for which deep, follow-up observations were conducted \citep{Kilpatrick+21,Ackley+20,Vieira+20,Morgan+20,Tucker+21,Gomez+19,Thakur+20,Andreoni+20}, the four events with the highest number of candidates have localizations of $>7500$ deg$^2$. Despite the initial classification of $>$99\% chance MassGap event \citep{GCN25324}, the relatively small localization and subsequent $>$99\% NSBH event classification (\citealt{GCN25333}; reported $\sim$11 hours later) resulted in extensive community follow-up of GW190814. The substantial targeted follow-up is also evidenced by the high fraction of GCN-reported candidates in comparison with other events.

\section{Vetting Candidates With Tools Available in Real-Time}
\label{sec:3}

We begin by cross-matching candidates with catalogs that provide critical contextual information, and searching public surveys for pre-merger detections to vet candidates with tools that would be available in real time. Motivated to find the most efficient means of ruling out kilonova imposters, we apply the first four conditions in parallel (point source/stellar catalogs, moving object catalog, quasar catalogs and pre-explosion detections; Sections~\ref{sec:ps}--\ref{sec:pre_expl_dets}) to every candidate in our sample, regardless of if it was classified in the GCNs or TNS. We apply the steps described in Section~\ref{subsec:host_dists}, host galaxy matching, to only the remaining candidates. In Figure~\ref{fig:sec3_pie} (left) we show the fraction of candidates eliminated by each tool in this section.

\subsection{Point Source and Variable Star Catalogs}
\label{sec:ps} % DONE

Variable stars are sources of contamination in transient surveys, and may be eliminated by cross-matching transients with the locations of known point sources. We query the \textit{Gaia} early Data Release 3 (eDR3; \citealt{GaiaMission16,Gaia_eDR3}) catalog, the Pan-STARRS (PS1) point source catalog \citep{TachibanaMiller18} and the ASAS-SN variable star catalog \citep{Jayasinghe+19} for objects within 2$''$ of each candidate. Though the astrometrical uncertainty varies by catalog, we elect to use a uniform cross-matching radius that is reflects the seeing and pixel size of the surveys. For \textit{Gaia}, a candidate is considered stellar if its match (i) has an absolute proper motion value $>3$ times the error in total proper motion, (ii) it is flagged as variable, or (iii) it has a parallax significance $>8$ (following \citealt{TachibanaMiller18}). The PS1 point source catalog assigns nearly all PS1 sources a score indicating their likelihood of being stellar based on a random forest machine-learning algorithm. We consider sources to be stellar if they have a single counterpart in the catalog whose point source score is greater than 0.83 (reflecting a true positive rate of $\sim$0.995, at a false positive rate of 0.005; \citealt{TachibanaMiller18}). Finally, we inspect the light curves of the four ASAS-SN variable star matches, finding that three are good matches to variable stars. The remaining source, AT\,2019rup or ASASSN-V J094204.78+234107.0, is classified as a Young Stellar Object (YSO) and is coincident with the nucleus of a galaxy. However, as its ASAS-SN light curve shows pre-merger variability, we rule this out as a viable candidate. In total, we conclude that 51 sources are stellar after cross-matching with three catalogs. Notably, four were reported in the GCNs as initially viable counterparts. The PS1 point source catalog eliminates the greatest number of candidates as kilonovae (38).

\subsection{Moving Object Catalog}
\label{sec:mo}

Near-Earth moving objects are another potential source of contaminants. We use a radius of 20$''$ to cross-match candidates with the IAU Minor Planet Center Orbit Database\footnote{\url{http://www.minorplanetcenter.net/iau/MPCORB.html}} (MPCORB). As many surveys already include moving object cross-matching in their pipeline, we expect few matches within our sample. Accordingly, we identify 10 candidates with a match in MPCORB.

We do not find any detections at the same coordinates prior to or following the initial discoveries of the 10 candidates in the public ZTF database, the SAGUARO database, TNS or using the ATLAS forced photometry tool (the use of these tools is further described in Section~\ref{sec:pre_expl_dets}). However, we note that for 3 candidates (AT\,2019nri, AT\,2019nsn, AT\,2019nsl) detections were reported $\sim$2--8~minutes apart in the $i$- and $z$-band filters, though these are consistent with the radial velocities of known moving objects. We eliminate all 10 candidates with matches in MPCORB as viable kilonovae.

\subsection{Quasar Catalogs}
\label{sec:qso}

AGN and quasars are known variable sources that may masquerade as real transients. We next query the Million Quasar Catalog (MILLIQUAS; \citealt{Flesch15,Flesch21}) for objects within $1''$ of candidates. We employ a smaller match radius than was used for point-source matching to avoid confusion between an offset candidate and an active host galaxy center. We accept candidates as quasars if their probability of being a quasar (calculated based on the association of photometric data with radio or X-ray detections) is $>97$\%, following \citet{Flesch15} which found this threshold to yield good agreement with confirmed quasars in the Sloan Digital Sky Survey DR16 Quasar Catalog (SDSS; \citealt{Lyke+20}). We also cross-match candidates to the SDSS Quasar catalog using the same matching radius. We consider objects marked by the catalog as questionable quasars to still be viable candidates. We also inspect the offsets of candidates to the host galaxy nucleus in the Legacy Survey viewer\footnote{\url{http://legacysurvey.org/viewer}} to ensure that quasars are not falsely attributed to real transients (for future, larger samples of candidates, calculating the offsets using catalogued coordinates may be prudent). Finally, we examine the candidates' light curves (further described in Section~\ref{sec:pre_expl_dets}) and find that 20 (none of which were reported in the GCNs) had credible pre-merger detections. Twenty-six candidates are marked as quasars, thus eliminating them from being viable kilonova candidates. In total, cross-matching to point source/stellar, moving object and quasar catalogs (Sections~\ref{sec:ps}--\ref{sec:qso}) results in the elimination of 77 candidates, or 11.8\% of our sample (Figure~\ref{fig:sec3_pie}, left).

\subsection{Pre-Merger Detections}
\label{sec:pre_expl_dets}

Compact object mergers are not typically expected to produce optical or IR emission prior to a GW event, allowing us to eliminate any candidate with a pre-merger detection. We search for and combine all available photometry of the \totcan unique candidates in our sample from TNS, the public Zwicky Transient Facility (ZTF; \citealt{Bellm+19}) database, the Asteroid Terrestrial-impact Last Alert System (ATLAS; \citealt{Tonry+18, ATLAS_smith+20}) forced photometry tool and our own observations from the SAGUARO database. As SAGUARO utilizes the Steward Observatory 1.5~m Mt. Lemmon telescope with its 5 deg$^2$ imager (operated by the Catalina Sky Survey; \citealt{CSS}) as its discovery engine, $\sim$3~years of observations at an average depth of 21.1~mag are available to search for pre-explosion detections (see \citealt{Lundquist+19,Paterson+21} for more details). In Section~\ref{subsec:53} we discuss additional surveys to search for pre-merger detections.

For ZTF, we gather image-subtracted photometry when available \citep{ZTFproducts19}. For ATLAS, we perform forced point-spread-function (PSF) photometry at the positions of candidates covered by the survey for 200 days preceding the GW events using the publicly-available service (simulating what would be computationally reasonable in real-time; \citealt{ATLAS_tonry+18,ATLAS_smith+20}). We also stack multiple ATLAS epochs of photometry in a given filter on a single night using the publicly-available script provided by the service\footnote{\url{https://gist.github.com/thespacedoctor/86777fa5a9567b7939e8d84fd8cf6a76}}. Finally, we query the SAGUARO database for candidates using a matching radius of $1''$, and obtain difference-imaged photometry or $5\sigma$ limits \citep{Lundquist+19,Paterson+21}. We convert all magnitudes to AB units and correct all detections for Milky Way extinction in the direction of the candidate \citep{SchlaflyFinkbeiner11}.

As not all difference images are publicly available, it is possible that poor galaxy subtractions can masquerade as pre-merger detections at or near candidate positions. Further, asteroid detections may contaminate the forced photometry tool. Thus, we take a conservative approach by requiring that any of the following criteria are met to eliminate a candidate as viable: (i) there is a pre-merger detection $<$12 days prior to the GW event that, upon manual inspection, follows a smooth trajectory consistent with the subsequent light curve, (ii) there are multiple pre-merger detections within 10 days of each other, (iii) there are multiple pre-merger detections at any time by different surveys, or (iv) there is a single pre-merger detection that does not meet any of the criteria above, but lacks a contaminating source within $\sim 5''$ of the candidate position in archival PS1 imaging. 

Ultimately, we rule out 186 candidates ($>$20\% of our initial sample; Figure~\ref{fig:sec3_pie}, left) sample based on pre-merger detections. Thirty-three of these eliminated candidates were reported in the GCNs. Combined with eliminations made in Sections~\ref{sec:ps}--\ref{sec:qso}, we eliminate 218 candidates, leaving 435 candidates to be carried to the following step.

\begin{deluxetable*}{llc}
\tabletypesize{\small}
\centering
\savetablenum{2}
\tablecolumns{3}
\tabcolsep0.15in
\tablecaption{Candidates Whose Hosts are within the GW Event Distance Range
\label{tab:gal_categories}}
\tablehead {
\colhead {Category}		&
\colhead {Description}		&
\colhead {\# of} \\
&
&
\colhead {Candidates}
}
\startdata
Platinum & Highly Confident Host Association and Spectroscopic Redshift within GW Distance Uncertainty & 19  \\ % add Pcc here?
Gold & Moderately Confident Host Association and Spectroscopic Redshift within GW Distance Error & 1 \\
Silver & Highly Confident Host Association and Photometric Redshift within GW Distance Error & 94 \\
Bronze & Moderately Confident Host Association and Photometric Redshift within GW Distance Error &  12  \\
Inconclusive & Either candidate region uncatalogued, no redshift of best galaxy or no confident association & 107 \\
Eliminated & Highly or Moderately Confident Host Association and Redshift outside GW Distance Error & 202 
\enddata
\tablecomments{Candidates remaining after cross-matching to catalogs and searching for pre-merger detections (Sections~\ref{sec:ps}-\ref{sec:pre_expl_dets}) separated into ctegories of host galaxy association confidence. Confidence that each candidate is associated with a host galaxy in the GW-inferred distance range descends from Platinum to Bronze.}
\end{deluxetable*}

\subsection{Host Galaxy Associations and Distances}
\label{subsec:host_dists}

We next attempt to associate each candidate to its most likely host galaxy and, if available, use the catalogued photometric or spectroscopic redshift to eliminate candidates whose hosts are outside the 95\% credible interval of the GW-inferred event distance. We search three catalogs for potential host galaxies: SDSS Data Release 12 (SDSS DR12; \citealt{Alam+15}), PanSTARRS Source Types and Redshifts with Machine Learning (PS1-STRM; \citealt{Chambers+16,Beck+21}) and Legacy Survey Data Release 9 (LS DR9; \citealt{Dey+19}). SDSS DR12 covers over 14,000 deg$^2$ to an average depth of $r>22.7$~AB mag and supplies spectroscopic redshifts of more than 1.4 million galaxies \citep{Alam+15}. The catalog presents $>$200 million photometric redshifts and includes star-galaxy probabilistic classifications of objects \citep{Beck+16}. The PS1-STRM catalog analyzes over 2.9 billion objects from PS1 DR1 (DR2 is unavailable), presenting star-galaxy classifications and a large catalog of photometric redshifts \citep{Beck+21}. PS1 DR1 covers $\sim$30,000 deg$^2$, reaching similar depths to SDSS \citep{Metcalfe+13,Chambers+16}. Finally, LS DR9 combines observations from the fourth BASS \citep{Zou+2017} and MzLS \citep{Silva+2016} data release and the seventh DECaLS \citep{Dey+2019} release. The combined survey covers $\approx$14,000 deg$^2$ to depths of $r>$~23.4~AB~mag \citep{Dey+19}. The LS DR9 photometric redshift catalog includes over 2.7 million objects which are identified as galaxies using color and magnitude cuts \citep{Zhou+21}. All three photometric redshift catalogs are trained on a sample of spectroscopically-classified galaxies spanning the range $0 \lesssim z \lesssim 0.8$. Together, these three catalogs provide nearly complete coverage of the candidates in our sample, with the positions of only 14 candidates, or 3.1\% of those considered in this step, not covered by the combined footprints of the surveys.

For host associations, we begin by searching for galaxies near each candidates' position in SDSS DR12 and PS1-STRM through Vizier\footnote{\url{https://vizier.cds.unistra.fr}} and in LS DR9 through NOIRLab's Data Lab\footnote{\url{https://datalab.noirlab.edu}}. Following \citet{Zhou+21}, we do not use LS DR9 photometric redshifts of sources $z < 21$~mag. Based on the observed offsets of short GRBs from the centers of their host galaxies (and thus the maximum observed offsets of NS mergers; \citealt{FongBerger13}), we determine that a 100~kpc offset between the transient and host center is a conservative search radius. At the nearest GW event distance in our sample (108 Mpc; Table~\ref{tab:event_cand}), this corresponds to a search radius of 3.138 arcminutes. We cross-match between the three catalogs using a radius of 2$''$ and remove sources identified as stellar by either SDSS or STRM. If a spectroscopic redshift is available, we consider this value as the galaxy redshift. Otherwise, we record any photometric redshifts reported in STRM, SDSS or LS DR9. 

Next, we compute the probability of chance coincidence (P$_{cc}$; \citealt{Bloom+02}) for each source. P$_{cc}$ calculates the probability of chance alignment between the transient and potential host galaxies in the field using the galaxy magnitudes and angular offsets from the candidate's position. A low P$_{cc}$ value, especially in comparison to the values of other galaxies in the field, indicates a more likely host. Using the hosts' P$_{cc}$ values, we sort the candidates into four categories based on our confidence in their host galaxy associations:

\begin{enumerate}
    \item \textbf{Highly Confident}: Across all galaxies considered, the minimum $P_{cc}$ value (P$_{\rm cc, min}$)$<0.01$, while the second smallest $P_{cc}\geq 3 \times P_{\rm cc,min}$.
    \item \textbf{Moderately Confident}: P$_{\rm cc,min}<0.15$ and the second smallest $P_{cc} \geq 3 \times P_{\rm cc,min}$.
    \item \textbf{Not Confident}: Sources were found within the vicinity of the candidate, but neither the ``Highly Confident'' nor ``Moderately Confident'' criteria were met.
    \item \textbf{Uncatalogued}: Candidate not covered by SDSS, STRM and LS DR9 footprint.
\end{enumerate}

\noindent We remove duplicate entries of a single host in multiple catalogs upon visual inspection and, when applicable, keep track of both photometric redshifts. 

In Table~\ref{tab:gal_categories} we define categories that combine our confidence in the host association (see above) and the relative robustness between a spectroscopic and photometric redshift in determining if a host is within the GW distance uncertainty. Candidates for which we are unable to make a confident association or are not covered by the footprint of the catalogs we query are marked inconclusive. In total, 364 candidates have ``Highly Confident'' associations. Of these, 37 have spectroscopic redshifts available (``Platinum'' associations) and 266 have photometric redshifts (``Silver''). Thirty-one candidates have ``Moderately Confident'' associations. One ``Moderately Confident'' host has a spectroscopic redshift (``Gold'') and the remaining have photometric redshifts (``Bronze''). As we cannot draw any firm conclusions for candidates with ``Not Confident'' associations, we mark the 107 ``Not Confident'' and uncatalogued candidates as ``inconclusive''.

Finally, we use the spectroscopic and photometric redshifts of the ``Highly Confident'' and ``Moderately Confident'' associations to determine if the candidates' host galaxy redshifts are consistent with the 95\% distance credible interval inferred from GWs (Table~\ref{tab:event_cand}), which is inherently position-dependent \citep{Singer+16_Supp}. We thus use the calculated GW distance uncertainties from the 3D localization maps at the position of each candidate for comparison. For hosts with photometric redshifts, we determine if the redshifts' $1\sigma$ error range falls within the 95\% GW credible interval. For hosts with spectroscopic redshifts, we simply use the central value, as spectroscopic redshift errors are negligible. Between the ``Highly Confident'' and ``Moderately Confident'' samples, we find hosts of 202 candidates (30.9\% of our initial sample; Figure~\ref{fig:sec3_pie}, left) that do not fall within the event distance range; thus these sources are ruled out as GW counterpart candidates. We state the number of remaining candidates in each of the Platinum, Gold, Silver and Bronze categories, as well as the number of inconclusive and eliminated candidates in Table~\ref{tab:gal_categories}.

\subsection{Luminosity Cut Using Host Distance}
\label{sec:lum_cut}

Finally, we attempt to eliminate candidates with luminosities inconsistent with kilonovae, calculated using the host galaxy distances determined in Section~\ref{subsec:host_dists} (which generally have lower errors than the GW event distances). We calculate the luminosity ($\nu L_\nu$) of each candidate with Platinum, Gold, Silver or Bronze associations using its host galaxy redshift derived in Section~\ref{subsec:host_dists}. For $\nu L_{\nu}$ we use the candidate's discovery magnitude, filter pivot wavelength and the 1$\sigma$ lower bound on the host galaxy distance, as this provides a lower bound on luminosity. We rule out any candidate which exceeds $\nu L_{\nu} > 10^{43}$~ergs~s$^{-1}$ (following the reasoning of Section~\ref{sec:gcn_tns_cand}). To rule out a candidate whose host has multiple photometric redshift measurements, the luminosities calculated from all inferred redshifts for a given host must exceed the cut. We rule out two candidates as viable kilonovae based on this criterion.

\begin{figure*}
\centering
\includegraphics[width=0.468\textwidth]{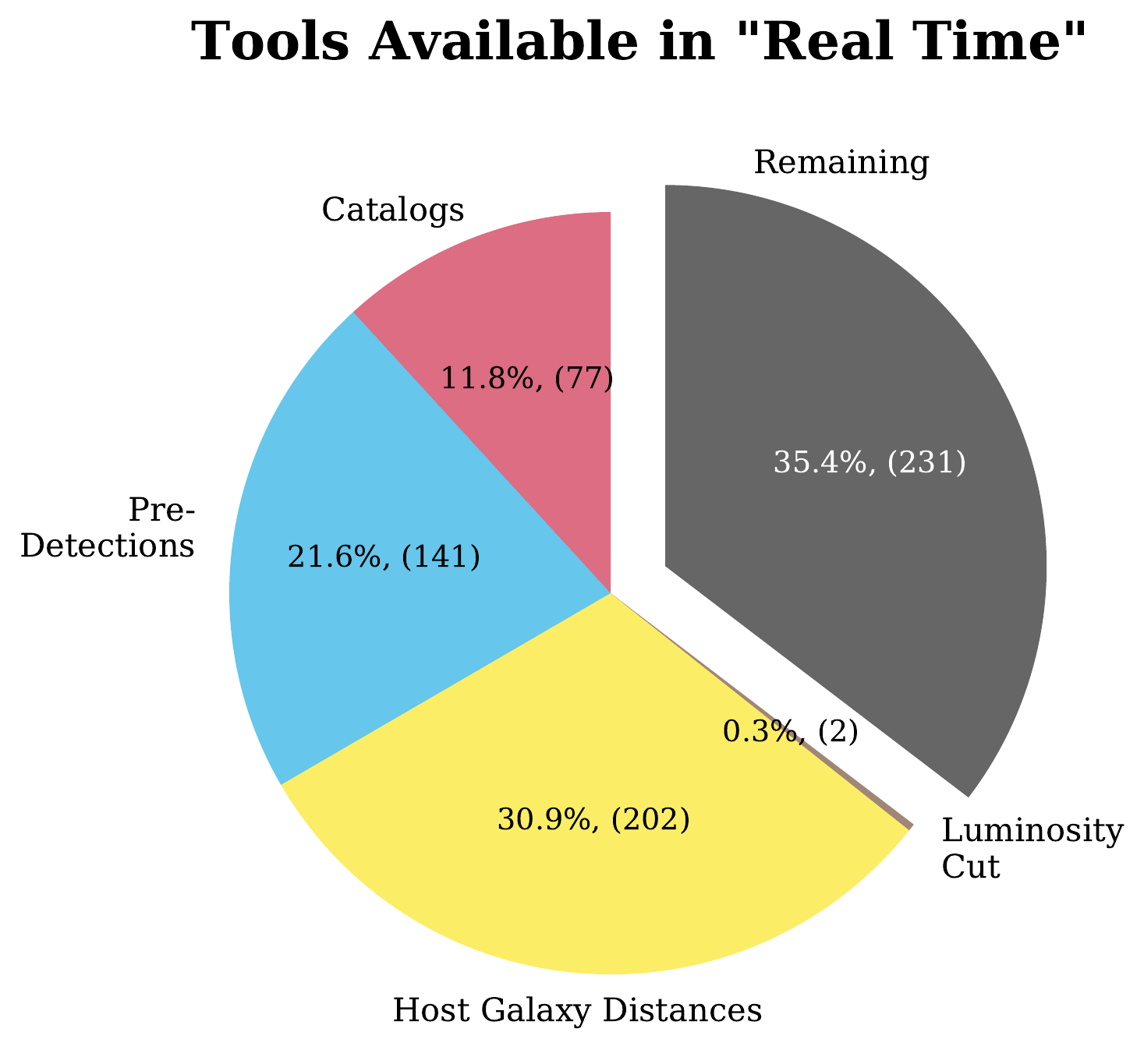}
\includegraphics[width=0.523\textwidth]{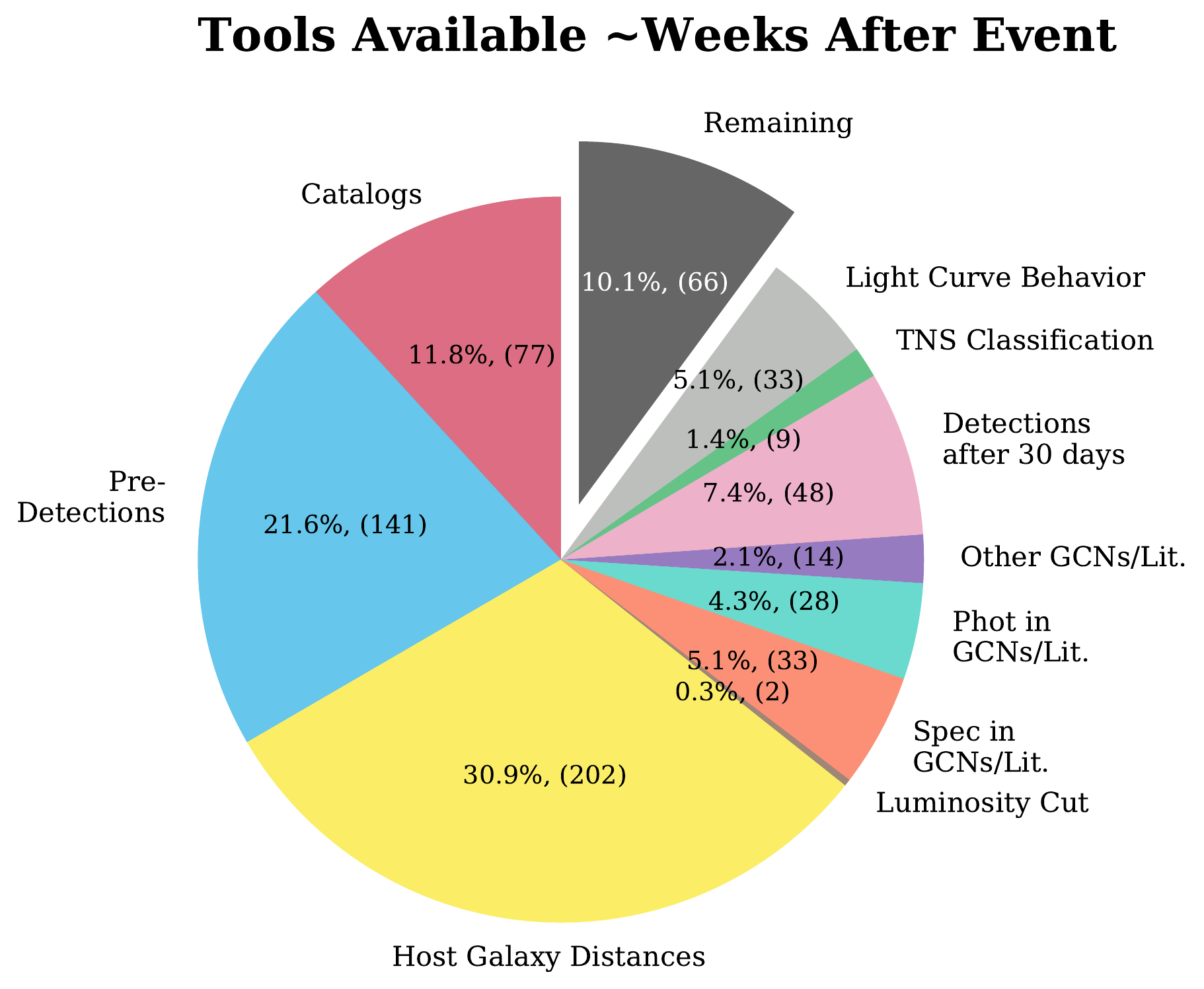}
\caption{\textbf{Left}: Pie chart demonstrating the fractions of candidates ruled out by cross-matching with stellar, quasar and moving object catalogs (Sections~\ref{sec:ps}--\ref{sec:mo}; red), photometric detections before the associated GW event (Section~\ref{sec:pre_expl_dets}; blue), host galaxies inconsistent with the GW event distance (Section~\ref{subsec:host_dists}; yellow) and a luminosity cut using the host galaxy distance (Section~\ref{sec:lum_cut}; brown). Each of these tools can be applied in real time during future observing runs. Together, host galaxy distances and pre-explosion detections rule out over half of the candidates in our sample, demonstrating their utility.\label{fig:sec3_pie} \textbf{Right}: Same as left, but with additional eliminations made using tools available $\sim$days--weeks after the event. Eliminations made in the GCNs and literature are divided into those made with spectroscopy (orange), photometry resources (turquoise), or other follow-up (mostly pre-detections, purple; Section~\ref{subsec:gcn_elim}). In addition, candidates ruled out as kilonovae due to detections after 30 days (Section~\ref{subsec:dt_30days}; pink), TNS classifications (Section~\ref{subsec:tns_elim}; green) and light curve behavior inconsistent with a kilonova (Section~\ref{subsec:lcbehavior}; grey) are shown. At the conclusion of the analysis, 66 candidates (10.1\% of the \totcan candidates in the original sample) remain viable kilonova candidates.\label{fig:sec4_pie}}
\end{figure*}

After applying the steps in Sections~\ref{sec:ps}--\ref{sec:lum_cut} in which only information available in real time is used, we have eliminated 422 of the \totcan candidates (64.6\%) in our total initial sample. Of the candidates we rule out using tools available in real time, 88 were reported in the GCNs, or 52.3\% of GCN-reported candidates in our sample. This demonstrates the power of using available contextual information for winnowing down kilonova candidates. For clarity, we break down exactly how these candidates were eliminated, using the methods in this section, in Figure~\ref{fig:sec3_pie} (left).

\section{Employing Information Available $\sim$Days--Weeks Post-Event}
\label{sec:4}

After eliminating candidates based on cross-matching to catalogs, pre-merger detections, host associations and luminosities in Section~\ref{sec:3}, we examine the remaining viable candidates (214) using tools or information available $\sim$days--weeks after the event. Through this process of further eliminating any of the 214 remaining candidates, we determine which (if any) candidate counterparts to the events in our sample remain viable. Exploring the sample of viable candidates in aggregate will inform what future tools would be most useful to eliminate them. In addition, tracking classifications from the GCNs and literature allows us to quantify what fraction of targeted follow-up would be considered redundant if all tools we apply in Section~\ref{sec:3} had been available in O3 (further discussed in Section~\ref{subsec:remaining}).

\subsection{GCN and Literature Follow-Up}
\label{subsec:gcn_elim}

We consider optical follow-up reported in the GCNs and literature (submitted or published papers) by numerous EM counterpart follow-up groups \citep{Andreoni+19,Coughlin+19,Goldstein+19,Gomez+19,Hosseinzadeh+19,Ackley+20,Andreoni+20,Antier+20a,Antier+20b,Garcia+20,Gompertz+20b,Kasliwal+20,Morgan+20,Thakur+20,Vieira+20,Watson+20,Anand+21,Becerra+21,Chang+21,Dichiara+21,Kilpatrick+21,Oates+21,Ohgami+21,Tucker+21}. We note that the majority of these works were focused on follow-up of GW190814, the most precisely-localized event potentially involving a NS throughout O3.

In total, we gather follow-up information of 199 of the \totcan candidates in our initial sample from the GCNs and literature. We find that 77 candidates (forty-eight and 29 reported to the GCNs and the literature, respectively) still considered viable after employing real-time tools (e.g., after Section~\ref{sec:3}) are eliminated. Twenty-eight and 11 of the GCN eliminations are due to spectroscopic classifications and photometric follow-up, respectively. Of the 26 candidates eliminated in the literature, two are reported as image artifacts, two as moving objects, one as an AGN based on PS1 imaging and five as stellar based on the \textit{Gaia}, PS1 or ASAS-SN catalogs. An additional fourteen transients are eliminated using serendipitous or targeted photometry of each candidate. Five candidates are eliminated by a spectroscopic classification. 

\citet{Kilpatrick+21} determine nine candidates cannot be kilonova counterparts to GW190814 based on host galaxy cross-matching. As we employ a different method of associating candidates with host galaxies and, in cases where multiple photometric redshifts are available, require all to be inconsistent with the GW event distance (Section~\ref{subsec:host_dists}), we retain the nine candidates eliminated by \citet{Kilpatrick+21} (and note them in Table~\ref{tab:remaining}). Thus, ignoring the 9 eliminations based on host distance, we rule out 29 candidates that have not already been rejected based on reasoning from the literature. In Figure~\ref{fig:sec3_pie} (right) we show the 71 candidates ruled out as viable kilonovae in the GCNs or literature, classified by the method of elimination. Later, we explore what fraction of the targeted follow-up could be considered redundant in light of eliminations made in Section~\ref{sec:3} in Section~\ref{subsec:redund_followup}.

\subsection{Detections After $\delta t =$ 30 days}
\label{subsec:dt_30days}

The optical and/or near-IR emission emitted by NS mergers is not predicted to be detectable at the GW distances considered in this work (see Table~\ref{tab:event_cand}) beyond $\delta t\sim 2$~weeks. Thus, we examine late-time light curves of the 144 remaining candidates, built using the same databases and process as described in Section~\ref{sec:pre_expl_dets}. We rule out transients as viable kilonovae if their light curves contain optical and/or near-IR detections beyond a conservative timescale of  $\delta t > 30$~days using criteria similar to those employed in Section~\ref{sec:pre_expl_dets}. In this case, to be eliminated the transient light curve must show either: (i) at least one detection past $\delta t = 30$~days that follows the shape of the preceding light curve, (ii) multiple detections past $\delta t = 30$~days within 10 days of each other, or (iii) multiple  detections past $\delta t = 30$~days by different telescopes. Forty-eight candidates meet this criterion, and are eliminated at this stage.

\subsection{TNS Classifications}
\label{subsec:tns_elim}

Next, we consider candidate classifications reported to TNS. Generally, these classifications are based on spectroscopic observations. Unlike in the GCNs, the follow-up and classifications reported to TNS are not necessarily performed with targeted kilonova or EM counterpart searches in mind. Rather, much of the TNS follow-up is focused on reporting newly-discovered transients from targeted or untargeted surveys often unconnected to GW follow-up. Nine candidates not eliminated in previous steps were classified in TNS as Type Ia, IIn or Ic supernovae \citep{TNS2019phq_pia,TNS2019pzb,TNS2019qem,TNS2019rta, TNS2019rwq,TNS2020pm}, leaving 99 remaining candidates at this stage.

\begin{figure*}
\label{fig:final_barchart}
\centering
\includegraphics[width=\textwidth]{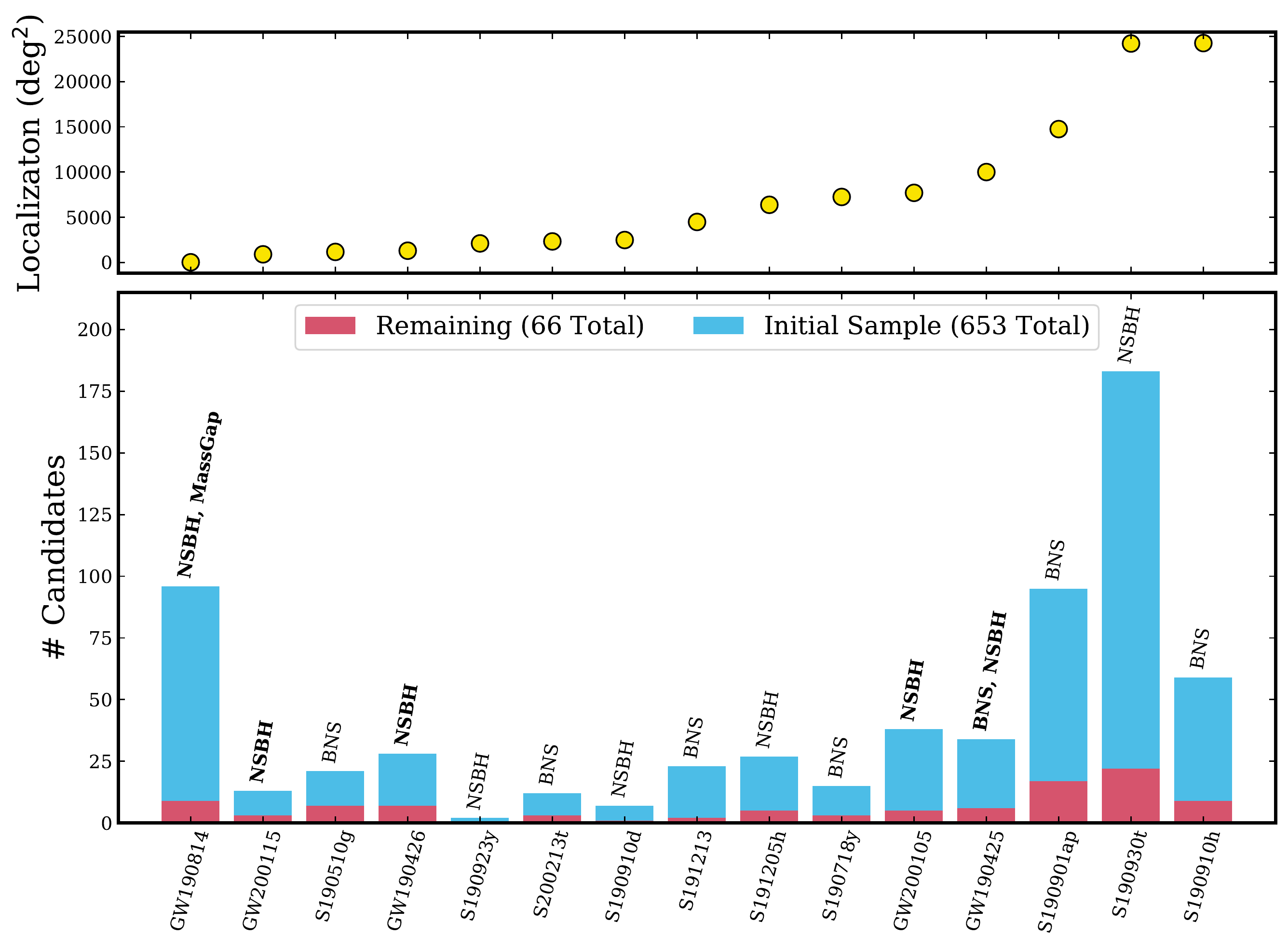}
\caption{\textbf{Bottom}: A bar chart showing the initial number of candidates per event (blue) and the remaining number (red) per event after our critical analysis discussed in Sections~\ref{sec:3} and \ref{sec:4}. The events are sorted by the 90\% confidence level localization size, shown on the \textbf{top}. The most likely event classifications are labeled at the top of each bar, with labels in bold showing candidates included in GWTC-2 or GWTC-3 \citep{GWTC2,GWTC-3}. With the exception of GW190814, the best localized and most followed-up event of O3, the initial and remaining sample sizes generally scale with the localization size.}
\label{fig:bar_chart_final}
\end{figure*}

\subsection{Analysis of Photometric Light Curves}
\label{subsec:lcbehavior}

Finally, we analyze the light curves of the remaining candidates and rule out those whose behaviors are inconsistent with predictions for kilonovae, even accounting for the most luminous kilonova models. We again employ photometric data gathered from TNS, the public ZTF stream, ATLAS forced photometry, and the SAGUARO database (cf., Section~\ref{sec:pre_expl_dets}).

Other O3 EM follow-up works analyzed light curves on a case-by-case basis, eliminating candidates whose light curves are flat \citep{Ackley+20} or decline slower than 0.3 mag day$^{-1}$ \citep{Kasliwal+20}. \citet{CowperthwaiteBerger15} established cuts of ($i-z$) $>$ 0.4~mag and $\delta t_{\rm{rise}} < 4$~days (where $\delta t_{\rm{rise}}$ is defined as the time it takes for a transient to rise from 1~mag pre-peak to peak in $z$-band) to discriminate between kilonovae, SNe Ia, other fast-evolving transients (e.g., \citealt{Drout+14,Margutti+19,Coppejans+20}) and other contaminants. 

We elect a more conservative approach grounded in the predicted luminosities and time evolution of several diverse kilonova models. The observed colors and luminosities of kilonovae are predicted to vary with both extrinsic (e.g., viewing angle) and intrinsic (primarily, the ejecta mass and composition) properties. Kilonovae viewed at an edge-on (pole-on) angle are expected to be redder (bluer) due to the lanthanide-rich tidal tails produced during the merger \citep{kasen+15,Chase+21,Korobkin+21}. The ejecta mass and composition are directly affected by the progenitor (BNS or NSBH) and remnant (e.g., BH, short-lived rotationally-supported NS, or magnetar) types. In particular, a bluer, slow-decaying kilonova is a predicted product of a long-lived NS or magnetar remnant due to the effects of neutrino irradiation unbinding the disk material surrounding the remnant \citep{metzgerfernandez14,kasen+15,Lippuner+17,Fong+21}. Taking into account predicted diversity, we consider an AT\,2017gfo-like model \citep{kasen+17}, a fiducial NSBH kilonova model \citep{Kawaguchi+20}, a stable NS remnant kilonova model \citep{kasen+15}, a model of the kilonova from massive progenitors \citep{Barbieri+20}, and a kilonova model for a magnetar remnant \citep{Fong+21}. We find that none rise in luminosity beyond $\delta t \approx 4$~days in a single band and thus employ this as our first criteria to eliminate candidates as kilonovae. We explain our use of detections in multiple filters below. In light of bluer kilonova models \citep{kasen+15}, the early, blue peak of AT\,2017gfo, and the lack of color measurements for most candidates, we do not employ any color criteria in eliminating candidates.

Eleven candidates show rising behavior in a single filter after $\delta t = 4$~days, and thus are not viable kilonovae. An additional three candidate light curves show nearly constant magnitudes in a single filter over periods of $\sim3$-9~days, inconsistent with any kilonova models. When enough data points were available, we find 13 candidate light curves decline by $\lesssim 1.5$~mag day$^{-1}$ over periods of 5 to 25 days in a single filter, again inconsistent with AT\,2017gfo or any kilonova model. We further eliminate five candidates whose light curves decline by $<1.5$~mag over 13 -- 18-day periods in ``adjacent'' (e.g., $g-$ and $o$-band) optical filters. Similarly unlike any expected kilonova, we eliminate one candidate that brightens over 6 days in adjacent filters. The majority of the remaining light curves contain only one detection, preventing us from making any conclusions about their behavior. In all, we determine 33 candidates are not kilonovae based on their light curves.

After applying all tools at our disposal, including those available in real time and those available $\sim$weeks after the merger, 66 candidates remain viable kilonovae. Our initial sample of \totcan is a fairly comprehensive compilation of candidates to O3 NS mergers, and this work eliminates 587 (90\%) of these candidates as kilonovae.

\begin{figure*}
\centering
\includegraphics[width=.8\textwidth]{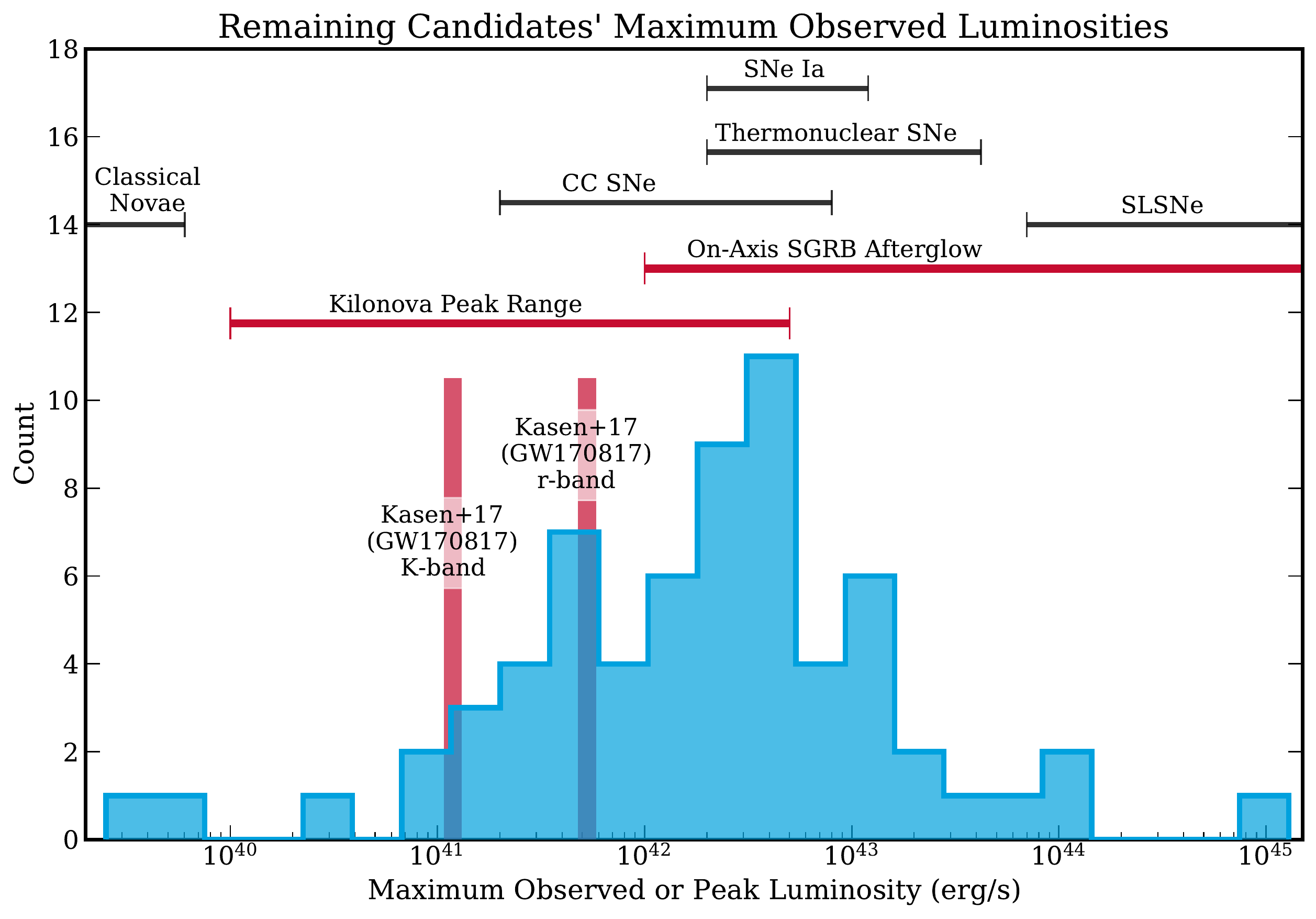}
\caption{Histogram of the peak luminosities of 66 kilonova candidates still considered viable after our analysis. Luminosities are calculated using the host distance if known from TNS or our host analysis (Section~\ref{subsec:host_dists}), and the GW event distance otherwise. Horizontal bars at the top of the figure demonstrate the range of peak luminosities of other optical transients (from \citealt{Bildsten+07,Darbha+10,Shen+10,Li+11,Kasliwal12,Cenko17}). The average luminosity of remaining candidates is $2.4 \times 10^{42}$ erg s$^{-1}$, on the upper end of the kilonova peak range, and consistent with the peak luminosities of short GRB afterglows and core-collapse, Type Ia, and thermonuclear supernovae.
\label{fig:lum_hist}}
\end{figure*}

\begin{figure*}
\centering
\includegraphics[width=.8\textwidth]{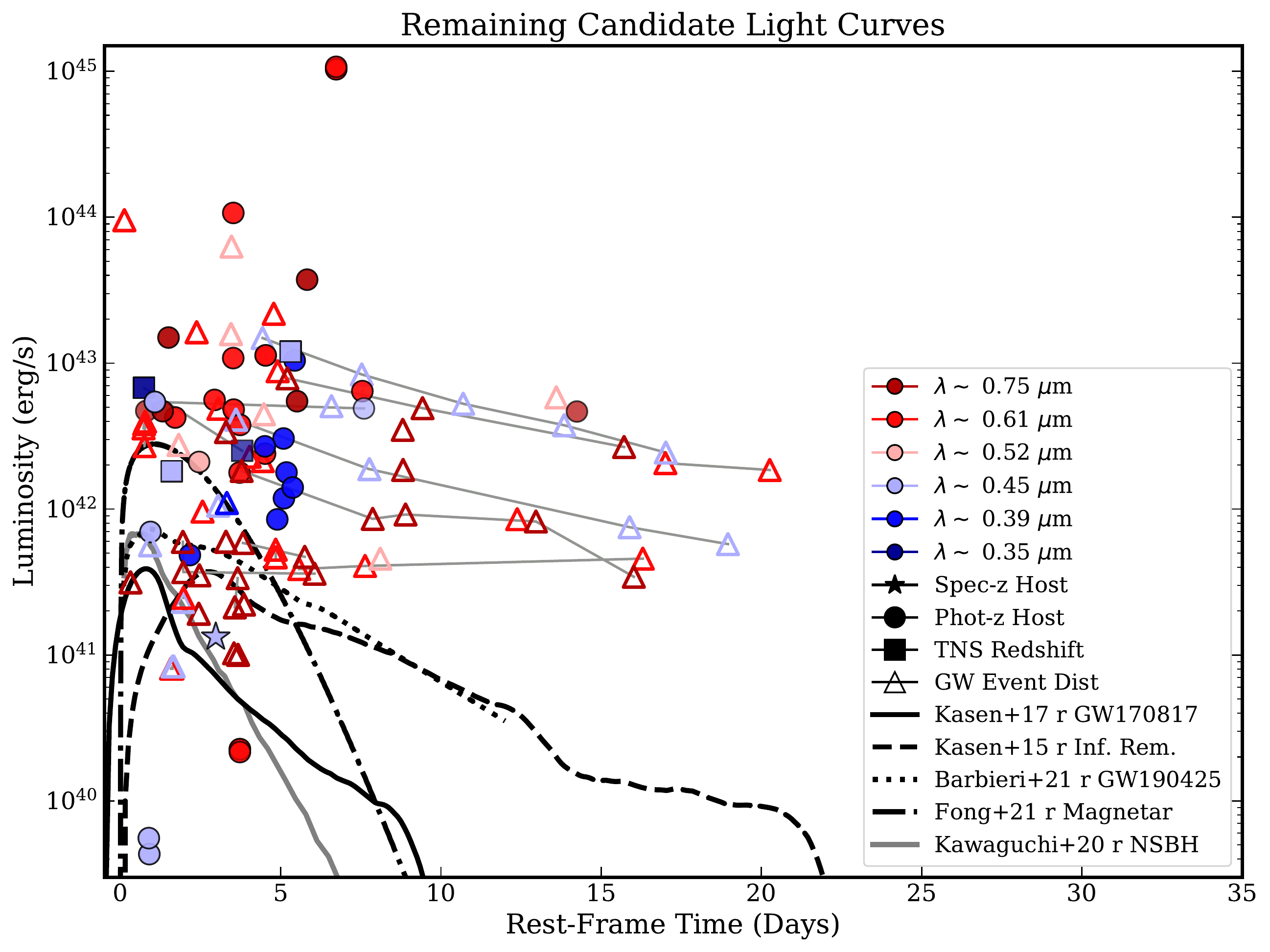}
\caption{Light curves of the remaining viable kilonova candidates after our vetting analysis in Sections~\ref{sec:3} and \ref{sec:4}. We build light curves using photometry from the ATLAS forced photometry tool, public ZTF data, TNS, and the SAGUARO database. We correct all photometry for Milky Way extinction and convert the observations to luminosity and rest-frame time using the TNS-reported redshift, the redshift of the associated galaxy (we average the distances if multiple are available; Section~\ref{subsec:host_dists}) or, if none is available, the GW event distance. Marker symbols denote the source of the distance and marker colors show the filter pivot wavelength. We also plot a diverse set of $r$-band ($\lambda \sim 0.61~\mu$m) kilonova models, includng those of an AT\,2017gfo-like kilonova (black solid line; \citealt{kasen+17}), a fiducial kilonova from a NSBH merger (grey solid line; \citealt{Kawaguchi+20}), an infinite-lifetime NS remnant kilonova (dashed line; \citealt{kasen+15}), a kilonova modeled for the event GW190425 (dotted line; \citealt{Barbieri+20}), and a magnetar-boosted kilonova (dashed-dotted line; \citealt{Fong+21}). \label{fig:lum_lcs}}
\end{figure*}

\section{Discussion}
\label{sec:discussion}

We now explore the results and implications of our analysis. We first determine if there is sufficient evidence to claim any of the remaining viable candidates as a real kilonova counterpart. The remaining viable candidates and their properties are summarized in Table~\ref{tab:remaining}. Next, we examine sources of redundancy in candidate follow-up, and make recommendations for improvements in future observing runs. Finally, we discuss the results of this analysis in the context of previous work and practices by the EM community.

\subsection{Aggregate Analysis of the Remaining Viable Candidates}
\label{subsec:remaining}

We now analyze the available information for the 66 remaining candidates still considered viable kilonovae after our vetting procedures, including their maximum observed luminosities and time evolution. We do not find convincing evidence that any of the remaining candidates are indeed kilonovae (though we cannot eliminate them using available information).

We show the initial and remaining number of candidates for each event, ordered by increasing localization area, in Figure~\ref{fig:bar_chart_final}. With the exception of GW190814, the three events with the largest localization areas have the highest numbers of remaining and initial candidates, as expected. Barring GW190814 and the three events with the largest localizations, the number of candidates (both initial and remaining) per event is roughly consistent. Notably, despite being the most precisely-localized event in the sample, GW190814 has the second greatest number of initial kilonova candidates (96). This indicates that well-localized, distant ($\gtrsim 200$ Mpc) events with a high astrophysical probability discovered in O4 will likely receive wide, deep coverage by the EM community, resulting in a much higher number of candidates per square degree than the average event. This is further demonstrated by the higher fraction of GCN-reported candidates for GW190814 compared to most other events (Figure~\ref{fig:cand_event_RO}).

In an effort to determine if our remaining candidates are either plausible kilonovae or clear contaminants, we next compare our sample in aggregate to the landscape of known transients. We first calculate the maximum observed luminosities ($\nu L_{\nu, {\rm max}}$) of remaining candidates based on the brightest observation in their photometric light curve and their host distance (determined based on TNS, redshift surveys, or the median GW event distance). Figure~\ref{fig:lum_hist} shows a histogram of $\nu L_{\nu, {\rm max}}$ values for the remaining candidates in blue. The peak $r$- and $K$-band luminosity of AT\,2017gfo, calculated based on best-fit models from \citet{kasen+17}, are shown as red vertical bars. The expected peak luminosity range of plausible kilonovae \citep{barneskasen13,tanaka+13,metzgerfernandez14,kasen+15,metzger19,Fong+21} and short GRB afterglows (based on observations from \citealt{fong+15}) are shown with red horizontal bars. The peak luminosity ranges of possible contaminating transients (including novae and various supernovae types) are overlaid in black bars at the top. Broadly, a majority of the remaining candidates' $\nu L_{\nu, {\rm max}}$ are consistent with either the expected kilonova peak range or the on-axis short GRB afterglow range. However, we note that no coincident short GRBs were detected during O3, despite at least partial coverage of all events in our sample by either the \textit{Swift Observatory} or \textit{Fermi Space Telescope}. In addition, many of the maximum observed luminosities of remaining candidates are consistent with that of core-collapse, Type Ia, thermonuclear or superluminous supernovae (Figure~\ref{fig:lum_hist}).

In Figure~\ref{fig:lum_lcs}, we plot the remaining candidates' light curves using all available information. We calculate luminosities and rest-frame times using the host redshift reported to TNS (points marked as squares), the spectroscopic (stars) or photometric (circles) redshift of a host found in Section~\ref{subsec:host_dists} (when multiple photometric redshifts are available, we take a weighted average) or the GW event distance (open triangles). As the majority of photometric points are in the optical (filter pivot wavelengths are denoted by the marker colors), we overplot five diverse kilonova models in the $r$-band. 

Figure~\ref{fig:lum_lcs} shows that the majority of light curves are more luminous than the most optimistic kilonova model plotted, although we caution that the large errors of the GW event distance and photometric redshifts (including some that are consistent with $z=0$) imply that the actual luminosity may vary widely. Additionally, 43 (71.7\%) of remaining candidates' light curves include only a single detection, indicating we have no knowledge of their behavior over time and their true peak luminosity may be much higher than what we observe. The majority of candidate light curves observed beyond $\sim5$~days decline more slowly than all models except that of the stable NS remnant \citep{kasen+15}. However, all of the slow-decaying light curves are $\sim$10 times more luminous than the model at the respective times. In general, the light curves of candidates which fall within the luminosity range of the plotted kilonova models consist of a single detection, preventing a comparison of their fading behavior or colors to the kilonova models. In addition, as demonstrated by the large fraction of triangle symbols in Figure~\ref{fig:lum_lcs}, we are unable to associate the majority (53.3\%) of the remaining candidates with a host galaxy catalogued in SDSS, LS DR9 or PS1-STRM. Finally, we note the lack of multiple bands of photometry (and thus color information) for nearly all candidates at early times, when color is a useful tool to distinguish kilonovae from other contaminants \citep{CowperthwaiteBerger15}.

Ultimately, we do not find sufficient evidence in the photometric light curves to claim that any of the remaining viable candidates is a true kilonova. Future observations of kilonovae from blind searches \citep{Doctor+17,Kasliwal+17,Smartt+17,Yang+17,Andreoni+20_ztfkn,Andreoni+21}, follow-up to short GRBs (see compilation studies of \citealt{Gompertz+18,Ascenzi+19,Rossi+20,Rastinejad+21}), or mergers detected by gravitational waves during O4 will further our knowledge on the acceptable range of kilonova luminosities, colors and decay timescales. 

\begin{figure}
\centering
\includegraphics[width=.5\textwidth]{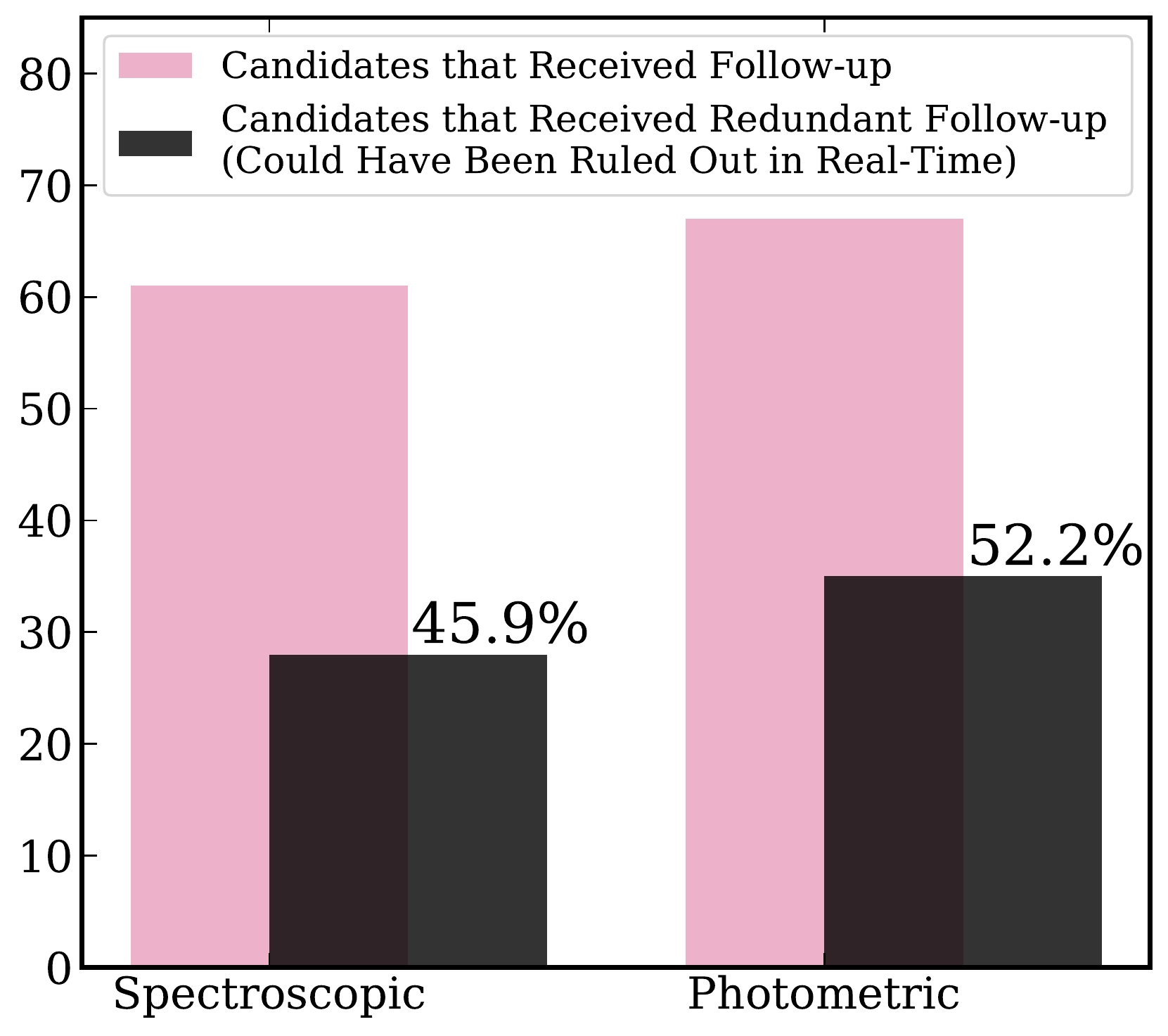}
\caption{Bar chart representing the number of candidates with spectroscopic and photometric follow-up reported to the GCNs and Literature (pink), and the number of these that we rule out with tools available in real time (black).
\label{fig:redundant}}
\end{figure}

\subsection{Examining Redundancy in Optical Candidate Follow-up}
\label{subsec:redund_followup}

With the aim of making recommendations to improve candidate follow-up in future observing runs, we investigate two sources of redundancy. First, we contrast the use of targeted spectroscopic and photometric follow-up of O3 kilonova candidates with eliminations made using the real-time tools we consider in Section~\ref{sec:3}. Second, we examine how often multiple spectra of a single object were reported to the GCNs and discuss methods to reduce the number of redundant observations during O4.

Both spectroscopic and multi-band imaging follow-up are time-intensive and expensive, and are best reserved for promising candidates one cannot eliminate with other methods. Generally, the GCNs and literature are a reflection of real-time follow-up in connection to GW events. However, many relevant follow-up observations are only reported in the literature (which is often not published until months following the event), implying that relying on real-time tools and reports is key. Sixty-one candidates in our initial sample of \totcan were followed up spectroscopically. Of these, 15 could have been eliminated based on pre-merger detections (although we note that the ATLAS forced photometry tool, of which the majority of these eliminations were based on, was not publicly available during O3). Moreover, an additional 13 could have been eliminated based on inconsistencies between the GW and host galaxy distances, all of which are ``highly confident'' host associations. Twelve eliminations were made based on photometric redshifts, while one was made based on a spectroscopic redshift. Our analysis utilizes the 95\% GW distance credible interval to eliminate candidates based on their host galaxy redshift. In cases where there are fewer candidate counterparts, follow-up groups may elect to use a wider interval. Though photometric redshifts are often inaccurate at low distances, they should still be used to prioritize follow-up spectra. We find that $\approx 46\%$ of candidates which received spectroscopic follow-up observations could have instead benefited from real-time archival information, making follow-up observations somewhat redundant (Figure~\ref{fig:redundant}). 

In a similar vein, sixty-seven candidates had photometric follow-up reported. Of these, one was a match to a star in PS1, 13 have pre-merger detections, and 21 have associated host galaxies with photometric redshifts outside of the GW event distance range. Thus, $\approx 51\%$ of candidates with reported photometric follow-up could have benefited from archival information. In Figure~\ref{fig:redundant} we summarize the number of candidates with GCN or literature-reported spectra (left) or photometry (right) in pink. The black bars show the fraction of this follow-up that is redundant or could have been avoided by employing all the real-time tools described in Section~\ref{sec:3}. 

We next examine another potential source of redundancy in follow-up resources: cases in which multiple follow-up spectra are taken of a single candidate. After examining the GCNs, we find seven initially promising candidates for which more than 3 spectra were reported including: AT\,2019dzk (SNIIn; 4 spectra taken), AT\,2019dzw (SNII; 5), AT\,2019ebq (Dust-reddened SNIb/c; 8), AT\,2019wqj (SNII; 3), AT\,2019wxt (SNIIb, 7), AT\,2020cja (blue, featureless continuum; 3), ZTFabvizsw (CV; 3). In general, all redundant spectra of a given candidate were obtained before any definitive classification results were reported. Notably, the majority of the candidates with multiple spectra were Type II supernovae. For AT\,2019ebq, the candidate with the most spectra taken, a near-IR spectrum was necessary to classify the dust-reddened supernova \citep{GCN2019ebq_Subaru,GCN2019ebq_Keck,GCN2019ebq_LBT,GCN2019ebq_Gemini,GCN2019ebq_Keck2}. Three out of seven of these candidates were potential counterparts to the NS merger GW190425. Generally, these well-followed candidates met some combination of the following criteria: (i) reported early in the search ($\delta t \lesssim$ 1~day), (ii) showed red colors and (iii) were associated with a host consistent with the GW distance. Overall, we find that obtaining multiple spectra of the same source was not a large sink on resources and was most notable for events of high interest (GW190425 and GW190814). Indeed, if the candidate was the true counterpart, high cadence early spectra would be essential for characterizing the kilonova (and would be critical for a comparison to the early, blue component of AT\,2017gfo; \citealt{Andreoni+17,Arcavi+17,Coulter+17,Cowperthwaite+17,Drout+17,Evans+17,Hu+17,Lipunov+17,Pian+17,Shappee+17,Smartt+17,Valenti+17,villar+17}) and its overall color evolution.

We note here several reasons for redundancy in follow-up and lessons learned. First, several of the tools that provide essential archival information were not available in O3, such as the ATLAS forced photometry tool and the LS DR9 photometric redshift catalog. The next era of EM follow-up to GW events will greatly benefit from the use of such tools. Additionally, we utilize the most updated GW localizations and distances in our analysis, which are generally closer to final values, thus allowing us to eliminate candidates that might have been viable kilonovae at the time of follow-up. For example, the photometric redshift of the host we associate with AT\,2019dzk in Section~\ref{subsec:host_dists} is inconsistent with the updated GW190425 distance but was consistent with the GW distance reported at the time follow-up spectra were taken. We note that some groups may choose to follow candidates which exhibit color and fading behavior similar to AT\,2017gfo, even if the photometric redshift is inconsistent (especially given that they become less robust for $z\lesssim0.1$). Further, we calculate the number of GCN and TNS candidates per event that meet our initial criteria (Section~\ref{sec:gcn_tns_cand}) using the localization maps first announced in the GCNs. We compare these numbers to the those found using the final localizations. Predictably, for events with large final localizations (e.g., S190901ap, S190930t) the number of candidates does not change significantly using the preliminary maps. However, for half the events in our sample, the number of candidates increases by $30-140\%$ when the preliminary localizations are utilized (GW190425, GW190426, GW190814, S190910d, GW200105, GW200115, S200213t), indicating the power of prompt localization updates in reducing the strain on follow-up resources. Together, these points highlight the importance of real-time updates from the GW community to the EM community, specifically, of localization maps and event distances. 

Overall, an examination of follow-up redundancies highlights the need for implementation of real-time tools that leverage archival information, improved organization of the community's follow-up plans and results, and updated key GW parameters (such as distance, localization maps, component masses and mass ratios) in future observing runs. Potential ways to assist community organization include reducing and reporting spectra as quickly as possible, reporting intended follow-up to the GCNs, or the creation of a database to report planned observations (similar to the Gravitational Wave Treasure Map, which aids the community by posting pointings; \citealt{Wyatt+20}), .

\subsection{Comparison to Other Kilonova Candidate Studies}
\label{subsec:53}

Finally, for context we comment on the methods used in our analysis compared to those employed by other GW follow-up groups. Many works focused on the candidates reported in their own searches, with the exception of some works which analyzed all publicly-reported candidates for a given event (e.g., \citealt{Hosseinzadeh+19,Kilpatrick+21}), and often the original samples of candidates were defined differently. For instance, \citet{Kilpatrick+21} took a more conservative approach in their study of GW190814 by including candidates reported to TNS within 14 days of the merger and within the 99\% localization. Their sample included 214 candidates, over twice the size of our initial GW190814 candidate sample. As we are analyzing a larger sample of GW events, we utilize the 90\% localization contour, a stricter cut for $\delta t$ motivated by kilonova light curve models, and make a preliminary luminosity cut (Section~\ref{sec:sample_coll}). 

We find that cross-matching to stellar catalogs, such as PS1, \textit{Gaia}, 2MASS, SDSS, USNO-B and DES, is a fairly ubiquitous practice in GW optical candidate vetting. In our analysis, the PS1 point source classification eliminated the greatest number of candidates (38) of the stellar catalogs that we considered. Cross-matching to AGN was not always standard practice. Two examples of strategies include cross-matching to the MILLIQUAS catalog \citep{Flesch15} and identifying AGN-like colors with Wide-field Infrared Survey Explorer (WISE; \citealt{WISE_Wright+10}) observations in combination with a nuclear position \citep{Ackley+20,Kasliwal+20}. Our analysis found 26 quasars by cross-matching (Section~\ref{sec:qso}), 20 of which also had pre-merger detections, indicating that the majority of quasars could be eliminated using the methods outlined in Section~\ref{sec:pre_expl_dets}.

The largest variation in candidate vetting is observed in the methods employed to associate candidates to host galaxies and the use of photometric or spectroscopic redshifts. Databases or catalogs used include the NASA/IPAC Extragalactic Database (NED), the Census of the Local Universe (CLU; \citealt{Cook+19}), Galaxy List for the Advanced Detector Era (GLADE; \citealt{Dalya+18}), LS DR9, PS1-STRM and DES Y3 (\citealt{Hartley+21}; \citealt{Ackley+20,Antier+20a,Kasliwal+17,Kilpatrick+21,Morgan+20,Paterson+21}). We find that SDSS DR12, PS1-STRM and LS DR9 footprints combined cover $>$97\% of kilonova candidates queried (Section~\ref{subsec:host_dists}). Precise association methods varied between searching web interfaces and determining hosts by eye \citep{Ackley+20}, calculating the galaxy with the minimum projected offset \citep{Kilpatrick+21}, and calculating probability based on angular offset and redshift (following the prescriptions of \citealt{Singer+16_Supp,Morgan+20}).  In our analysis, the SDSS, PS1 and LS DR9 redshift catalogs are the most effective tool at eliminating candidates, although we note that at low distances, photometric redshifts are not always robust. Future spectroscopic redshift surveys (e.g., DESI, Subaru; \citealt{Takada+14,DESI16}) will be valuable tools for eliminating candidates.

Photometric detections were also frequently used in the literature to eliminate candidates based on their light curves, although it was not clear if this is standard practice in real time. Some works examined their own datastreams (e.g., GRAWITA, ZTF, SAGUARO/CSS) and some utilized public datastreams (e.g, the PS1 Detection catalog, TNS, ZTF, the VISTA archive; \citealt{Ackley+20,Kasliwal+20,Kilpatrick+21,Paterson+21}). Others obtained follow-up observations at later times to examine their candidates' late-time light curves \citep{Morgan+20}. This tool was the second most effective  at eliminating candidates in our analysis. Looking forward, we recommend that surveys make these available as they promise to be an invaluable tool in eliminating candidates in real time, especially as the number of detected NS mergers grows in subsequent years. Eventually, publicly-available, deep, multi-band observations from Vera C.\ Rubin Observatory will transform the search for pre-explosion detections, at least in the southern hemisphere. 

Overall, we find that several of the steps we apply in Section~\ref{sec:3} are not standard practice amongst the O3 literature and can signficantly improve community candidate vetting in future observing runs. The expansion of new surveys and resources will also certainly enhance the candidate vetting process in future observing runs.

\section{Conclusion \& Future Prospects}
\label{sec:conclusion}

We analyze \totcan optical candidate counterparts to 15 O3 GW mergers involving at least one NS using a combination of information available at the time of the GW event (``real-time'') and that available days to weeks afterward. The number of candidates in our initial sample per event roughly scales with the size of the 90\% c.l.\ localization. The notable exception to this is GW190814, the best-localized multi-messenger prospect in O3, indicating that similarly well-localized O4 events will also result in large numbers of candidates.  At the conclusion of our analysis, we find that only 66 candidates remain viable, none of which have sufficient information to be claimed as a real kilonova. We also review the GCNs and literature and make recommendations for avoiding redundant observations to classify a candidate in O4. Our main conclusions are as follows:

\begin{enumerate}
    \item Employment of the real-time tools (including pre-merger detections and cross-matching with catalogs) which use archival information eliminated 65\% of the original candidate sample as viable kilonovae. In particular, pre-merger detections in public surveys account for $>20\%$ of eliminations alone, and 15 of these still received follow-up observations. Availability and incorporation of these tools into follow-up of future GW events will allow the community to focus limited follow-up resources and reduce redundancy.
    \item The most effective real-time tool at eliminating candidates as viable kilonovae was association to host galaxies in public photometric or spectroscopic redshift surveys outside the 95\% GW event distance. The combination of PS1-STRM, SDSS and LS DR9 covered the footprint of $>$97\% of candidates queried. Future spectroscopic redshift surveys will increase the robustness of host galaxy redshifts. Meanwhile, photometric redshifts are an important tool for prioritizing classification resources.
    \item At the conclusion of our analysis, 66 candidates remain viable as kilonovae, although the majority have insufficient information to be considered otherwise (single data point, unidentified host). The remaining candidates with light curves and redshifts would be particularly luminous if they were kilonovae, although given the diversity of model luminosities, cannot be confidently eliminated as such.
    \item Increased collaboration and transparency between and within the GW and EM communities would facilitate the search for EM counterparts. For instance, tools that can reduce redundancy or increase transparency among EM follow-up groups should be more widely adopted. Moreover, the prompt release of updated localization maps and distance measurements in particular would reduce the number of kilonova candidates that pass initial vetting, and updated component masses would help to prioritize the use of limited follow-up resources. 
\end{enumerate}

Looking forward, the larger volumes probed by GW detectors makes the issue of candidate contamination increasingly urgent. It is imperative to take advantage of any available tool that leverages the wealth of existing or follow-up data, as well as build tools which facilitate community follow-up \citep{Wyatt+20,SciMMA,TACH}. 

In tandem with observational strides, theoretical works predict a wide diversity in the timescales, colors and peak luminosities of kilonovae (e.g., \citealt{lipaczynski98,metzgerfernandez14,Lippuner+17,Shibata+19,Kawaguchi+20b}). With concurrent GW observations, it will be possible to connect each kilonova and its $r$-process abundance to the observed population of NSs and BHs. Each successive GW observing run has brought new and exciting discoveries; the methods presented in this work along with many other developments in GW-EM astronomy set the stage for novel, multi-messenger revelations in forthcoming observing runs. 

\facilities{SO:1.5m}

\software{astropy \citep{Astropy2013,Astropy2018}, Numpy \citep{numpy},
The IDL Astronomy User's Library \citep{IDLforever}, SCAMP \citep{scamp,scamp2}, SWarp \citep{swarp}, IRAF \citep{Tody86,Tody93},
          SExtractor \citep{Bertin1996}, ZOGY (\url{https://github.com/pmvreeswijk/ZOGY})
          }

\startlongtable
\begin{deluxetable*}{crrcccccCccCr}
\tabletypesize{\footnotesize}
\centering
\savetablenum{3}
\tablecolumns{13}
\tabcolsep0.02in
\tablecaption{Remaining Viable Kilonova Candidates \& Their Properties
\label{tab:remaining}}
\tablehead {
\colhead{Name} &
\colhead{RA} &
\colhead{Dec} &
\colhead{Source} & 
\colhead{Event} &
\colhead{Event Dist} &
\colhead{$\delta$t$^{\dagger}$} & 
\colhead{Mag.$^{\dagger}$} & 
\colhead{Filt.$^{\dagger}$} & 
% \colhead{Prob} & 
\colhead{Host$^{\ddagger}$} & 
\colhead{Host $z$} &
\colhead{$\nu L_{\nu}$} &
\colhead{Notes} \\
&
\colhead{(Deg)} &
\colhead{(Deg)} &
&
& 
\colhead{(Mpc)} &
\colhead{(Days)} & 
\colhead{(AB Mag)} & 
\colhead{} & 
% \colhead{(\%)} & 
&
\colhead{} &
\colhead{(erg/s)} &
\colhead{} 
}
\startdata
AT\,2019eig & 9.92875 & -31.99236 & Both & GW190425 & 157$\pm$70 & 4.29 & 18.5 & G & N &  & 2.1 \times 10^{42} & $^{b}$ \\
AT\,2019efr & 246.73323 & 10.93689 & TNS & GW190425 & 157$\pm$70 & 2.10 & 20.6 & w & S & 0.035$^{+0.124}_{-0.027}$ & 4.6 \times 10^{41} & $^{b}$ \\
AT\,2019frd & 180.68045 & -15.08255 & TNS & GW190426 & 377$\pm$160 & 4.69 & 20.7 & w & S & 0.086$^{+0.006}_{-0.006}$ & 4.5 \times 10^{41} & $^{b}$ \\
AT\,2019fxg & 165.75082 & -7.19200 & TNS & GW190426 & 377$\pm$160 & 4.62 & 22.1 & w & S & 0.102$^{+0.009}_{-0.006}$ & 1.2 \times 10^{41} &  \\
AT\,2019ioy & 172.08144 & -11.04771 & TNS & GW190426 & 377$\pm$160 & 4.68 & 21.8 & w & S & 0.107$^{+0.036}_{-0.036}$ & 1.7 \times 10^{41} & $^{b}$ \\
AT\,2019jry & 173.97281 & -12.71867 & TNS & GW190426 & 377$\pm$160 & 4.68 & 20.8 & w & S & 0.161$^{+0.048}_{-0.048}$ & 4.0 \times 10^{41} & $^{b}$ \\
AT\,2019jsv & 172.73276 & -11.73411 & TNS & GW190426 & 377$\pm$160 & 4.68 & 20.6 & w & S & 0.045$^{+0.022}_{-0.022}$ & 5.0 \times 10^{41} & $^{b}$ \\
 &  &  &  &  &  &  &  &  & S & 0.108$^{+0.011}_{-0.011}$ & 6.9 \times 10^{41} &  \\
DG19ouub & 171.47329 & -9.48849 & GCN & GW190426 & 377$\pm$160 & 0.33 & 21.6 & z & S & 0.133$^{+0.036}_{-0.032}$ & 8.2 \times 10^{40} &  \\
 &  &  &  &  &  &  &  &  & S & 0.104$^{+0.035}_{-0.035}$ & 4.7 \times 10^{41} &  \\
DG19zdwb & 167.29677 & -2.26828 & GCN & GW190426 & 377$\pm$160 & 0.49 & 22.0 & z & S & 0.231$^{+0.046}_{-0.043}$ & 5.7 \times 10^{40} & $^{c}$ \\
 &  &  &  &  &  &  &  &  & S & 0.213$^{+0.089}_{-0.089}$ & 3.3 \times 10^{41} &  \\
AT\,2019fhw & 92.78350 & -18.11934 & TNS & S190510g & 227$\pm$92 & 1.64 & 18.4 & G & S & 0.044$^{+0.009}_{-0.009}$ & 2.5 \times 10^{42} & $^{b}$ \\
AT\,2019flq & 88.20863 & -30.38138 & TNS & S190510g & 227$\pm$92 & 0.88 & 21.6 & g & S & 0.071$^{+0.008}_{-0.011}$ & 1.6 \times 10^{41} & $^{b}$ \\
AT\,2019fng & 89.21019 & -38.86941 & TNS & S190510g & 227$\pm$92 & 0.90 & 21.6 & g & S & 0.006$^{+0.010}_{-0.002}$ & 1.6 \times 10^{41} & $^{b}$ \\
AT\,2019fnr & 92.02175 & -35.88393 & TNS & S190510g & 227$\pm$92 & 0.88 & 21.1 & g & N &  & 5.2 \times 10^{41} & $^{b}$ \\
AT\,2019fnu & 90.41562 & -31.13027 & TNS & S190510g & 227$\pm$92 & 0.88 & 21.0 & g & S & 0.005$^{+0.002}_{-0.002}$ & 2.6 \times 10^{41} & $^{b}$ \\
desgw-190510f & 92.29446 & -34.88468 & GCN & S190510g & 227$\pm$92 & 0.86 & 21.3 & r & N &  & 3.3 \times 10^{41} & $^{b}$ \\
desgw-190510g & 92.46892 & -34.08657 & GCN & S190510g & 227$\pm$92 & 0.86 & 21.9 & r & N &  & 1.9 \times 10^{41} & $^{b}$ \\
AT\,2019lro & 346.89671 & -4.51006 & TNS & S190718y & 227$\pm$165 & 4.55 & 16.8 & G & N &  & 2.1 \times 10^{43} & $^{b}$ \\
AT\,2019nvb & 11.71320 & -25.42759 & TNS & GW190814 & 241$\pm$50 & 3.66 & 21.7 & z & B & 0.896$^{+0.288}_{-0.186}$ & 7.4 \times 10^{40} & $^{a,b}$ \\
 &  &  &  &  &  &  &  &  & B & 0.285$^{+0.238}_{-0.238}$ & 4.3 \times 10^{41} &  \\
AT\,2019nut & 10.44606 & -24.30040 & TNS & GW190814 & 241$\pm$50 & 3.39 & 21.7 & i & N &  & 2.1 \times 10^{41} & $^{a,b}$ \\
AT\,2019nxd & 10.68582 & -24.95565 & Both & GW190814 & 241$\pm$50 & 2.32 & 21.8 & i & N &  & 1.9 \times 10^{41} & $^{a}$ \\
AT\,2019pnr & 37.51050 & -58.24664 & TNS & S190901ap & 241$\pm$79 & 3.58 & 19.0 & G & S & 0.039$^{+0.018}_{-0.016}$ & 1.4 \times 10^{42} & $^{b}$ \\
AT\,2019pns & 57.21954 & -55.42570 & TNS & S190901ap & 241$\pm$79 & 2.82 & 17.9 & G & S & 0.042$^{+0.009}_{-0.005}$ & 3.9 \times 10^{42} & $^{b}$ \\
AT\,2019pqc & 43.59650 & -53.69043 & TNS & S190901ap & 241$\pm$79 & 4.32 & 18.9 & G & S & 0.044$^{+0.021}_{-0.013}$ & 1.5 \times 10^{42} & $^{b}$ \\
AT\,2019pqe & 279.59279 & 8.11615 & Both & S190901ap & 241$\pm$79 & 2.90 & 18.9 & G & N &  & 3.7 \times 10^{42} &  \\
AT\,2019pqr & 33.56754 & -61.82602 & TNS & S190901ap & 241$\pm$79 & 2.83 & 18.8 & G & S & 0.245$^{+0.663}_{-0.214}$ & 1.7 \times 10^{42} & $^{b}$ \\
AT\,2019pqw & 39.78525 & -56.65426 & TNS & S190901ap & 241$\pm$79 & 3.83 & 19.4 & G & N &  & 2.2 \times 10^{42} & $^{b}$ \\
AT\,2019pra & 274.56783 & 6.03930 & Both & S190901ap & 241$\pm$79 & 4.66 & 18.1 & G & N &  & 7.2 \times 10^{42} &  \\
AT\,2019psg & 40.96500 & -58.29228 & TNS & S190901ap & 241$\pm$79 & 3.33 & 18.9 & G & S & 0.063$^{+0.012}_{-0.012}$ & 1.4 \times 10^{42} & $^{b}$ \\
AT\,2019qkd & 67.96767 & -68.60806 & TNS & S190901ap & 241$\pm$79 & 2.34 & 21.2 & I & N &  & 3.3 \times 10^{41} & $^{b}$ \\
AT\,2019qle & 357.17813 & -60.35636 & TNS & S190901ap & 241$\pm$79 & 3.59 & 18.2 & G & S & 0.040$^{+0.014}_{-0.009}$ & 3.0 \times 10^{42} & $^{b}$ \\
AT\,2019rfa & 52.91621 & -54.71872 & TNS & S190901ap & 241$\pm$79 & 3.32 & 18.0 & G & S & 0.060$^{+0.018}_{-0.014}$ & 3.6 \times 10^{42} & $^{b}$ \\
AT\,2019yca & 5.74979 & -69.57372 & TNS & S190901ap & 241$\pm$79 & 0.30 & 21.3 & I & N &  & 3.1 \times 10^{41} & $^{b}$ \\
AT\,2019qci & 45.86922 & 9.48785 & TNS & S190910h & 230$\pm$88 & 4.12 & 19.2 & r & B & 0.099$^{+0.033}_{-0.033}$ & 1.1 \times 10^{42} &  \\
AT\,2019qcy & 280.67417 & -31.35883 & TNS & S190910h & 230$\pm$88 & 2.26 & 17.3 & G & N &  & 1.4 \times 10^{43} &  \\
AT\,2019qko & 36.14908 & 32.09166 & TNS & S190910h & 230$\pm$88 & 1.15 & 20.6 & i & S & 0.151$^{+0.020}_{-0.016}$ & 2.4 \times 10^{41} & $^{b}$ \\
 &  &  &  &  &  &  &  &  & S & 0.119$^{+0.030}_{-0.030}$ & 1.4 \times 10^{42} &  \\
AT\,2019qlc & 45.81797 & -13.07055 & TNS & S190910h & 230$\pm$88 & 1.20 & 20.8 & i & S & 0.253$^{+0.052}_{-0.034}$ & 2.1 \times 10^{41} & $^{b}$ \\
 &  &  &  &  &  &  &  &  & S & 0.513$^{+0.686}_{-0.686}$ & 1.2 \times 10^{42} &  \\
AT\,2019aabc & 160.33252 & 71.95680 & TNS & S190930t & 108$\pm$38 & 1.90 & 20.4 & g & N &  & 2.2 \times 10^{41} & $^{b}$ \\
AT\,2019aabl & 321.84832 & 9.10231 & TNS & S190930t & 108$\pm$38 & 1.54 & 21.3 & r & N &  & 7.8e \times 10^{40} &  \\
AT\,2019aaif & 303.94083 & -7.92903 & TNS & S190930t & 108$\pm$38 & 2.50 & 18.6 & G & N &  & 9.2 \times 10^{41} &  \\
AT\,2019rqt & 296.30119 & -55.34081 & TNS & S190930t & 108$\pm$38 & 1.57 & 17.1 & g & N &  & 4.7 \times 10^{42} & $^{b}$ \\
 &  &  &  &  &  &  &  &  & T & 0.015 & 5.7 \times 10^{43} &  \\
AT\,2019rst & 77.63845 & 40.07212 & TNS & S190930t & 108$\pm$38 & 1.90 & 19.4 & o & N &  & 3.9 \times 10^{41} &  \\
AT\,2019rvn & 355.74622 & -6.35195 & TNS & S190930t & 108$\pm$38 & 3.69 & 19.4 & r & B & 0.826$^{+0.866}_{-0.866}$ & 8.9 \times 10^{41} &  \\
AT\,2019rvo & 13.25551 & -13.49827 & TNS & S190930t & 108$\pm$38 & 3.71 & 19.2 & r & B & 0.005$^{+0.010}_{-0.002}$ & 1.1 \times 10^{42} &  \\
AT\,2019rwi & 334.09021 & -36.42172 & TNS & S190930t & 108$\pm$38 & 4.37 & 17.0 & Clear & N &  & 4.4 \times 10^{42} & $^{b}$ \\
AT\,2019rwj & 357.49801 & -20.71066 & TNS & S190930t & 108$\pm$38 & 4.73 & 19.3 & r & N &  & 4.6 \times 10^{41} &  \\
AT\,2019sbk & 341.82542 & -58.24744 & Both & S190930t & 108$\pm$38 & 0.70 & 18.8 & U & S & 0.036$^{+0.008}_{-0.006}$ & 2.8 \times 10^{42} &  \\
 &  &  &  &  &  &  &  &  & T & 0.054 & 2.8 \times 10^{42} &  \\
AT\,2019sim & 286.57788 & -8.17877 & TNS & S190930t & 108$\pm$38 & 3.52 & 17.6 & g & N &  & 2.9 \times 10^{42} &  \\
AT\,2019tkf & 7.67617 & -69.20519 & TNS & S190930t & 108$\pm$38 & 3.58 & 20.8 & I & N &  & 9.9 \times 10^{40} & $^{b}$ \\
AT\,2019vvl & 355.48451 & 12.57208 & TNS & S190930t & 108$\pm$38 & 4.68 & 21.0 & i & S & 0.319$^{+0.155}_{-0.104}$ & 1.7 \times 10^{41} &  \\
 &  &  &  &  &  &  &  &  & S & 0.106$^{+0.093}_{-0.093}$ & 1.0 \times 10^{42}&  \\
AT\,2019xnq & 75.30739 & -6.31813 & TNS & S190930t & 108$\pm$38 & 5.00 & 21.2 & w & B & 0.087$^{+0.016}_{-0.018}$ & 2.7 \times 10^{41} &  \\
 &  &  &  &  &  &  &  &  & B & 0.040$^{+0.030}_{-0.030}$ & 1.6 \times 10^{42}&  \\
M205329.99+224421.2 & 313.37496 & 22.73922 & GCN & S190930t & 108$\pm$38 & 0.29 & 18.1 & Clear & N &  & 1.6 \times 10^{42} & $^{b}$ \\
AT\,2019wjb & 150.80922 & 25.28472 & TNS & S191205ah & 385$\pm$164 & 4.64 & 20.2 & g & S & 0.202$^{+0.028}_{-0.022}$ & 5.9 \times 10^{41} &  \\
 &  &  &  &  &  &  &  &  & S & 0.145$^{+0.040}_{-0.040}$ & 3.4 \times 10^{42}&  \\
 &  &  &  &  &  &  &  &  & T & 0.145 & 3.4 \times 10^{42}&  \\
AT\,2019zwe & 142.84725 & -52.77048 & TNS & S191205ah & 385$\pm$164 & 0.11 & 18.2 & G & N &  & 1.7 \times 10^{43} &  \\
AT\,2019xkk & 36.66231 & 33.81442 & TNS & S191213g & 201$\pm$81 & 3.15 & 18.4 & i & N &  & 3.2 \times 10^{42}&  \\
AT\,2019yjg & 33.36393 & 33.84112 & TNS & S191213g & 201$\pm$81 & 3.15 & 20.2 & i & N &  & 5.7 \times 10^{41} & $^{b}$ \\
AT\,2020aqx & 219.21566 & 29.75517 & TNS & GW200105 & 280$\pm$110 & 2.87 & 21.8 & g & P & 0.035 & 1.3 \times 10^{41} & $^{b}$ \\
AT\,2020bnv & 219.73419 & 46.68873 & TNS & GW200105 & 280$\pm$110 & 2.87 & 20.8 & g & N &  & 1.0 \times 10^{42}& $^{b}$ \\
AT\,2020dzt & 117.25912 & 12.49079 & TNS & GW200105 & 280$\pm$110 & 0.71 & 19.5 & r & N &  & 2.6 \times 10^{42}&  \\
AT\,2020qk & 117.70990 & 11.86319 & TNS & GW200105 & 280$\pm$110 & 0.68 & 19.2 & r & N &  & 3.5 \times 10^{42}&  \\
AT\,2020rz & 42.22675 & -18.34336 & TNS & GW200105 & 280$\pm$110 & 3.25 & 17.7 & Clear & N &  & 1.5 \times 10^{43} &  \\
AT\,2020ajz & 42.25907 & 4.98657 & Both & GW200115 & 300$\pm$100 & 3.11 & 21.2 & w & N &  & 9.8 \times 10^{41} & $^{b}$ \\
AT\,2020akb & 47.52255 & 5.98790 & Both & GW200115 & 300$\pm$100 & 4.14 & 21.2 & w & S & 0.089$^{+0.016}_{-0.016}$ & 2.8 \times 10^{41} & $^{b}$ \\
AT\,2020cph & 70.03650 & -65.21731 & TNS & S200213t & 201$\pm$80 & 1.74 & 18.9 & Clear & N &  & 2.6 \times 10^{42}&  \\
AT\,2020cqi & 10.70342 & 41.31194 & TNS & S200213t & 201$\pm$80 & 2.37 & 19.2 & Clear & S & 0.034$^{+0.034}_{-0.034}$ & 1.2 \times 10^{42}& $^{b}$ \\
AT\,2020cxw & 34.37870 & 22.29686 & TNS & S200213t & 201$\pm$80 & 0.99 & 19.9 & g & S & 0.051$^{+0.019}_{-0.036}$ & 7.3 \times 10^{41} &  \\
 &  &  &  &  &  &  &  &  & S & 0.091$^{+0.008}_{-0.008}$ & 4.2 \times 10^{42} &  \\
\enddata
\tablecomments{$^\dagger$Time, magnitude and filter of discovery. \\
$^{\ddagger}$Host Galaxy Association Class, where P, G, S and B are abbreviations for the Platinum, Gold, Silver and Bronze classes (as defined in Table~\ref{tab:gal_categories}), respectively. T indicates a redshift reported to TNS, and N indicates no host galaxy can be confidently associated with the candidate. \\
$^{a}$ \citet{Kilpatrick+21} eliminate candidate based on an inconsistent host distance. As we employ different methods of eliminating candidates with host galaxy redshifts (cf. Section~\ref{subsec:host_dists}), this candidate remains viable in our analysis. \\
$^{b}$ Candidate's light curve (based on photometry gathered from TNS, ATLAS, ZTF and SAGUARO, further described in Section~\ref{sec:pre_expl_dets}) consists of a single detection. \\
$^{c}$ Time of discovery is not included in reporting GCN and object is not included in reporting group's published work summarizing candidates from this event \citep{GCN_DECam190426,Goldstein+19}. We approximate its discovery time with those of candidates reported in the literature by the same discovery group.}
\end{deluxetable*}

\begin{acknowledgments}

The authors gratefully acknowledge Charlie Kilpatrick, Chase Kimball, Zoheyr Doctor, Adam Miller, Stefano Valenti, Stephanie Juneau and Dustin Lang for their helpful discussions in preparing this manuscript. SAGUARO is supported by the National Science Foundation (NSF) under Award Nos.\ AST-1909358 and AST-1908972.

Time domain research by D.J.S. is supported by NSF grants AST-1821987, 1813466 \& 2108032, and by the Heising-Simons Foundation under grant \#2020-1864. The Fong Group at Northwestern acknowledges support by the National Science Foundation under grant Nos. AST-1814782 and CAREER grant No. AST-2047919. W.F. gratefully acknowledges support by the David and Lucile Packard Foundation. S. Y. has been supported by the research project grant “Understanding the Dynamic Universe” funded by the Knut and Alice Wallenberg Foundation under Dnr KAW 2018.0067, and the G.R.E.A.T research environment, funded by {\em Vetenskapsr\aa det}, the Swedish Research Council, project number 2016-06012, and the Wenner-Gren Foundations.

The operation of the facilities of Steward Observatory are supported in part by the state of Arizona.

This work has made use of data from the European Space Agency (ESA) mission {\it Gaia} (\url{https://www.cosmos.esa.int/gaia}), processed by the {\it Gaia} Data Processing and Analysis Consortium (DPAC, \url{https://www.cosmos.esa.int/web/gaia/dpac/consortium}). Funding for the DPAC has been provided by national institutions, in particular the institutions participating in the {\it Gaia} Multilateral Agreement.

This research has made use of data and/or services provided by the International Astronomical Union's Minor Planet Center.

This work has made use of data from the Asteroid Terrestrial-impact Last Alert System (ATLAS) project. The Asteroid Terrestrial-impact Last Alert System (ATLAS) project is primarily funded to search for near earth asteroids through NASA grants NN12AR55G, 80NSSC18K0284, and 80NSSC18K1575; byproducts of the NEO search include images and catalogs from the survey area. This work was partially funded by Kepler/K2 grant J1944/80NSSC19K0112 and HST GO-15889, and STFC grants ST/T000198/1 and ST/S006109/1. The ATLAS science products have been made possible through the contributions of the University of Hawaii Institute for Astronomy, the Queen’s University Belfast, the Space Telescope Science Institute, the South African Astronomical Observatory, and The Millennium Institute of Astrophysics (MAS), Chile.

The Legacy Surveys consist of three individual and complementary projects: the Dark Energy Camera Legacy Survey (DECaLS; Proposal ID \#2014B-0404; PIs: David Schlegel and Arjun Dey), the Beijing-Arizona Sky Survey (BASS; NOAO Prop. ID \#2015A-0801; PIs: Zhou Xu and Xiaohui Fan), and the Mayall z-band Legacy Survey (MzLS; Prop. ID \#2016A-0453; PI: Arjun Dey). DECaLS, BASS and MzLS together include data obtained, respectively, at the Blanco telescope, Cerro Tololo Inter-American Observatory, NSF's NOIRLab; the Bok telescope, Steward Observatory, University of Arizona; and the Mayall telescope, Kitt Peak National Observatory, NOIRLab. The Legacy Surveys project is honored to be permitted to conduct astronomical research on Iolkam Du'ag (Kitt Peak), a mountain with particular significance to the Tohono O'odham Nation.

NOIRLab is operated by the Association of Universities for Research in Astronomy (AURA) under a cooperative agreement with the National Science Foundation.

This project used data obtained with the Dark Energy Camera (DECam), which was constructed by the Dark Energy Survey (DES) collaboration. Funding for the DES Projects has been provided by the U.S. Department of Energy, the U.S. National Science Foundation, the Ministry of Science and Education of Spain, the Science and Technology Facilities Council of the United Kingdom, the Higher Education Funding Council for England, the National Center for Supercomputing Applications at the University of Illinois at Urbana-Champaign, the Kavli Institute of Cosmological Physics at the University of Chicago, Center for Cosmology and Astro-Particle Physics at the Ohio State University, the Mitchell Institute for Fundamental Physics and Astronomy at Texas A\&M University, Financiadora de Estudos e Projetos, Fundacao Carlos Chagas Filho de Amparo, Financiadora de Estudos e Projetos, Fundacao Carlos Chagas Filho de Amparo a Pesquisa do Estado do Rio de Janeiro, Conselho Nacional de Desenvolvimento Cientifico e Tecnologico and the Ministerio da Ciencia, Tecnologia e Inovacao, the Deutsche Forschungsgemeinschaft and the Collaborating Institutions in the Dark Energy Survey. The Collaborating Institutions are Argonne National Laboratory, the University of California at Santa Cruz, the University of Cambridge, Centro de Investigaciones Energeticas, Medioambientales y Tecnologicas-Madrid, the University of Chicago, University College London, the DES-Brazil Consortium, the University of Edinburgh, the Eidgenossische Technische Hochschule (ETH) Zurich, Fermi National Accelerator Laboratory, the University of Illinois at Urbana-Champaign, the Institut de Ciencies de l'Espai (IEEC/CSIC), the Institut de Fisica d'Altes Energies, Lawrence Berkeley National Laboratory, the Ludwig Maximilians Universitat Munchen and the associated Excellence Cluster Universe, the University of Michigan, NSF's NOIRLab, the University of Nottingham, the Ohio State University, the University of Pennsylvania, the University of Portsmouth, SLAC National Accelerator Laboratory, Stanford University, the University of Sussex, and Texas A\&M University.

BASS is a key project of the Telescope Access Program (TAP), which has been funded by the National Astronomical Observatories of China, the Chinese Academy of Sciences (the Strategic Priority Research Program ``The Emergence of Cosmological Structures'' Grant \# XDB09000000), and the Special Fund for Astronomy from the Ministry of Finance. The BASS is also supported by the External Cooperation Program of Chinese Academy of Sciences (Grant \# 114A11KYSB20160057), and Chinese National Natural Science Foundation (Grant \# 11433005).

The Legacy Survey team makes use of data products from the Near-Earth Object Wide-field Infrared Survey Explorer (NEOWISE), which is a project of the Jet Propulsion Laboratory/California Institute of Technology. NEOWISE is funded by the National Aeronautics and Space Administration.

The Legacy Surveys imaging of the DESI footprint is supported by the Director, Office of Science, Office of High Energy Physics of the U.S. Department of Energy under Contract No. DE-AC02-05CH1123, by the National Energy Research Scientific Computing Center, a DOE Office of Science User Facility under the same contract; and by the U.S. National Science Foundation, Division of Astronomical Sciences under Contract No. AST-0950945 to NOAO.
\end{acknowledgments}

\bibliographystyle{aasjournal}
\bibliography{refs}

\begin{thebibliography}{}
\expandafter\ifx\csname natexlab\endcsname\relax\def\natexlab#1{#1}\fi
\providecommand{\url}[1]{\href{#1}{#1}}
\providecommand{\dodoi}[1]{doi:~\href{http://doi.org/#1}{\nolinkurl{#1}}}
\providecommand{\doeprint}[1]{\href{http://ascl.net/#1}{\nolinkurl{http://ascl.net/#1}}}
\providecommand{\doarXiv}[1]{\href{https://arxiv.org/abs/#1}{\nolinkurl{https://arxiv.org/abs/#1}}}

\bibitem[{{Abbott} {et~al.}(2016{\natexlab{a}}){Abbott}, {Abbott}, {Abbott},
  {Abernathy}, {Acernese}, {Ackley}, {Adams}, {Adams}, {Addesso}, {Adhikari},
  {Adya}, {Affeldt}, {Agathos}, {Agatsuma}, {Aggarwal}, {Aguiar}, {Aiello},
  {Ain}, {Ajith}, {Allen}, {Allocca}, {Altin}, {Anderson}, {Anderson}, {Arai},
  {Arain}, {Araya}, {Arceneaux}, {Areeda}, {Arnaud}, {Arun}, {Ascenzi}, {LIGO
  Scientific Collaboration}, \& {Virgo Collaboration}}]{LVC16}
{Abbott}, B.~P., {Abbott}, R., {Abbott}, T.~D., {et~al.} 2016{\natexlab{a}},
  \prl, 116, 061102, \dodoi{10.1103/PhysRevLett.116.061102}

\bibitem[{{Abbott} {et~al.}(2016{\natexlab{b}}){Abbott}, {Abbott}, {Abbott},
  {Abernathy}, {Acernese}, {Ackley}, {Adams}, {Adams}, {Addesso}, {Adhikari},
  \& et~al.}]{lvc_2016_LRR}
---. 2016{\natexlab{b}}, Living Reviews in Relativity, 19,
  \dodoi{10.1007/lrr-2016-1}

\bibitem[{{Abbott} {et~al.}(2017{\natexlab{a}}){Abbott}, {Abbott}, {Abbott},
  {Acernese}, {Ackley}, {Adams}, {Adams}, {Addesso}, {Adhikari}, {Adya},
  {Affeldt}, {Afrough}, {Agarwal}, {Agathos}, {Agatsuma}, {Aggarwal}, {Aguiar},
  {Aiello}, \& et~al.}]{Abbott+17a}
---. 2017{\natexlab{a}}, \prl, 119, 161101,
  \dodoi{10.1103/PhysRevLett.119.161101}

\bibitem[{{Abbott} {et~al.}(2017{\natexlab{b}}){Abbott}, {Abbott}, {Abbott},
  {Acernese}, {Ackley}, {Adams}, {Adams}, {Addesso}, {Adhikari}, {Adya},
  {Affeldt}, {Afrough}, {Agarwal}, {Agathos}, {Agatsuma}, {Aggarwal}, {Aguiar},
  {Aiello}, {Ain}, {Ajith}, {Allen}, {Allen}, {Allocca}, {Altin}, {Amato},
  {Ananyeva}, \& {Anderson}}]{LVC_GW170817_H0}
---. 2017{\natexlab{b}}, \nat, 551, 85, \dodoi{10.1038/nature24471}

\bibitem[{{Abbott} {et~al.}(2019){Abbott}, {Abbott}, {Abbott}, {Abraham},
  {Acernese}, {Ackley}, {Adams}, {Adhikari}, {Adya}, {Affeldt}, {LIGO
  Scientific Collaboration}, \& {Virgo Collaboration}}]{GWTC-1}
---. 2019, Physical Review X, 9, 031040, \dodoi{10.1103/PhysRevX.9.031040}

\bibitem[{{Abbott} {et~al.}(2020{\natexlab{a}}){Abbott}, {Abbott}, {Abbott},
  {Abraham}, {Acernese}, {Ackley}, {Adams}, {Adhikari}, {Adya}, {Affeldt},
  {Agathos}, {Agatsuma}, {Aggarwal}, {Aguiar}, {Aiello}, {Ain}, {Ajith},
  {Allen}, {Allocca}, {Aloy}, {Altin}, {Amato}, {Anand}, {Ananyeva},
  {Anderson}, {Anderson}, {Angelova}, {Antier}, {Appert}, {Arai}, {Araya},
  {Areeda}, {Ar{\`e}ne}, {Arnaud}, {Aronson}, {Arun}, {Ascenzi}, {Ashton},
  {Aston}, {Astone}, {Aubin}, {Aufmuth}, {AultONeal}, {Austin}, {Avendano},
  {Avila-Alvarez}, {Babak}, {Bacon}, {Badaracco}, {Bader}, {Bae}, {Baird},
  {Baker}, {Baldaccini}, {Ballardin}, {Ballmer}, {Bals}, {Banagiri},
  {Barayoga}, {Barbieri}, {Barclay}, {Barish}, {Barker}, {Barkett}, {Barnum},
  {Barone}, {Barr}, {Barsotti}, {Barsuglia}, {Barta}, {Bartlett}, {Bartos},
  {Bassiri}, {Basti}, {Bawaj}, {Bayley}, {Baylor}, {Bazzan}, {B{\'e}csy},
  {Bejger}, {Belahcene}, {Bell}, {Beniwal}, {Benjamin}, {Berger}, {Bergmann},
  {Bernuzzi}, \& {Berry}}]{LVC_GW190425}
---. 2020{\natexlab{a}}, \apjl, 892, L3, \dodoi{10.3847/2041-8213/ab75f5}

\bibitem[{{Abbott} {et~al.}(2020{\natexlab{b}}){Abbott}, {Abbott}, {Abraham},
  {Acernese}, {Ackley}, {Adams}, {Adhikari}, {Adya}, {Affeldt}, {Agathos},
  {Agatsuma}, {Aggarwal}, {Aguiar}, {Aich}, {Aiello}, {Ain}, {Ajith}, {Akcay},
  {Allen}, {Allocca}, {Altin}, {Amato}, {Anand}, {Ananyeva}, {Anderson},
  {Anderson}, {Angelova}, {Ansoldi}, {Antier}, {Appert}, {Arai}, {Araya},
  {Areeda}, {Ar{\`e}ne}, {Arnaud}, {Aronson}, {Arun}, {Asali}, {Ascenzi},
  {Ashton}, {Aston}, {Astone}, {Aubin}, {Aufmuth}, {AultONeal}, {Austin},
  {Avendano}, {Babak}, {Bacon}, {Badaracco}, {Bader}, {Bae}, {Baer}, {Baird},
  {Baldaccini}, {Ballardin}, {Ballmer}, {Bals}, {Balsamo}, {Baltus},
  {Banagiri}, {Bankar}, {Bankar}, {Barayoga}, {Barbieri}, {Barish}, {Barker},
  \& {Barkett}}]{LVC_GW190814}
{Abbott}, R., {Abbott}, T.~D., {Abraham}, S., {et~al.} 2020{\natexlab{b}},
  \apjl, 896, L44, \dodoi{10.3847/2041-8213/ab960f}

\bibitem[{{Abbott} {et~al.}(2021{\natexlab{a}}){Abbott}, {Abbott}, {Abraham},
  {Acernese}, {Ackley}, {Adams}, {Adams}, {Adhikari}, {Adya}, {Affeldt},
  {Agathos}, {Agatsuma}, {Aggarwal}, {Aguiar}, {Aiello}, {Ain}, {Ajith},
  {Akcay}, {Allen}, {Allocca}, {Altin}, {Amato}, {Anand}, {Ananyeva},
  {Anderson}, {Anderson}, {Angelova}, {Ansoldi}, {Antelis}, {Antier}, {Appert},
  {Arai}, {Araya}, {Areeda}, {Ar{\`e}ne}, {Arnaud}, {Aronson}, {Arun}, {Asali},
  {Ascenzi}, {Ashton}, {Aston}, {Astone}, {Aubin}, {Aufmuth}, {AultONeal},
  {Austin}, {Avendano}, {Babak}, {Badaracco}, {Bader}, {Bae}, {Baer},
  {Bagnasco}, {Baird}, {Ball}, {Ballardin}, {Ballmer}, {Bals}, {Balsamo},
  {Baltus}, {Banagiri}, {Bankar}, {Bankar}, {Barayoga}, {Barbieri}, {Barish},
  {Barker}, {Barneo}, {Barnum}, {Barone}, {Barr}, {Barsotti}, {Barsuglia},
  {Barta}, {Bartlett}, {Bartos}, {Bassiri}, {Basti}, {Bawaj}, {Bayley},
  {Bazzan}, {Becher}, {B{\'e}csy}, {Bedakihale}, {Bejger}, {Belahcene},
  {Beniwal}, {Benjamin}, {Bennett}, {Bentley}, {Bergamin}, {Berger},
  {Bergmann}, {Bernuzzi}, {Berry}, {Bersanetti}, {Bertolini}, {Betzwieser},
  {Bhandare}, {Bhandari}, {Bhattacharjee}, {Bidler}, {Bilenko}, {Billingsley},
  {Birney}, {Birnholtz}, {Biscans}, {Bischi}, {Biscoveanu}, {Bisht}, {Bitossi},
  {Bizouard}, {Blackburn}, {Blackman}, {Blair}, {Blair}, {Blair}, {Blanch},
  {Bobba}, {Bode}, {Boer}, {Boetzel}, {Bogaert}, {Boldrini}, {Bondu},
  {Bonilla}, {Bonnand}, {Booker}, {Boom}, {Bork}, {Boschi}, {Bose},
  {Bossilkov}, {Boudart}, {Bouffanais}, {Bozzi}, {Bradaschia}, {Brady},
  {Bramley}, {Branchesi}, {Brau}, {Breschi}, {Briant}, {Briggs}, {Brighenti},
  {Brillet}, {Brinkmann}, {Brockill}, {Brooks}, {Brooks}, {Brown}, {Brunett},
  {Bruno}, {Bruntz}, {Buikema}, {Bulik}, {Bulten}, {Buonanno}, {Buscicchio},
  {Buskulic}, {Byer}, {Cabero}, {Cadonati}, {Caesar}, {Cagnoli}, {Cahillane},
  {Calder{\'o}n Bustillo}, {Callaghan}, {Callister}, {Calloni}, {Camp},
  {Canepa}, {Cannon}, {Cao}, {Cao}, {Carapella}, {Carbognani}, {Carney},
  {Carpinelli}, {Carullo}, {Carver}, {Casanueva Diaz}, {Casentini}, {Caudill},
  {Cavagli{\`a}}, {Cavalier}, {Cavalieri}, {Cella}, {Cerd{\'a}-Dur{\'a}n},
  {Cesarini}, {Chaibi}, {Chakravarti}, {Chan}, {Chan}, {Chandra}, {Chanial},
  {Chao}, {Charlton}, {Chase}, {Chassande-Mottin}, {Chatterjee},
  {Chattopadhyay}, {Chaturvedi}, {Chatziioannou}, {Chen}, {Chen}, {Chen},
  {Chen}, {Cheng}, {Cheong}, {Chia}, {Chiadini}, {Chierici}, {Chincarini},
  {Chiummo}, {Cho}, {Cho}, {Cho}, {Choate}, {Christensen}, {Chu}, {Chua},
  {Chung}, {Chung}, {Ciani}, {Ciecielag}, {Cie{\'s}lar}, {Cifaldi}, {Ciobanu},
  {Ciolfi}, {Cipriano}, {Cirone}, {Clara}, {Clark}, {Clark}, {Clarke},
  {Clearwater}, {Clesse}, {Cleva}, {Coccia}, {Cohadon}, {Cohen}, {Colleoni},
  {Collette}, {Collins}, {Colpi}, {Constancio}, {Conti}, {Cooper}, {Corban},
  {Corbitt}, {Cordero-Carri{\'o}n}, {Corezzi}, {Corley}, {Cornish}, {Corre},
  {Corsi}, {Cortese}, {Costa}, {Cotesta}, {Coughlin}, {Coughlin}, {Coulon},
  {Countryman}, {Cousins}, {Couvares}, {Covas}, {Coward}, {Cowart}, {Coyne},
  {Coyne}, {Creighton}, {Creighton}, {Croquette}, {Crowder}, {Cudell},
  {Cullen}, {Cumming}, {Cummings}, {Cunningham}, {Cuoco}, {Cury{\l}o},
  {Canton}, {D{\'a}lya}, {Dana}, {DaneshgaranBajastani}, {D'Angelo}, {Danila},
  {Danilishin}, {D'Antonio}, {Danzmann}, {Darsow-Fromm}, {Dasgupta}, {Datrier},
  {Dattilo}, {Dave}, {Davier}, {Davies}, {Davis}, {Daw}, {Dean}, {DeBra},
  {Deenadayalan}, {Degallaix}, {De Laurentis}, {Del{\'e}glise}, {Del Favero},
  {De Lillo}, {De Lillo}, {Del Pozzo}, {DeMarchi}, {De Matteis}, {D'Emilio},
  {Demos}, {Denker}, {Dent}, {Depasse}, {De Pietri}, {De Rosa}, {De Rossi},
  {DeSalvo}, {de Varona}, {Dhurandhar}, {D{\'\i}az}, {Diaz-Ortiz}, {Didio},
  {Dietrich}, {Di Fiore}, {DiFronzo}, {Di Giorgio}, {Di Giovanni}, {Di
  Giovanni}, {Di Girolamo}, {Di Lieto}, {Ding}, {Di Pace}, {Di Palma}, {Di
  Renzo}, {Divakarla}, {Dmitriev}, {Doctor}, {D'Onofrio}, {Donovan}, {Dooley},
  {Doravari}, {Dorrington}, {Downes}, {Drago}, {Driggers}, {Du}, {Ducoin},
  {Dupej}, {Durante}, {D'Urso}, {Duverne}, {Dwyer}, {Easter}, {Eddolls},
  {Edelman}, {Edo}, {Edy}, {Effler}, {Eichholz}, {Eikenberry}, {Eisenmann},
  {Eisenstein}, {Ejlli}, {Errico}, {Essick}, {Estell{\'e}s}, {Estevez},
  {Etienne}, {Etzel}, {Evans}, {Evans}, {Ewing}, {Fafone}, {Fair}, {Fairhurst},
  {Fan}, {Farah}, {Farinon}, {Farr}, {Farr}, {Fauchon-Jones}, {Favata}, {Fays},
  {Fazio}, {Feicht}, {Fejer}, {Feng}, {Fenyvesi}, {Ferguson},
  {Fernandez-Galiana}, {Ferrante}, {Ferreira}, {Fidecaro}, {Figura}, {Fiori},
  {Fiorucci}, {Fishbach}, {Fisher}, {Fishner}, {Fittipaldi}, {Fitz-Axen},
  {Fiumara}, {Flaminio}, {Floden}, {Flynn}, {Fong}, {Font}, {Forsyth},
  {Fournier}, {Frasca}, {Frasconi}, {Frei}, {Freise}, {Frey}, {Frey},
  {Fritschel}, {Frolov}, {Fronz{\'e}}, {Fulda}, {Fyffe}, {Gabbard}, {Gadre},
  {Gaebel}, {Gair}, {Gais}, {Galaudage}, {Gamba}, {Ganapathy}, {Ganguly},
  {Gaonkar}, {Garaventa}, {Garc{\'\i}a-Quir{\'o}s}, {Garufi}, {Gateley},
  {Gaudio}, {Gayathri}, {Gemme}, {Gennai}, {George}, {George}, {George},
  {Gergely}, {Ghonge}, {Ghosh}, {Ghosh}, {Ghosh}, {Giacomazzo}, {Giacoppo},
  {Giaime}, {Giardina}, {Gibson}, {Gier}, {Gill}, {Giri}, {Glanzer}, {Gleckl},
  {Godwin}, {Goetz}, {Goetz}, {Gohlke}, {Goncharov}, {Gonz{\'a}lez},
  {Gopakumar}, {Gossan}, {Gosselin}, {Gouaty}, {Grace}, {Grado}, {Granata},
  {Granata}, {Grant}, {Gras}, {Grassia}, {Gray}, {Gray}, {Greco}, {Green},
  {Green}, {Gretarsson}, {Griggs}, {Grignani}, {Grimaldi}, {Grimes}, {Grimm},
  {Grote}, {Grunewald}, {Gruning}, {Guerrero}, {Guidi}, {Guimaraes},
  {Guix{\'e}}, {Gulati}, {Guo}, {Gupta}, {Gupta}, {Gupta}, {Gustafson},
  {Gustafson}, {Guzman}, {Haegel}, {Halim}, {Hall}, {Hamilton}, {Hammond},
  {Haney}, {Hanke}, {Hanks}, {Hanna}, {Hannam}, {Hannuksela}, {Hannuksela},
  {Hansen}, {Hansen}, {Hanson}, {Harder}, {Hardwick}, {Haris}, {Harms},
  {Harry}, {Harry}, {Hartwig}, {Hasskew}, {Haster}, {Haughian}, {Hayes},
  {Healy}, {Heidmann}, {Heintze}, {Heinze}, {Heinzel}, {Heitmann}, {Hellman},
  {Hello}, {Helmling-Cornell}, {Hemming}, {Hendry}, {Heng}, {Hennes}, {Hennig},
  {Hennig}, {Hernandez Vivanco}, {Heurs}, {Hild}, {Hill}, {Hines}, {Hochheim},
  {Hofgard}, {Hofman}, {Hohmann}, {Holgado}, {Holland}, {Hollows}, {Holmes},
  {Holt}, {Holz}, {Hopkins}, {Horst}, {Hough}, {Howell}, {Hoy}, {Hoyland},
  {Huang}, {H{\"u}bner}, {Huddart}, {Huerta}, {Hughey}, {Hui}, {Husa},
  {Huttner}, {Hutzler}, {Huxford}, {Huynh-Dinh}, {Idzkowski}, {Iess},
  {Imperato}, {Inchauspe}, {Ingram}, {Intini}, {Isi}, {Iyer},
  {JaberianHamedan}, {Jacqmin}, {Jadhav}, {Jadhav}, {James}, {Jani},
  {Janssens}, {Janthalur}, {Jaranowski}, {Jariwala}, {Jaume}, {Jenkins},
  {Jeunon}, {Jiang}, {Johns}, {Johnson-McDaniel}, {Jones}, {Jones}, {Jones},
  {Jones}, {Jones}, {Jonker}, {Ju}, {Junker}, {Kalaghatgi}, {Kalogera},
  {Kamai}, {Kandhasamy}, {Kang}, {Kanner}, {Kapadia}, {Kapasi}, {Karathanasis},
  {Karki}, {Kashyap}, {Kasprzack}, {Kastaun}, {Katsanevas}, {Katsavounidis},
  {Katzman}, {Kawabe}, {K{\'e}f{\'e}lian}, {Keitel}, {Key}, {Khadka},
  {Khalili}, {Khan}, {Khan}, {Khazanov}, {Khetan}, {Khursheed}, {Kijbunchoo},
  {Kim}, {Kim}, {Kim}, {Kim}, {Kim}, {Kim}, {Kimball}, {King}, {Kinley-Hanlon},
  {Kirchhoff}, {Kissel}, {Kleybolte}, {Klimenko}, {Knowles}, {Knyazev}, {Koch},
  {Koehlenbeck}, {Koekoek}, {Koley}, {Kolstein}, {Komori}, {Kondrashov},
  {Kontos}, {Koper}, {Korobko}, {Korth}, {Kovalam}, {Kozak}, {Kr{\"a}mer},
  {Kringel}, {Krishnendu}, {Kr{\'o}lak}, {Kuehn}, {Kumar}, {Kumar}, {Kumar},
  {Kumar}, {Kuns}, {Kwang}, {Lackey}, {Laghi}, {Lalande}, {Lam}, {Lamberts},
  {Landry}, {Lane}, {Lang}, {Lange}, {Lantz}, {Lanza}, {La Rosa},
  {Lartaux-Vollard}, {Lasky}, {Laxen}, {Lazzarini}, {Lazzaro}, {Leaci},
  {Leavey}, {Lecoeuche}, {Lee}, {Lee}, {Lee}, {Lee}, {Lehmann}, {Leon},
  {Leroy}, {Letendre}, {Levin}, {Li}, {Li}, {Li}, {Li}, {Li}, {Linde},
  {Linker}, {Linley}, {Littenberg}, {Liu}, {Liu}, {Llorens-Monteagudo}, {Lo},
  {Lockwood}, {London}, {Longo}, {Lorenzini}, {Loriette}, {Lormand}, {Losurdo},
  {Lough}, {Lousto}, {Lovelace}, {L{\"u}ck}, {Lumaca}, {Lundgren}, {Ma},
  {Macas}, {MacInnis}, {Macleod}, {MacMillan}, {Macquet}, {Maga{\~n}a
  Hernandez}, {Maga{\~n}a-Sandoval}, {Magazz{\`u}}, {Magee}, {Majorana},
  {Maksimovic}, {Maliakal}, {Malik}, {Man}, {Mandic}, {Mangano}, {Mansell},
  {Manske}, {Mantovani}, {Mapelli}, {Marchesoni}, {Marion}, {M{\'a}rka},
  {M{\'a}rka}, {Markakis}, {Markosyan}, {Markowitz}, {Maros}, {Marquina},
  {Marsat}, {Martelli}, {Martin}, {Martin}, {Martinez}, {Martinez}, {Martynov},
  {Masalehdan}, {Mason}, {Massera}, {Masserot}, {Massinger}, {Masso-Reid},
  {Mastrogiovanni}, {Matas}, {Mateu-Lucena}, {Matichard}, {Matiushechkina},
  {Mavalvala}, {Maynard}, {McCann}, {McCarthy}, {McClelland}, {McCormick},
  {McCuller}, {McGuire}, {McIsaac}, {McIver}, {McManus}, {McRae}, {McWilliams},
  {Meacher}, {Meadors}, {Mehmet}, {Mehta}, {Melatos}, {Melchor}, {Mendell},
  {Menendez-Vazquez}, {Mercer}, {Mereni}, {Merfeld}, {Merilh}, {Merritt},
  {Merzougui}, {Meshkov}, {Messenger}, {Messick}, {Metzdorff}, {Meyers},
  {Meylahn}, {Mhaske}, {Miani}, {Miao}, {Michaloliakos}, {Michel}, {Middleton},
  {Milano}, {Miller}, {Millhouse}, {Mills}, {Milotti}, {Milovich-Goff},
  {Minazzoli}, {Minenkov}, {Mir}, {Mishkin}, {Mishra}, {Mistry}, {Mitra},
  {Mitrofanov}, {Mitselmakher}, {Mittleman}, {Mo}, {Mogushi}, {Mohapatra},
  {Mohite}, {Molina}, {Molina-Ruiz}, {Mondin}, {Montani}, {Moore}, {Moraru},
  {Morawski}, {Moreno}, {Morisaki}, {Mours}, {Mow-Lowry}, {Mozzon},
  {Muciaccia}, {Mukherjee}, {Mukherjee}, {Mukherjee}, {Mukherjee}, {Mukund},
  {Mullavey}, {Munch}, {Mu{\~n}iz}, {Murray}, {Nadji}, {Nagar}, {Nardecchia},
  {Naticchioni}, {Nayak}, {Neil}, {Neilson}, {Nelemans}, {Nelson}, {Nery},
  {Neunzert}, {Nitz}, {Ng}, {Ng}, {Nguyen}, {Nguyen}, {Nguyen}, {Nichols},
  {Nissanke}, {Nocera}, {Noh}, {North}, {Nothard}, {Nuttall}, {Oberling},
  {O'Brien}, {O'Dell}, {Oganesyan}, {Ogin}, {Oh}, {Oh}, {Ohme}, {Ohta},
  {Okada}, {Olivetto}, {Oppermann}, {Oram}, {O'Reilly}, {Ormiston}, {Ortega},
  {O'Shaughnessy}, {Ossokine}, {Osthelder}, {Ottaway}, {Overmier}, {Owen},
  {Pace}, {Pagano}, {Page}, {Pagliaroli}, {Pai}, {Pai}, {Palamos}, {Palashov},
  {Palomba}, {Pan}, {Panda}, {Pang}, {Pankow}, {Pannarale}, {Pant}, {Paoletti},
  {Paoli}, {Paolone}, {Parker}, {Pascucci}, {Pasqualetti}, {Passaquieti},
  {Passuello}, {Patel}, {Patricelli}, {Payne}, {Pechsiri}, {Pedraza},
  {Pegoraro}, {Pele}, {Penn}, {Perego}, {Perez}, {P{\'e}rigois}, {Perreca},
  {Perri{\`e}s}, {Petermann}, {Petterson}, {Pfeiffer}, {Pham}, {Phukon},
  {Piccinni}, {Pichot}, {Piendibene}, {Piergiovanni}, {Pierini}, {Pierro},
  {Pillant}, {Pilo}, {Pinard}, {Pinto}, {Piotrzkowski}, {Pirello}, {Pitkin},
  {Placidi}, {Plastino}, {Pluchar}, {Poggiani}, {Polini}, {Pong}, {Ponrathnam},
  {Popolizio}, {Porter}, {Poverman}, {Powell}, {Pracchia}, {Prajapati},
  {Prasai}, {Prasanna}, {Pratten}, {Prestegard}, {Principe}, {Prodi},
  {Prokhorov}, {Prosposito}, {Prudenzi}, {Puecher}, {Punturo}, {Puosi},
  {Puppo}, {P{\"u}rrer}, {Qi}, {Quetschke}, {Quinonez}, {Quitzow-James},
  {Raab}, {Raaijmakers}, {Radkins}, {Radulesco}, {Raffai}, {Rafferty}, {Rail},
  {Raja}, {Rajan}, {Rajbhandari}, {Rakhmanov}, {Ramirez}, {Ramirez},
  {Ramos-Buades}, {Rana}, {Rao}, {Rapagnani}, {Rapol}, {Ratto}, {Raymond},
  {Razzano}, {Read}, {Regimbau}, {Rei}, {Reid}, {Reitze}, {Rettegno}, {Ricci},
  {Richardson}, {Richardson}, {Richardson}, {Ricker}, {Riemenschneider},
  {Riles}, {Rizzo}, {Robertson}, {Robinet}, {Rocchi}, {Rocha}, {Rodriguez},
  {Rodriguez-Soto}, {Rolland}, {Rollins}, {Roma}, {Romanelli}, {Romano},
  {Romel}, {Romero}, {Romero-Shaw}, {Romie}, {Ronchini}, {Rose}, {Rose},
  {Rose}, {Rosell}, {Rosi{\'n}ska}, {Rosofsky}, {Ross}, {Rowan}, {Rowlinson},
  {Roy}, {Roy}, {Ruggi}, {Ryan}, {Sachdev}, {Sadecki}, {Sadiq},
  {Sakellariadou}, {Salafia}, {Salconi}, {Saleem}, {Samajdar}, {Sanchez},
  {Sanchez}, {Sanchez}, {Sanchis-Gual}, {Sanders}, {Sandles}, {Santiago},
  {Santos}, {Saravanan}, {Sarin}, {Sassolas}, {Sathyaprakash}, {Sauter},
  {Savage}, {Savant}, {Sawant}, {Sayah}, {Schaetzl}, {Schale}, {Scheel},
  {Scheuer}, {Schindler-Tyka}, {Schmidt}, {Schnabel}, {Schofield},
  {Sch{\"o}nbeck}, {Schreiber}, {Schulte}, {Schutz}, {Schwarm}, {Schwartz},
  {Scott}, {Scott}, {Seglar-Arroyo}, {Seidel}, {Sellers}, {Sengupta},
  {Sennett}, {Sentenac}, {Sequino}, {Sergeev}, {Setyawati}, {Shaffer},
  {Shahriar}, {Sharifi}, {Sharma}, {Sharma}, {Shawhan}, {Shen}, {Shikauchi},
  {Shink}, {Shoemaker}, {Shoemaker}, {Shukla}, {ShyamSundar}, {Sieniawska},
  {Sigg}, {Singer}, {Singh}, {Singh}, {Singha}, {Singhal}, {Sintes}, {Sipala},
  {Skliris}, {Slagmolen}, {Slaven-Blair}, {Smetana}, {Smith}, {Smith},
  {Somala}, {Son}, {Soni}, {Soni}, {Sorazu}, {Sordini}, {Sorrentino},
  {Sorrentino}, {Soulard}, {Souradeep}, {Sowell}, {Spencer}, {Spera},
  {Srivastava}, {Srivastava}, {Staats}, {Stachie}, {Steer}, {Steinhoff},
  {Steinke}, {Steinlechner}, {Steinlechner}, {Steinmeyer}, {Stevenson},
  {Stolle-McAllister}, {Stops}, {Stover}, {Strain}, {Stratta}, {Strunk},
  {Sturani}, {Stuver}, {S{\"u}dbeck}, {Sudhagar}, {Sudhir}, {Suh},
  {Summerscales}, {Sun}, {Sun}, {Sunil}, {Sur}, {Suresh}, {Sutton}, {Swinkels},
  {Szczepa{\'n}czyk}, {Tacca}, {Tait}, {Talbot}, {Tanasijczuk}, {Tanner},
  {Tao}, {Tapia}, {Tapia San Martin}, {Tasson}, {Taylor}, {Tenorio},
  {Terkowski}, {Thirugnanasambandam}, {Thomas}, {Thomas}, {Thomas}, {Thompson},
  {Thondapu}, {Thorne}, {Thrane}, {Tiwari}, {Tiwari}, {Tiwari}, {Toland},
  {Tolley}, {Tonelli}, {Tornasi}, {Torres-Forn{\'e}}, {Torrie}, {e Melo},
  {T{\"o}yr{\"a}}, {Tran}, {Trapananti}, {Travasso}, {Traylor}, {Tringali},
  {Tripathee}, {Trovato}, {Trudeau}, {Tsai}, {Tsang}, {Tse}, {Tso}, {Tsukada},
  {Tsuna}, {Tsutsui}, {Turconi}, {Ubhi}, {Udall}, {Ueno}, {Ugolini},
  {Unnikrishnan}, {Urban}, {Usman}, {Utina}, {Vahlbruch}, {Vajente}, {Vajpeyi},
  {Valdes}, {Valentini}, {Valsan}, {van Bakel}, {van Beuzekom}, {van den
  Brand}, {Van Den Broeck}, {Vander-Hyde}, {van der Schaaf}, {van Heijningen},
  {Vardaro}, {Vargas}, {Varma}, {Vass}, {Vas{\'u}th}, {Vecchio}, {Vedovato},
  {Veitch}, {Veitch}, {Venkateswara}, {Venneberg}, {Venugopalan}, {Verkindt},
  {Verma}, {Veske}, {Vetrano}, {Vicer{\'e}}, {Viets}, {Vijaykumar},
  {Villa-Ortega}, {Vinet}, {Vitale}, {Vo}, {Vocca}, {Vorvick}, {Vyatchanin},
  {Wade}, {Wade}, {Wade}, {Walet}, {Walker}, {Wallace}, {Wallace}, {Walsh},
  {Wang}, {Wang}, {Wang}, {Wang}, {Ward}, {Warner}, {Was}, {Washington},
  {Watchi}, {Weaver}, {Wei}, {Weinert}, {Weinstein}, {Weiss}, {Wellmann},
  {Wen}, {We{\ss}els}, {Westhouse}, {Wette}, {Whelan}, {White}, {White},
  {Whiting}, {Whittle}, {Wilken}, {Williams}, {Williams}, {Williamson},
  {Willis}, {Willke}, {Wilson}, {Wimmer}, {Winkler}, {Wipf}, {Woan}, {Woehler},
  {Wofford}, {Wong}, {Wrangel}, {Wright}, {Wu}, {Wysocki}, {Xiao}, {Yamamoto},
  {Yang}, {Yang}, {Yang}, {Yap}, {Yeeles}, {Yoon}, {Yu}, {Yu}, {Yuen},
  {Zadro{\.Z}ny}, {Zanolin}, {Zelenova}, {Zendri}, {Zevin}, {Zhang}, {Zhang},
  {Zhang}, {Zhang}, {Zhao}, {Zhao}, {Zheng}, {Zhou}, {Zhou}, {Zhu},
  {Zimmerman}, {Zlochower}, {Zucker}, {Zweizig}, {LIGO Scientific
  Collaboration}, \& {Virgo Collaboration}}]{GWTC2}
---. 2021{\natexlab{a}}, Physical Review X, 11, 021053,
  \dodoi{10.1103/PhysRevX.11.021053}

\bibitem[{{Abbott} {et~al.}(2021{\natexlab{b}}){Abbott}, {Abbott}, {Abraham},
  {Acernese}, {Ackley}, {Adams}, {Adams}, {Adhikari}, {Adya}, {Affeldt},
  {Agarwal}, {Agathos}, {Agatsuma}, {Aggarwal}, {Aguiar}, {Aiello}, {Ain},
  {Ajith}, {Akutsu}, {Aleman}, {Allen}, {Allocca}, {Altin}, {Amato}, {Anand},
  {Ananyeva}, {Anderson}, {Anderson}, {Ando}, {Ligo Scientific Collaboration},
  {VIRGO Collaboration}, \& {KAGRA Collaboration}}]{Abbott+21_NSBH}
---. 2021{\natexlab{b}}, \apjl, 915, L5, \dodoi{10.3847/2041-8213/ac082e}

\bibitem[{{Ackley} {et~al.}(2020){Ackley}, {Amati}, {Barbieri}, {Bauer},
  {Benetti}, {Bernardini}, {Bhirombhakdi}, {Botticella}, {Branchesi},
  {Brocato}, {Bruun}, {Bulla}, {Campana}, {Cappellaro}, {Castro-Tirado},
  {Chambers}, {Chaty}, {Chen}, {Ciolfi}, {Coleiro}, {Copperwheat}, {Covino},
  {Cutter}, {D'Ammando}, {D'Avanzo}, {De Cesare}, {D'Elia}, {Della Valle},
  {Denneau}, {De Pasquale}, {Dhillon}, {Dyer}, {Elias-Rosa}, {Evans},
  {Eyles-Ferris}, {Fiore}, {Fraser}, {Fruchter}, {Fynbo}, {Galbany}, {Gall},
  {Galloway}, {Getman}, {Ghirlanda}, {Gillanders}, {Gomboc}, {Gompertz},
  {Gonz{\'a}lez-Fern{\'a}ndez}, {Gonz{\'a}lez-Gait{\'a}n}, {Grado}, {Greco},
  {Gromadzki}, {Groot}, {Guti{\'e}rrez}, {Heikkil{\"a}}, {Heintz}, {Hjorth},
  {Hu}, {Huber}, {Inserra}, {Izzo}, {Japelj}, {Jerkstrand}, {Jin}, {Jonker},
  {Kankare}, {Kann}, {Kennedy}, {Kim}, {Klose}, {Kool}, {Kotak},
  {Kuncarayakti}, {Lamb}, {Leloudas}, {Levan}, {Longo}, {Lowe}, {Lyman},
  {Magnier}, {Maguire}, {Maiorano}, {Mandel}, {Mapelli}, {Mattila}, {McBrien},
  {Melandri}, {Micha{\l}owski}, {Milvang-Jensen}, {Moran}, {Nicastro},
  {Nicholl}, {Nicuesa Guelbenzu}, {Nuttal}, {Oates}, {O'Brien}, {Onori},
  {Palazzi}, {Patricelli}, {Perego}, {Torres}, {Perley}, {Pian}, {Pignata},
  {Piranomonte}, {Poshyachinda}, {Possenti}, {Pumo}, {Quirola-V{\'a}squez},
  {Ragosta}, {Ramsay}, {Rau}, {Rest}, {Reynolds}, {Rosetti}, {Rossi},
  {Rosswog}, {Sabha}, {Sagu{\'e}s Carracedo}, {Salafia}, {Salmon},
  {Salvaterra}, {Savaglio}, {Sbordone}, {Schady}, {Schipani}, {Schultz},
  {Schweyer}, {Smartt}, {Smith}, {Smith}, {Sollerman}, {Srivastav}, {Stanway},
  {Starling}, {Steeghs}, {Stratta}, {Stubbs}, {Tanvir}, {Testa}, {Thrane},
  {Tonry}, {Turatto}, {Ulaczyk}, {van der Horst}, {Vergani}, {Walton},
  {Watson}, {Wiersema}, {Wiik}, {Wyrzykowski}, {Yang}, {Yi}, \&
  {Young}}]{Ackley+20}
{Ackley}, K., {Amati}, L., {Barbieri}, C., {et~al.} 2020, \aap, 643, A113,
  \dodoi{10.1051/0004-6361/202037669}

\bibitem[{{Alam} {et~al.}(2015){Alam}, {Albareti}, {Allende Prieto}, {Anders},
  {Anderson}, {Anderton}, {Andrews}, {Armengaud}, {Aubourg}, {Bailey}, {Basu},
  {Bautista}, {Beaton}, {Beers}, {Bender}, {Berlind}, {Beutler}, {Bhardwaj},
  {Bird}, {Bizyaev}, {Blake}, {Blanton}, {Blomqvist}, {Bochanski}, {Bolton},
  {Bovy}, {Shelden Bradley}, {Brandt}, {Brauer}, {Brinkmann}, {Brown},
  {Brownstein}, {Burden}, {Burtin}, {Busca}, {Cai}, {Capozzi}, {Carnero
  Rosell}, {Carr}, {Carrera}, {Chambers}, {Chaplin}, {Chen}, {Chiappini},
  {Chojnowski}, {Chuang}, {Clerc}, {Comparat}, {Covey}, {Croft}, {Cuesta},
  {Cunha}, {da Costa}, {Da Rio}, {Davenport}, {Dawson}, {De Lee}, {Delubac},
  {Deshpande}, {Dhital}, {Dutra-Ferreira}, {Dwelly}, {Ealet}, {Ebelke},
  {Edmondson}, {Eisenstein}, {Ellsworth}, {Elsworth}, {Epstein}, {Eracleous},
  {Escoffier}, {Esposito}, {Evans}, {Fan}, {Fern{\'a}ndez-Alvar}, {Feuillet},
  {Filiz Ak}, {Finley}, {Finoguenov}, {Flaherty}, {Fleming}, {Font-Ribera},
  {Foster}, {Frinchaboy}, {Galbraith-Frew}, {Garc{\'\i}a},
  {Garc{\'\i}a-Hern{\'a}ndez}, {Garc{\'\i}a P{\'e}rez}, {Gaulme}, {Ge},
  {G{\'e}nova-Santos}, {Georgakakis}, {Ghezzi}, {Gillespie}, {Girardi},
  {Goddard}, {Gontcho}, {Gonz{\'a}lez Hern{\'a}ndez}, {Grebel}, {Green},
  {Grieb}, {Grieves}, {Gunn}, {Guo}, {Harding}, {Hasselquist}, {Hawley},
  {Hayden}, {Hearty}, {Hekker}, {Ho}, {Hogg}, {Holley-Bockelmann}, {Holtzman},
  {Honscheid}, {Huber}, {Huehnerhoff}, {Ivans}, {Jiang}, {Johnson},
  {Kinemuchi}, {Kirkby}, {Kitaura}, {Klaene}, {Knapp}, {Kneib}, {Koenig},
  {Lam}, {Lan}, {Lang}, {Laurent}, {Le Goff}, {Leauthaud}, {Lee}, {Lee},
  {Licquia}, {Liu}, {Long}, {L{\'o}pez-Corredoira}, {Lorenzo-Oliveira},
  {Lucatello}, {Lundgren}, {Lupton}, {Mack}, {Mahadevan}, {Maia}, {Majewski},
  {Malanushenko}, {Malanushenko}, {Manchado}, {Manera}, {Mao}, {Maraston},
  {Marchwinski}, {Margala}, {Martell}, {Martig}, {Masters}, {Mathur},
  {McBride}, {McGehee}, {McGreer}, {McMahon}, {M{\'e}nard}, {Menzel},
  {Merloni}, {M{\'e}sz{\'a}ros}, {Miller}, {Miralda-Escud{\'e}}, {Miyatake},
  {Montero-Dorta}, {More}, {Morganson}, {Morice-Atkinson}, {Morrison},
  {Mosser}, {Muna}, {Myers}, {Nandra}, {Newman}, {Neyrinck}, {Nguyen},
  {Nichol}, {Nidever}, {Noterdaeme}, {Nuza}, {O'Connell}, {O'Connell},
  {O'Connell}, {Ogando}, {Olmstead}, {Oravetz}, {Oravetz}, {Osumi}, {Owen},
  {Padgett}, {Padmanabhan}, {Paegert}, {Palanque-Delabrouille}, {Pan},
  {Parejko}, {P{\^a}ris}, {Park}, {Pattarakijwanich}, {Pellejero-Ibanez},
  {Pepper}, {Percival}, {P{\'e}rez-Fournon}, {P{\'e}rez-R{\`a}fols},
  {Petitjean}, {Pieri}, {Pinsonneault}, {Porto de Mello}, {Prada}, {Prakash},
  {Price-Whelan}, {Protopapas}, {Raddick}, {Rahman}, {Reid}, {Rich}, {Rix},
  {Robin}, {Rockosi}, {Rodrigues}, {Rodr{\'\i}guez-Torres}, {Roe}, {Ross},
  {Ross}, {Rossi}, {Ruan}, {Rubi{\~n}o-Mart{\'\i}n}, {Rykoff},
  {Salazar-Albornoz}, {Salvato}, {Samushia}, {S{\'a}nchez}, {Santiago},
  {Sayres}, {Schiavon}, {Schlegel}, {Schmidt}, {Schneider}, {Schultheis},
  {Schwope}, {Sc{\'o}ccola}, {Scott}, {Sellgren}, {Seo}, {Serenelli}, {Shane},
  {Shen}, {Shetrone}, {Shu}, {Silva Aguirre}, {Sivarani}, {Skrutskie},
  {Slosar}, {Smith}, {Sobreira}, {Souto}, {Stassun}, {Steinmetz}, {Stello},
  {Strauss}, {Streblyanska}, {Suzuki}, {Swanson}, {Tan}, {Tayar}, {Terrien},
  {Thakar}, {Thomas}, {Thomas}, {Thompson}, {Tinker}, {Tojeiro}, {Troup},
  {Vargas-Maga{\~n}a}, {Vazquez}, {Verde}, {Viel}, {Vogt}, {Wake}, {Wang},
  {Weaver}, {Weinberg}, {Weiner}, {White}, {Wilson}, {Wisniewski},
  {Wood-Vasey}, {Ye`che}, {York}, {Zakamska}, {Zamora}, {Zasowski}, {Zehavi},
  {Zhao}, {Zheng}, {Zhou}, {Zhou}, {Zou}, \& {Zhu}}]{Alam+15}
{Alam}, S., {Albareti}, F.~D., {Allende Prieto}, C., {et~al.} 2015, \apjs, 219,
  12, \dodoi{10.1088/0067-0049/219/1/12}

\bibitem[{{Alexander} {et~al.}(2021){Alexander}, {Schroeder}, {Paterson},
  {Fong}, {Cowperthwaite}, {Gomez}, {Margalit}, {Margutti}, {Berger},
  {Blanchard}, {Chornock}, {Eftekhari}, {Laskar}, {Metzger}, {Nicholl},
  {Villar}, \& {Williams}}]{Alexander+21}
{Alexander}, K.~D., {Schroeder}, G., {Paterson}, K., {et~al.} 2021, \apj, 923,
  66, \dodoi{10.3847/1538-4357/ac281a}

\bibitem[{{Anand} {et~al.}(2021){Anand}, {Coughlin}, {Kasliwal}, {Bulla},
  {Ahumada}, {Sagu{\'e}s Carracedo}, {Almualla}, {Andreoni}, {Stein},
  {Foucart}, {Singer}, {Sollerman}, {Bellm}, {Bolin}, {Caballero-Garc{\'\i}a},
  {Castro-Tirado}, {Cenko}, {De}, {Dekany}, {Duev}, {Feeney}, {Fremling},
  {Goldstein}, {Golkhou}, {Graham}, {Guessoum}, {Hankins}, {Hu}, {Kong},
  {Kool}, {Kulkarni}, {Kumar}, {Laher}, {Masci}, {Mr{\'o}z}, {Nissanke},
  {Porter}, {Reusch}, {Riddle}, {Rosnet}, {Rusholme}, {Serabyn},
  {S{\'a}nchez-Ram{\'\i}rez}, {Rigault}, {Shupe}, {Smith}, {Soumagnac},
  {Walters}, \& {Valeev}}]{Anand+21}
{Anand}, S., {Coughlin}, M.~W., {Kasliwal}, M.~M., {et~al.} 2021, Nature
  Astronomy, 5, 46, \dodoi{10.1038/s41550-020-1183-3}

\bibitem[{{Andreoni} {et~al.}(2017){Andreoni}, {Ackley}, {Cooke}, {Acharyya},
  {Allison}, {Anderson}, {Ashley}, {Baade}, {Bailes}, {Bannister}, {Beardsley},
  {Bessell}, {Bian}, {Bland}, {Boer}, {Booler}, {Brandeker}, {Brown},
  {Buckley}, {Chang}, {Coward}, {Crawford}, {Crisp}, {Crosse}, {Cucchiara},
  {Cup{\'a}k}, {de Gois}, {Deller}, {Devillepoix}, {Dobie}, {Elmer}, {Emrich},
  {Farah}, {Farrell}, {Franzen}, {Gaensler}, {Galloway}, {Gendre}, {Giblin},
  {Goobar}, {Green}, {Hancock}, {Hartig}, {Howell}, {Horsley}, {Hotan},
  {Howie}, {Hu}, {Hu}, {James}, {Johnston}, {Johnston-Hollitt}, {Kaplan},
  {Kasliwal}, {Keane}, {Kenney}, {Klotz}, {Lau}, {Laugier}, {Lenc}, {Li},
  {Liang}, {Lidman}, {Luvaul}, {Lynch}, {Ma}, {Macpherson}, {Mao},
  {McClelland}, {McCully}, {M{\"o}ller}, {Morales}, {Morris}, {Murphy},
  {Noysena}, {Onken}, {Orange}, {Os{\l}owski}, {Pallot}, {Paxman}, {Potter},
  {Pritchard}, {Raja}, {Ridden-Harper}, {Romero-Colmenero}, {Sadler}, {Sansom},
  {Scalzo}, {Schmidt}, {Scott}, {Seghouani}, {Shang}, {Shannon}, {Shao},
  {Shara}, {Sharp}, {Sokolowski}, {Sollerman}, {Staff}, {Steele}, {Sun},
  {Suntzeff}, {Tao}, {Tingay}, {Towner}, {Thierry}, {Trott}, {Tucker},
  {V{\"a}is{\"a}nen}, {Krishnan}, {Walker}, {Wang}, {Wang}, {Wayth}, {Whiting},
  {Williams}, {Williams}, {Wolf}, {Wu}, {Wu}, {Yang}, {Yuan}, {Zhang}, {Zhou},
  \& {Zovaro}}]{Andreoni+17}
{Andreoni}, I., {Ackley}, K., {Cooke}, J., {et~al.} 2017, \pasa, 34, e069,
  \dodoi{10.1017/pasa.2017.65}

\bibitem[{{Andreoni} {et~al.}(2019{\natexlab{a}}){Andreoni}, {Goldstein},
  {Anand}, {Coughlin}, {Singer}, {Ahumada}, {Medford}, {Kool}, {Webb}, {Bulla},
  {Bloom}, {Kasliwal}, {Nugent}, {Bagdasaryan}, {Barnes}, {Cook}, {Cooke},
  {Duev}, {Fremling}, {Gatkine}, {Golkhou}, {Kong}, {Mahabal},
  {Mart{\'\i}nez-Palomera}, {Tao}, \& {Zhang}}]{Andreoni+19}
{Andreoni}, I., {Goldstein}, D.~A., {Anand}, S., {et~al.} 2019{\natexlab{a}},
  \apjl, 881, L16, \dodoi{10.3847/2041-8213/ab3399}

\bibitem[{{Andreoni} {et~al.}(2019{\natexlab{b}}){Andreoni}, {Goldstein},
  {Coughlin}, {Kasliwal}, {Nugent}, {Zhang}, {Palomera}, {Anand}, {Bloom},
  {Cenko}, {Cooke}, \& {Singer}}]{GCN_DECam190426}
{Andreoni}, I., {Goldstein}, D.~A., {Coughlin}, M., {et~al.}
  2019{\natexlab{b}}, GRB Coordinates Network, 24268, 1

\bibitem[{{Andreoni} {et~al.}(2020{\natexlab{a}}){Andreoni}, {Goldstein},
  {Kasliwal}, {Nugent}, {Zhou}, {Newman}, {Bulla}, {Foucart}, {Hotokezaka},
  {Nakar}, {Nissanke}, {Raaijmakers}, {Bloom}, {De}, {Jencson}, {Ward},
  {Ahumada}, {Anand}, {Buckley}, {Caballero-Garc{\'\i}a}, {Castro-Tirado},
  {Copperwheat}, {Coughlin}, {Cenko}, {Gromadzki}, {Hu}, {Karambelkar},
  {Perley}, {Sharma}, {Valeev}, {Cook}, {Fremling}, {Kumar}, {Taggart},
  {Bagdasaryan}, {Cooke}, {Dahiwale}, {Dhawan}, {Dobie}, {Gatkine}, {Golkhou},
  {Goobar}, {Chaves}, {Hankins}, {Kaplan}, {Kong}, {Kool}, {Mohite},
  {Sollerman}, {Tzanidakis}, {Webb}, \& {Zhang}}]{Andreoni+20}
{Andreoni}, I., {Goldstein}, D.~A., {Kasliwal}, M.~M., {et~al.}
  2020{\natexlab{a}}, \apj, 890, 131, \dodoi{10.3847/1538-4357/ab6a1b}

\bibitem[{{Andreoni} {et~al.}(2020{\natexlab{b}}){Andreoni}, {Kool},
  {Sagu{\'e}s Carracedo}, {Kasliwal}, {Bulla}, {Ahumada}, {Coughlin}, {Anand},
  {Sollerman}, {Goobar}, {Kaplan}, {Loveridge}, {Karambelkar}, {Cooke},
  {Bagdasaryan}, {Bellm}, {Cenko}, {Cook}, {De}, {Dekany}, {Delacroix},
  {Drake}, {Duev}, {Fremling}, {Golkhou}, {Graham}, {Hale}, {Kulkarni},
  {Kupfer}, {Laher}, {Mahabal}, {Masci}, {Rusholme}, {Smith}, {Tzanidakis},
  {Van Sistine}, \& {Yao}}]{Andreoni+20_ztfkn}
{Andreoni}, I., {Kool}, E.~C., {Sagu{\'e}s Carracedo}, A., {et~al.}
  2020{\natexlab{b}}, \apj, 904, 155, \dodoi{10.3847/1538-4357/abbf4c}

\bibitem[{{Andreoni} {et~al.}(2021){Andreoni}, {Coughlin}, {Kool}, {Kasliwal},
  {Kumar}, {Bhalerao}, {Carracedo}, {Ho}, {Pang}, {Saraogi}, {Sharma},
  {Shenoy}, {Burns}, {Ahumada}, {Anand}, {Singer}, {Perley}, {De}, {Fremling},
  {Bellm}, {Bulla}, {Crellin-Quick}, {Dietrich}, {Drake}, {Duev}, {Goobar},
  {Graham}, {Kaplan}, {Kulkarni}, {Laher}, {Mahabal}, {Shupe}, {Sollerman},
  {Walters}, \& {Yao}}]{Andreoni+21}
{Andreoni}, I., {Coughlin}, M.~W., {Kool}, E.~C., {et~al.} 2021, \apj, 918, 63,
  \dodoi{10.3847/1538-4357/ac0bc7}

\bibitem[{{Antier} {et~al.}(2020{\natexlab{a}}){Antier}, {Agayeva}, {Aivazyan},
  {Alishov}, {Arbouch}, {Baransky}, {Barynova}, {Bai}, {Basa}, {Beradze},
  {Bertin}, {Berthier}, {Bla{\v{z}}ek}, {Bo{\"e}r}, {Burkhonov}, {Burrell},
  {Cailleau}, {Chabert}, {Chen}, {Christensen}, {Coleiro}, {Cordier}, {Corre},
  {Coughlin}, {Coward}, {Crisp}, {Delattre}, {Dietrich}, {Ducoin}, {Duverne},
  {Marchal-Duval}, {Gendre}, {Eymar}, {Fock-Hang}, {Han}, {Hello}, {Howell},
  {Inasaridze}, {Ismailov}, {Kann}, {Kapanadze}, {Klotz}, {Kochiashvili},
  {Lachaud}, {Leroy}, {Le Van Su}, {Lin}, {Li}, {Lognone}, {Marron}, {Mo},
  {Moore}, {Natsvlishvili}, {Noysena}, {Perrigault}, {Peyrot}, {Samadov},
  {Sadibekova}, {Simon}, {Stachie}, {Teng}, {Thierry}, {Th{\"o}ne}, {Tillayev},
  {Turpin}, {de Ugarte Postigo}, {Vachier}, {Vardosanidze}, {Vasylenko},
  {Vidadi}, {Wang}, {Wang}, {Wei}, {Yan}, {Zhang}, {Zhang}, \&
  {Zhang}}]{Antier+20a}
{Antier}, S., {Agayeva}, S., {Aivazyan}, V., {et~al.} 2020{\natexlab{a}},
  \mnras, 492, 3904, \dodoi{10.1093/mnras/stz3142}

\bibitem[{{Antier} {et~al.}(2020{\natexlab{b}}){Antier}, {Agayeva}, {Almualla},
  {Awiphan}, {Baransky}, {Barynova}, {Beradze}, {Bla{\v{z}}ek}, {Bo{\"e}r},
  {Burkhonov}, {Christensen}, {Coleiro}, {Corre}, {Coughlin}, {Crisp},
  {Dietrich}, {Ducoin}, {Duverne}, {Marchal-Duval}, {Gendre}, {Gokuldass},
  {Eggenstein}, {Eymar}, {Hello}, {Howell}, {Ismailov}, {Kann}, {Karpov},
  {Klotz}, {Kochiashvili}, {Lachaud}, {Leroy}, {Lin}, {Li}, {Ma{\v{s}}ek},
  {Mo}, {Menard}, {Morris}, {Noysena}, {Orange}, {Prouza}, {Rattanamala},
  {Sadibekova}, {Saint-Gelais}, {Serrau}, {Simon}, {Stachie}, {Th{\"o}ne},
  {Tillayev}, {Turpin}, {Postigo}, {Vasylenko}, {Vidadi}, {Was}, {Wang},
  {Zhang}, {Zhang}, \& {Zhang}}]{Antier+20b}
{Antier}, S., {Agayeva}, S., {Almualla}, M., {et~al.} 2020{\natexlab{b}},
  \mnras, 497, 5518, \dodoi{10.1093/mnras/staa1846}

\bibitem[{{Arcavi} {et~al.}(2017){Arcavi}, {Hosseinzadeh}, {Howell}, {McCully},
  {Poznanski}, {Kasen}, {Barnes}, {Zaltzman}, {Vasylyev}, {Maoz}, \&
  {Valenti}}]{Arcavi+17}
{Arcavi}, I., {Hosseinzadeh}, G., {Howell}, D.~A., {et~al.} 2017, \nat, 551,
  64, \dodoi{10.1038/nature24291}

\bibitem[{{Ascenzi} {et~al.}(2019){Ascenzi}, {Coughlin}, {Dietrich}, {Foley},
  {Ramirez-Ruiz}, {Piranomonte}, {Mockler}, {Murguia-Berthier}, {Fryer},
  {Lloyd-Ronning}, \& {Rosswog}}]{Ascenzi+19}
{Ascenzi}, S., {Coughlin}, M.~W., {Dietrich}, T., {et~al.} 2019, \mnras, 486,
  672, \dodoi{10.1093/mnras/stz891}

\bibitem[{{Ashton} {et~al.}(2021){Ashton}, {Ackley}, {Hernandez}, \&
  {Piotrzkowski}}]{Ashton+20}
{Ashton}, G., {Ackley}, K., {Hernandez}, I.~M., \& {Piotrzkowski}, B. 2021,
  Classical and Quantum Gravity, 38, 235004, \dodoi{10.1088/1361-6382/ac33bb}

\bibitem[{{Astropy Collaboration} {et~al.}(2013){Astropy Collaboration},
  {Robitaille}, {Tollerud}, {Greenfield}, {Droettboom}, {Bray}, {Aldcroft},
  {Davis}, {Ginsburg}, {Price-Whelan}, {Kerzendorf}, {Conley}, {Crighton},
  {Barbary}, {Muna}, {Ferguson}, {Grollier}, {Parikh}, {Nair}, {Unther},
  {Deil}, {Woillez}, {Conseil}, {Kramer}, {Turner}, {Singer}, {Fox}, {Weaver},
  {Zabalza}, {Edwards}, {Azalee Bostroem}, {Burke}, {Casey}, {Crawford},
  {Dencheva}, {Ely}, {Jenness}, {Labrie}, {Lim}, {Pierfederici}, {Pontzen},
  {Ptak}, {Refsdal}, {Servillat}, \& {Streicher}}]{Astropy2013}
{Astropy Collaboration}, {Robitaille}, T.~P., {Tollerud}, E.~J., {et~al.} 2013,
  \aap, 558, A33, \dodoi{10.1051/0004-6361/201322068}

\bibitem[{{Astropy Collaboration} {et~al.}(2018){Astropy Collaboration},
  {Price-Whelan}, {Sip{\H{o}}cz}, {G{\"u}nther}, {Lim}, {Crawford}, {Conseil},
  {Shupe}, {Craig}, {Dencheva}, {Ginsburg}, {VanderPlas}, {Bradley},
  {P{\'e}rez-Su{\'a}rez}, {de Val-Borro}, {Aldcroft}, {Cruz}, {Robitaille},
  {Tollerud}, {Ardelean}, {Babej}, {Bach}, {Bachetti}, {Bakanov}, {Bamford},
  {Barentsen}, {Barmby}, {Baumbach}, {Berry}, {Biscani}, {Boquien}, {Bostroem},
  {Bouma}, {Brammer}, {Bray}, {Breytenbach}, {Buddelmeijer}, {Burke},
  {Calderone}, {Cano Rodr{\'\i}guez}, {Cara}, {Cardoso}, {Cheedella}, {Copin},
  {Corrales}, {Crichton}, {D'Avella}, {Deil}, {Depagne}, {Dietrich}, {Donath},
  {Droettboom}, {Earl}, {Erben}, {Fabbro}, {Ferreira}, {Finethy}, {Fox},
  {Garrison}, {Gibbons}, {Goldstein}, {Gommers}, {Greco}, {Greenfield},
  {Groener}, {Grollier}, {Hagen}, {Hirst}, {Homeier}, {Horton}, {Hosseinzadeh},
  {Hu}, {Hunkeler}, {Ivezi{\'c}}, {Jain}, {Jenness}, {Kanarek}, {Kendrew},
  {Kern}, {Kerzendorf}, {Khvalko}, {King}, {Kirkby}, {Kulkarni}, {Kumar},
  {Lee}, {Lenz}, {Littlefair}, {Ma}, {Macleod}, {Mastropietro}, {McCully},
  {Montagnac}, {Morris}, {Mueller}, {Mumford}, {Muna}, {Murphy}, {Nelson},
  {Nguyen}, {Ninan}, {N{\"o}the}, {Ogaz}, {Oh}, {Parejko}, {Parley}, {Pascual},
  {Patil}, {Patil}, {Plunkett}, {Prochaska}, {Rastogi}, {Reddy Janga},
  {Sabater}, {Sakurikar}, {Seifert}, {Sherbert}, {Sherwood-Taylor}, {Shih},
  {Sick}, {Silbiger}, {Singanamalla}, {Singer}, {Sladen}, {Sooley},
  {Sornarajah}, {Streicher}, {Teuben}, {Thomas}, {Tremblay}, {Turner},
  {Terr{\'o}n}, {van Kerkwijk}, {de la Vega}, {Watkins}, {Weaver}, {Whitmore},
  {Woillez}, {Zabalza}, \& {Astropy Contributors}}]{Astropy2018}
{Astropy Collaboration}, {Price-Whelan}, A.~M., {Sip{\H{o}}cz}, B.~M., {et~al.}
  2018, \aj, 156, 123, \dodoi{10.3847/1538-3881/aabc4f}

\bibitem[{{Barbieri} {et~al.}(2021){Barbieri}, {Salafia}, {Colpi}, {Ghirlanda},
  \& {Perego}}]{Barbieri+20}
{Barbieri}, C., {Salafia}, O.~S., {Colpi}, M., {Ghirlanda}, G., \& {Perego}, A.
  2021, \aap, 654, A12, \dodoi{10.1051/0004-6361/202037778}

\bibitem[{{Barnes} \& {Kasen}(2013)}]{barneskasen13}
{Barnes}, J., \& {Kasen}, D. 2013, \apj, 775, 18,
  \dodoi{10.1088/0004-637X/775/1/18}

\bibitem[{{Becerra} {et~al.}(2021){Becerra}, {Dichiara}, {Watson}, {Troja},
  {Butler}, {Pereyra}, {Moreno M{\'e}ndez}, {De Colle}, {Lee}, {Kutyrev}, \&
  {L{\'o}pez}}]{Becerra+21}
{Becerra}, R.~L., {Dichiara}, S., {Watson}, A.~M., {et~al.} 2021, \mnras, 507,
  1401, \dodoi{10.1093/mnras/stab2086}

\bibitem[{{Beck} {et~al.}(2016){Beck}, {Dobos}, {Budav{\'a}ri}, {Szalay}, \&
  {Csabai}}]{Beck+16}
{Beck}, R., {Dobos}, L., {Budav{\'a}ri}, T., {Szalay}, A.~S., \& {Csabai}, I.
  2016, \mnras, 460, 1371, \dodoi{10.1093/mnras/stw1009}

\bibitem[{{Beck} {et~al.}(2021){Beck}, {Szapudi}, {Flewelling}, {Holmberg},
  {Magnier}, \& {Chambers}}]{Beck+21}
{Beck}, R., {Szapudi}, I., {Flewelling}, H., {et~al.} 2021, \mnras, 500, 1633,
  \dodoi{10.1093/mnras/staa2587}

\bibitem[{{Bellm} {et~al.}(2019){Bellm}, {Kulkarni}, {Graham}, {Dekany},
  {Smith}, {Riddle}, {Masci}, {Helou}, {Prince}, {Adams}, {Barbarino},
  {Barlow}, {Bauer}, {Beck}, {Belicki}, {Biswas}, {Blagorodnova}, {Bodewits},
  {Bolin}, {Brinnel}, {Brooke}, {Bue}, {Bulla}, {Burruss}, {Cenko}, {Chang},
  {Connolly}, {Coughlin}, {Cromer}, {Cunningham}, {De}, {Delacroix}, {Desai},
  {Duev}, {Eadie}, {Farnham}, {Feeney}, {Feindt}, {Flynn}, {Franckowiak},
  {Frederick}, {Fremling}, {Gal-Yam}, {Gezari}, {Giomi}, {Goldstein},
  {Golkhou}, {Goobar}, {Groom}, {Hacopians}, {Hale}, {Henning}, {Ho}, {Hover},
  {Howell}, {Hung}, {Huppenkothen}, {Imel}, {Ip}, {Ivezi{\'c}}, {Jackson},
  {Jones}, {Juric}, {Kasliwal}, {Kaspi}, {Kaye}, {Kelley}, {Kowalski},
  {Kramer}, {Kupfer}, {Landry}, {Laher}, {Lee}, {Lin}, {Lin}, {Lunnan},
  {Giomi}, {Mahabal}, {Mao}, {Miller}, {Monkewitz}, {Murphy}, {Ngeow},
  {Nordin}, {Nugent}, {Ofek}, {Patterson}, {Penprase}, {Porter}, {Rauch},
  {Rebbapragada}, {Reiley}, {Rigault}, {Rodriguez}, {van Roestel}, {Rusholme},
  {van Santen}, {Schulze}, {Shupe}, {Singer}, {Soumagnac}, {Stein}, {Surace},
  {Sollerman}, {Szkody}, {Taddia}, {Terek}, {Van Sistine}, {van Velzen},
  {Vestrand}, {Walters}, {Ward}, {Ye}, {Yu}, {Yan}, \& {Zolkower}}]{Bellm+19}
{Bellm}, E.~C., {Kulkarni}, S.~R., {Graham}, M.~J., {et~al.} 2019, \pasp, 131,
  018002, \dodoi{10.1088/1538-3873/aaecbe}

\bibitem[{{Bennett} {et~al.}(2014){Bennett}, {Larson}, {Weiland}, \&
  {Hinshaw}}]{Bennett+14}
{Bennett}, C.~L., {Larson}, D., {Weiland}, J.~L., \& {Hinshaw}, G. 2014, \apj,
  794, 135, \dodoi{10.1088/0004-637X/794/2/135}

\bibitem[{{Bertin}(2006)}]{scamp}
{Bertin}, E. 2006, in Astronomical Society of the Pacific Conference Series,
  Vol. 351, Astronomical Data Analysis Software and Systems XV, ed.
  C.~{Gabriel}, C.~{Arviset}, D.~{Ponz}, \& S.~{Enrique}, 112

\bibitem[{{Bertin}(2010{\natexlab{a}})}]{scamp2}
{Bertin}, E. 2010{\natexlab{a}}, {SCAMP: Automatic Astrometric and Photometric
  Calibration}.
\newblock \doeprint{1010.063}

\bibitem[{{Bertin}(2010{\natexlab{b}})}]{swarp}
---. 2010{\natexlab{b}}, {SWarp: Resampling and Co-adding FITS Images
  Together}.
\newblock \doeprint{1010.068}

\bibitem[{{Bertin} \& {Arnouts}(1996)}]{Bertin1996}
{Bertin}, E., \& {Arnouts}, S. 1996, \aaps, 117, 393,
  \dodoi{10.1051/aas:1996164}

\bibitem[{{Bhakta} {et~al.}(2021){Bhakta}, {Mooley}, {Corsi},
  {Balasubramanian}, {Dobie}, {Frail}, {Hallinan}, {Kaplan}, {Myers}, \&
  {Singer}}]{Bhakta+21}
{Bhakta}, D., {Mooley}, K.~P., {Corsi}, A., {et~al.} 2021, \apj, 911, 77,
  \dodoi{10.3847/1538-4357/abeaa8}

\bibitem[{{Bildsten} {et~al.}(2007){Bildsten}, {Shen}, {Weinberg}, \&
  {Nelemans}}]{Bildsten+07}
{Bildsten}, L., {Shen}, K.~J., {Weinberg}, N.~N., \& {Nelemans}, G. 2007,
  \apjl, 662, L95, \dodoi{10.1086/519489}

\bibitem[{{Bloom} {et~al.}(2002){Bloom}, {Kulkarni}, \&
  {Djorgovski}}]{Bloom+02}
{Bloom}, J.~S., {Kulkarni}, S.~R., \& {Djorgovski}, S.~G. 2002, \aj, 123, 1111,
  \dodoi{10.1086/338893}

\bibitem[{{Brennan} {et~al.}(2019){Brennan}, {Callis}, {Ihanec}, {Gromadzki},
  \& {Irani}}]{TNS2019phq_pia}
{Brennan}, S., {Callis}, E., {Ihanec}, N., {Gromadzki}, M., \& {Irani}, I.
  2019, Transient Name Server Classification Report, 2019-1773, 1

\bibitem[{{Broekgaarden} {et~al.}(2021){Broekgaarden}, {Berger}, {Neijssel},
  {Vigna-G{\'o}mez}, {Chattopadhyay}, {Stevenson}, {Chruslinska}, {Justham},
  {de Mink}, \& {Mandel}}]{Broekgaarden+21}
{Broekgaarden}, F.~S., {Berger}, E., {Neijssel}, C.~J., {et~al.} 2021, \mnras,
  508, 5028, \dodoi{10.1093/mnras/stab2716}

\bibitem[{{Carini} {et~al.}(2019){Carini}, {Izzo}, {Palazzi}, {Sand}, {Rossi},
  {Cassara}, {Gargiulo}, {D'Avanzo}, {Melandri}, {Tomasella}, {Benetti},
  {Botticella}, {Branchesi}, {D'Elia}, {Brocato}, {Valle}, {Greco}, {Kuhn},
  {Shivaei}, {Fan}, {Andrews}, {Fong}, {Paterson}, {Lundquist}, \&
  {Cusano}}]{GCN2019ebq_LBT}
{Carini}, R., {Izzo}, L., {Palazzi}, E., {et~al.} 2019, GRB Coordinates
  Network, 24252, 1

\bibitem[{{Cenko}(2017)}]{Cenko17}
{Cenko}, S.~B. 2017, Nature Astronomy, 1, 0008, \dodoi{10.1038/s41550-016-0008}

\bibitem[{{Chambers} {et~al.}(2016){Chambers}, {Magnier}, {Metcalfe},
  {Flewelling}, {Huber}, {Waters}, {Denneau}, {Draper}, {Farrow}, {Finkbeiner},
  {Holmberg}, {Koppenhoefer}, {Price}, {Rest}, {Saglia}, {Schlafly}, {Smartt},
  {Sweeney}, {Wainscoat}, {Burgett}, {Chastel}, {Grav}, {Heasley}, {Hodapp},
  {Jedicke}, {Kaiser}, {Kudritzki}, {Luppino}, {Lupton}, {Monet}, {Morgan},
  {Onaka}, {Shiao}, {Stubbs}, {Tonry}, {White}, {Ba{\~n}ados}, {Bell},
  {Bender}, {Bernard}, {Boegner}, {Boffi}, {Botticella}, {Calamida},
  {Casertano}, {Chen}, {Chen}, {Cole}, {Deacon}, {Frenk}, {Fitzsimmons},
  {Gezari}, {Gibbs}, {Goessl}, {Goggia}, {Gourgue}, {Goldman}, {Grant},
  {Grebel}, {Hambly}, {Hasinger}, {Heavens}, {Heckman}, {Henderson}, {Henning},
  {Holman}, {Hopp}, {Ip}, {Isani}, {Jackson}, {Keyes}, {Koekemoer}, {Kotak},
  {Le}, {Liska}, {Long}, {Lucey}, {Liu}, {Martin}, {Masci}, {McLean}, {Mindel},
  {Misra}, {Morganson}, {Murphy}, {Obaika}, {Narayan}, {Nieto-Santisteban},
  {Norberg}, {Peacock}, {Pier}, {Postman}, {Primak}, {Rae}, {Rai}, {Riess},
  {Riffeser}, {Rix}, {R{\"o}ser}, {Russel}, {Rutz}, {Schilbach}, {Schultz},
  {Scolnic}, {Strolger}, {Szalay}, {Seitz}, {Small}, {Smith}, {Soderblom},
  {Taylor}, {Thomson}, {Taylor}, {Thakar}, {Thiel}, {Thilker}, {Unger},
  {Urata}, {Valenti}, {Wagner}, {Walder}, {Walter}, {Watters}, {Werner},
  {Wood-Vasey}, \& {Wyse}}]{Chambers+16}
{Chambers}, K.~C., {Magnier}, E.~A., {Metcalfe}, N., {et~al.} 2016, arXiv
  e-prints, arXiv:1612.05560.
\newblock \doarXiv{1612.05560}

\bibitem[{{Chang} {et~al.}(2019){Chang}, {Allen}, {Anderson}, {Bianco},
  {Bloom}, {Brady}, {Brazier}, {Cenko}, {Couch}, {DeYoung}, {Deelman},
  {Etienne}, {Foley}, {Fox}, {Golkhou}, {Grant}, {Hanna}, {Holley-Bockelmann},
  {Howell}, {Huerta}, {Johnson}, {Juric}, {Kaplan}, {Katz}, {Keivani},
  {Kerzendorf}, {Kopper}, {Lam}, {Lehner}, {Marka}, {Marka}, {Nabrzyski},
  {Narayan}, {O'Shea}, {Petravick}, {Quick}, {Street}, {Taboada}, {Timmes},
  {Turk}, {Weltman}, \& {Zhang}}]{SciMMA}
{Chang}, P., {Allen}, G., {Anderson}, W., {et~al.} 2019, \baas, 51, 436.
\newblock \doarXiv{1903.04590}

\bibitem[{{Chang} {et~al.}(2021){Chang}, {Onken}, {Wolf}, {Luvaul},
  {M{\"o}ller}, {Scalzo}, {Schmidt}, {Scott}, {Sura}, \& {Yuan}}]{Chang+21}
{Chang}, S.-W., {Onken}, C.~A., {Wolf}, C., {et~al.} 2021, \pasa, 38, e024,
  \dodoi{10.1017/pasa.2021.17}

\bibitem[{{Chase} {et~al.}(2021){Chase}, {O'Connor}, {Fryer}, {Troja},
  {Korobkin}, {Wollaeger}, {Ristic}, {Fontes}, {Hungerford}, \&
  {Herring}}]{Chase+21}
{Chase}, E.~A., {O'Connor}, B., {Fryer}, C.~L., {et~al.} 2021, arXiv e-prints,
  arXiv:2105.12268.
\newblock \doarXiv{2105.12268}

\bibitem[{{Christensen} {et~al.}(2018){Christensen}, {Africano}, {Farneth},
  {Fuls}, {Gibbs}, {Grauer}, {Groeller}, {Johnson}, {Kowalski}, {Larson},
  {Leonard}, {Seaman}, \& {Shelly}}]{CSS}
{Christensen}, E., {Africano}, B., {Farneth}, G., {et~al.} 2018, in
  AAS/Division for Planetary Sciences Meeting Abstracts, 310.10

\bibitem[{{Cook} {et~al.}(2019){Cook}, {Kasliwal}, {Van Sistine}, {Kaplan},
  {Sutter}, {Kupfer}, {Shupe}, {Laher}, {Masci}, {Dale}, {Sesar}, {Brady},
  {Yan}, {Ofek}, {Reitze}, \& {Kulkarni}}]{Cook+19}
{Cook}, D.~O., {Kasliwal}, M.~M., {Van Sistine}, A., {et~al.} 2019, \apj, 880,
  7, \dodoi{10.3847/1538-4357/ab2131}

\bibitem[{{Coppejans} {et~al.}(2020){Coppejans}, {Margutti}, {Terreran},
  {Nayana}, {Coughlin}, {Laskar}, {Alexander}, {Bietenholz}, {Caprioli},
  {Chandra}, {Drout}, {Frederiks}, {Frohmaier}, {Hurley}, {Kochanek},
  {MacLeod}, {Meisner}, {Nugent}, {Ridnaia}, {Sand}, {Svinkin}, {Ward}, {Yang},
  {Baldeschi}, {Chilingarian}, {Dong}, {Esquivia}, {Fong}, {Guidorzi},
  {Lundqvist}, {Milisavljevic}, {Paterson}, {Reichart}, {Shappee}, {Stroh},
  {Valenti}, {Zauderer}, \& {Zhang}}]{Coppejans+20}
{Coppejans}, D.~L., {Margutti}, R., {Terreran}, G., {et~al.} 2020, \apjl, 895,
  L23, \dodoi{10.3847/2041-8213/ab8cc7}

\bibitem[{{Coughlin} {et~al.}(2019){Coughlin}, {Ahumada}, {Anand}, {De},
  {Hankins}, {Kasliwal}, {Singer}, {Bellm}, {Andreoni}, {Cenko}, {Cooke},
  {Copperwheat}, {Dugas}, {Jencson}, {Perley}, {Yu}, {Bhalerao}, {Kumar},
  {Bloom}, {Anupama}, {Ashley}, {Bagdasaryan}, {Biswas}, {Buckley}, {Burdge},
  {Cook}, {Cromer}, {Cunningham}, {D'A{\`\i}}, {Dekany}, {Delacroix},
  {Dichiara}, {Duev}, {Dutta}, {Feeney}, {Frederick}, {Gatkine}, {Ghosh},
  {Goldstein}, {Golkhou}, {Goobar}, {Graham}, {Hanayama}, {Horiuchi}, {Hung},
  {Jha}, {Kong}, {Giomi}, {Kaplan}, {Karambelkar}, {Kowalski}, {Kulkarni},
  {Kupfer}, {Masci}, {Mazzali}, {Moore}, {Mogotsi}, {Neill}, {Ngeow},
  {Mart{\'\i}nez-Palomera}, {La Parola}, {Pavana}, {Ofek}, {Patil}, {Riddle},
  {Rigault}, {Rusholme}, {Serabyn}, {Shupe}, {Sharma}, {Singh}, {Sollerman},
  {Soon}, {Staats}, {Taggart}, {Tan}, {Travouillon}, {Troja}, {Waratkar}, \&
  {Yatsu}}]{Coughlin+19}
{Coughlin}, M.~W., {Ahumada}, T., {Anand}, S., {et~al.} 2019, \apjl, 885, L19,
  \dodoi{10.3847/2041-8213/ab4ad8}

\bibitem[{{Coulter} {et~al.}(2017){Coulter}, {Foley}, {Kilpatrick}, {Drout},
  {Piro}, {Shappee}, {Siebert}, {Simon}, {Ulloa}, {Kasen}, {Madore},
  {Murguia-Berthier}, {Pan}, {Prochaska}, {Ramirez-Ruiz}, {Rest}, \&
  {Rojas-Bravo}}]{Coulter+17}
{Coulter}, D.~A., {Foley}, R.~J., {Kilpatrick}, C.~D., {et~al.} 2017, Science,
  358, 1556, \dodoi{10.1126/science.aap9811}

\bibitem[{{Cowperthwaite} \& {Berger}(2015)}]{CowperthwaiteBerger15}
{Cowperthwaite}, P.~S., \& {Berger}, E. 2015, \apj, 814, 25,
  \dodoi{10.1088/0004-637X/814/1/25}

\bibitem[{{Cowperthwaite} {et~al.}(2017){Cowperthwaite}, {Berger}, {Villar},
  {Metzger}, {Nicholl}, {Chornock}, {Blanchard}, {Fong}, {Margutti},
  {Soares-Santos}, {Alexander}, {Allam}, {Annis}, {Brout}, {Brown}, {Butler},
  {Chen}, {Diehl}, {Doctor}, {Drout}, {Eftekhari}, {Farr}, {Finley}, {Foley},
  {Frieman}, {Fryer}, {Garc{\'{\i}}a-Bellido}, {Gill}, {Guillochon}, {Herner},
  {Holz}, {Kasen}, {Kessler}, {Marriner}, {Matheson}, {Neilsen}, {Quataert},
  {Palmese}, {Rest}, {Sako}, {Scolnic}, {Smith}, {Tucker}, {Williams},
  {Balbinot}, {Carlin}, {Cook}, {Durret}, {Li}, {Lopes}, {Louren{\c c}o},
  {Marshall}, {Medina}, {Muir}, {Mu{\~n}oz}, {Sauseda}, {Schlegel}, {Secco},
  {Vivas}, {Wester}, {Zenteno}, {Zhang}, {Abbott}, {Banerji}, {Bechtol},
  {Benoit-L{\'e}vy}, {Bertin}, {Buckley-Geer}, {Burke}, {Capozzi}, {Carnero
  Rosell}, {Carrasco Kind}, {Castander}, {Crocce}, {Cunha}, {D'Andrea}, {da
  Costa}, {Davis}, {DePoy}, {Desai}, {Dietrich}, {Drlica-Wagner}, {Eifler},
  {Evrard}, {Fernandez}, {Flaugher}, {Fosalba}, {Gaztanaga}, {Gerdes},
  {Giannantonio}, {Goldstein}, {Gruen}, {Gruendl}, {Gutierrez}, {Honscheid},
  {Jain}, {James}, {Jeltema}, {Johnson}, {Johnson}, {Kent}, {Krause}, {Kron},
  {Kuehn}, {Nuropatkin}, {Lahav}, {Lima}, {Lin}, {Maia}, {March}, {Martini},
  {McMahon}, {Menanteau}, {Miller}, {Miquel}, {Mohr}, {Neilsen}, {Nichol},
  {Ogando}, {Plazas}, {Roe}, {Romer}, {Roodman}, {Rykoff}, {Sanchez},
  {Scarpine}, {Schindler}, {Schubnell}, {Sevilla-Noarbe}, {Smith}, {Smith},
  {Sobreira}, {Suchyta}, {Swanson}, {Tarle}, {Thomas}, {Thomas}, {Troxel},
  {Vikram}, {Walker}, {Wechsler}, {Weller}, {Yanny}, \&
  {Zuntz}}]{Cowperthwaite+17}
{Cowperthwaite}, P.~S., {Berger}, E., {Villar}, V.~A., {et~al.} 2017, \apjl,
  848, L17, \dodoi{10.3847/2041-8213/aa8fc7}

\bibitem[{{Dahiwale} \& {Fremling}(2019)}]{TNS2019rta}
{Dahiwale}, A., \& {Fremling}, C. 2019, Transient Name Server Classification
  Report, 2019-2407, 1

\bibitem[{{D{\'a}lya} {et~al.}(2018){D{\'a}lya}, {Galg{\'o}czi}, {Dobos},
  {Frei}, {Heng}, {Macas}, {Messenger}, {Raffai}, \& {de Souza}}]{Dalya+18}
{D{\'a}lya}, G., {Galg{\'o}czi}, G., {Dobos}, L., {et~al.} 2018, \mnras, 479,
  2374, \dodoi{10.1093/mnras/sty1703}

\bibitem[{{Darbha} {et~al.}(2010){Darbha}, {Metzger}, {Quataert}, {Kasen},
  {Nugent}, \& {Thomas}}]{Darbha+10}
{Darbha}, S., {Metzger}, B.~D., {Quataert}, E., {et~al.} 2010, \mnras, 409,
  846, \dodoi{10.1111/j.1365-2966.2010.17353.x}

\bibitem[{{de Jaeger} {et~al.}(2022){de Jaeger}, {Shappee}, {Kochanek},
  {Stanek}, {Beacom}, {Holoien}, {Thompson}, {Franckowiak}, \&
  {Holmbo}}]{deJaeger+22}
{de Jaeger}, T., {Shappee}, B.~J., {Kochanek}, C.~S., {et~al.} 2022, \mnras,
  509, 3427, \dodoi{10.1093/mnras/stab3141}

\bibitem[{{de Wet} {et~al.}(2021){de Wet}, {Groot}, {Bloemen}, {Le Poole},
  {Klein-Wolt}, {K{\"o}rding}, {McBride}, {Paterson}, {Pieterse}, {Vreeswijk},
  \& {Woudt}}]{de_Wet+21}
{de Wet}, S., {Groot}, P.~J., {Bloemen}, S., {et~al.} 2021, \aap, 649, A72,
  \dodoi{10.1051/0004-6361/202040231}

\bibitem[{{DESI Collaboration} {et~al.}(2016){DESI Collaboration}, {Aghamousa},
  {Aguilar}, {Ahlen}, {Alam}, {Allen}, {Allende Prieto}, {Annis}, {Bailey},
  {Balland}, {Ballester}, {Baltay}, {Beaufore}, {Bebek}, {Beers}, {Bell},
  {Bernal}, {Besuner}, {Beutler}, {Blake}, {Bleuler}, {Blomqvist}, {Blum},
  {Bolton}, {Briceno}, {Brooks}, {Brownstein}, {Buckley-Geer}, {Burden},
  {Burtin}, {Busca}, {Cahn}, {Cai}, {Cardiel-Sas}, {Carlberg}, {Carton},
  {Casas}, {Castander}, {Cervantes-Cota}, {Claybaugh}, {Close}, {Coker},
  {Cole}, {Comparat}, {Cooper}, {Cousinou}, {Crocce}, {Cuby}, {Cunningham},
  {Davis}, {Dawson}, {de la Macorra}, {De Vicente}, {Delubac}, {Derwent},
  {Dey}, {Dhungana}, {Ding}, {Doel}, {Duan}, {Ealet}, {Edelstein},
  {Eftekharzadeh}, {Eisenstein}, {Elliott}, {Escoffier}, {Evatt}, {Fagrelius},
  {Fan}, {Fanning}, {Farahi}, {Farihi}, {Favole}, {Feng}, {Fernandez},
  {Findlay}, {Finkbeiner}, {Fitzpatrick}, {Flaugher}, {Flender}, {Font-Ribera},
  {Forero-Romero}, {Fosalba}, {Frenk}, {Fumagalli}, {Gaensicke}, {Gallo},
  {Garcia-Bellido}, {Gaztanaga}, {Pietro Gentile Fusillo}, {Gerard},
  {Gershkovich}, {Giannantonio}, {Gillet}, {Gonzalez-de-Rivera},
  {Gonzalez-Perez}, {Gott}, {Graur}, {Gutierrez}, {Guy}, {Habib}, {Heetderks},
  {Heetderks}, {Heitmann}, {Hellwing}, {Herrera}, {Ho}, {Holland}, {Honscheid},
  {Huff}, {Hutchinson}, {Huterer}, {Hwang}, {Illa Laguna}, {Ishikawa},
  {Jacobs}, {Jeffrey}, {Jelinsky}, {Jennings}, {Jiang}, {Jimenez}, {Johnson},
  {Joyce}, {Jullo}, {Juneau}, {Kama}, {Karcher}, {Karkar}, {Kehoe}, {Kennamer},
  {Kent}, {Kilbinger}, {Kim}, {Kirkby}, {Kisner}, {Kitanidis}, {Kneib},
  {Koposov}, {Kovacs}, {Koyama}, {Kremin}, {Kron}, {Kronig}, {Kueter-Young},
  {Lacey}, {Lafever}, {Lahav}, {Lambert}, {Lampton}, {Landriau}, {Lang},
  {Lauer}, {Le Goff}, {Le Guillou}, {Le Van Suu}, {Lee}, {Lee}, {Leitner},
  {Lesser}, {Levi}, {L'Huillier}, {Li}, {Liang}, {Lin}, {Linder}, {Loebman},
  {Luki{\'c}}, {Ma}, {MacCrann}, {Magneville}, {Makarem}, {Manera}, {Manser},
  {Marshall}, {Martini}, {Massey}, {Matheson}, {McCauley}, {McDonald},
  {McGreer}, {Meisner}, {Metcalfe}, {Miller}, {Miquel}, {Moustakas}, {Myers},
  {Naik}, {Newman}, {Nichol}, {Nicola}, {Nicolati da Costa}, {Nie}, {Niz},
  {Norberg}, {Nord}, {Norman}, {Nugent}, {O'Brien}, {Oh}, {Olsen}, {Padilla},
  {Padmanabhan}, {Padmanabhan}, {Palanque-Delabrouille}, {Palmese},
  {Pappalardo}, {P{\^a}ris}, {Park}, {Patej}, {Peacock}, {Peiris}, {Peng},
  {Percival}, {Perruchot}, {Pieri}, {Pogge}, {Pollack}, {Poppett}, {Prada},
  {Prakash}, {Probst}, {Rabinowitz}, {Raichoor}, {Ree}, {Refregier}, {Regal},
  {Reid}, {Reil}, {Rezaie}, {Rockosi}, {Roe}, {Ronayette}, {Roodman}, {Ross},
  {Ross}, {Rossi}, {Rozo}, {Ruhlmann-Kleider}, {Rykoff}, {Sabiu}, {Samushia},
  {Sanchez}, {Sanchez}, {Schlegel}, {Schneider}, {Schubnell}, {Secroun},
  {Seljak}, {Seo}, {Serrano}, {Shafieloo}, {Shan}, {Sharples}, {Sholl},
  {Shourt}, {Silber}, {Silva}, {Sirk}, {Slosar}, {Smith}, {Smoot}, {Som},
  {Song}, {Sprayberry}, {Staten}, {Stefanik}, {Tarle}, {Sien Tie}, {Tinker},
  {Tojeiro}, {Valdes}, {Valenzuela}, {Valluri}, {Vargas-Magana}, {Verde},
  {Walker}, {Wang}, {Wang}, {Weaver}, {Weaverdyck}, {Wechsler}, {Weinberg},
  {White}, {Yang}, {Yeche}, {Zhang}, {Zhao}, {Zheng}, {Zhou}, {Zhou}, {Zhu},
  {Zou}, \& {Zu}}]{DESI16}
{DESI Collaboration}, {Aghamousa}, A., {Aguilar}, J., {et~al.} 2016, arXiv
  e-prints, arXiv:1611.00036.
\newblock \doarXiv{1611.00036}

\bibitem[{{Dey} {et~al.}(2019{\natexlab{a}}){Dey}, {Schlegel}, {Lang}, {Blum},
  {Burleigh}, {Fan}, {Findlay}, {Finkbeiner}, {Herrera}, {Juneau}, {Landriau},
  {Levi}, {McGreer}, {Meisner}, {Myers}, {Moustakas}, {Nugent}, {Patej},
  {Schlafly}, {Walker}, {Valdes}, {Weaver}, {Y{\`e}che}, {Zou}, {Zhou},
  {Abareshi}, {Abbott}, {Abolfathi}, {Aguilera}, {Alam}, {Allen}, {Alvarez},
  {Annis}, {Ansarinejad}, {Aubert}, {Beechert}, {Bell}, {BenZvi}, {Beutler},
  {Bielby}, {Bolton}, {Brice{\~n}o}, {Buckley-Geer}, {Butler}, {Calamida},
  {Carlberg}, {Carter}, {Casas}, {Castander}, {Choi}, {Comparat},
  {Cukanovaite}, {Delubac}, {DeVries}, {Dey}, {Dhungana}, {Dickinson}, {Ding},
  {Donaldson}, {Duan}, {Duckworth}, {Eftekharzadeh}, {Eisenstein}, {Etourneau},
  {Fagrelius}, {Farihi}, {Fitzpatrick}, {Font-Ribera}, {Fulmer},
  {G{\"a}nsicke}, {Gaztanaga}, {George}, {Gerdes}, {Gontcho}, {Gorgoni},
  {Green}, {Guy}, {Harmer}, {Hernandez}, {Honscheid}, {Huang}, {James},
  {Jannuzi}, {Jiang}, {Joyce}, {Karcher}, {Karkar}, {Kehoe}, {Kneib},
  {Kueter-Young}, {Lan}, {Lauer}, {Le Guillou}, {Le Van Suu}, {Lee}, {Lesser},
  {Perreault Levasseur}, {Li}, {Mann}, {Marshall}, {Mart{\'\i}nez-V{\'a}zquez},
  {Martini}, {du Mas des Bourboux}, {McManus}, {Meier}, {M{\'e}nard},
  {Metcalfe}, {Mu{\~n}oz-Guti{\'e}rrez}, {Najita}, {Napier}, {Narayan},
  {Newman}, {Nie}, {Nord}, {Norman}, {Olsen}, {Paat}, {Palanque-Delabrouille},
  {Peng}, {Poppett}, {Poremba}, {Prakash}, {Rabinowitz}, {Raichoor}, {Rezaie},
  {Robertson}, {Roe}, {Ross}, {Ross}, {Rudnick}, {Safonova}, {Saha},
  {S{\'a}nchez}, {Savary}, {Schweiker}, {Scott}, {Seo}, {Shan}, {Silva},
  {Slepian}, {Soto}, {Sprayberry}, {Staten}, {Stillman}, {Stupak}, {Summers},
  {Sien Tie}, {Tirado}, {Vargas-Maga{\~n}a}, {Vivas}, {Wechsler}, {Williams},
  {Yang}, {Yang}, {Yapici}, {Zaritsky}, {Zenteno}, {Zhang}, {Zhang}, {Zhou}, \&
  {Zhou}}]{Dey+19}
{Dey}, A., {Schlegel}, D.~J., {Lang}, D., {et~al.} 2019{\natexlab{a}}, \aj,
  157, 168, \dodoi{10.3847/1538-3881/ab089d}

\bibitem[{{Dey} {et~al.}(2019{\natexlab{b}}){Dey}, {Schlegel}, {Lang}, {Blum},
  {Burleigh}, {Fan}, {Findlay}, {Finkbeiner}, {Herrera}, {Juneau}, {Landriau},
  {Levi}, {McGreer}, {Meisner}, {Myers}, {Moustakas}, {Nugent}, {Patej},
  {Schlafly}, {Walker}, {Valdes}, {Weaver}, {Y{\`e}che}, {Zou}, {Zhou},
  {Abareshi}, {Abbott}, {Abolfathi}, {Aguilera}, {Alam}, {Allen}, {Alvarez},
  {Annis}, {Ansarinejad}, {Aubert}, {Beechert}, {Bell}, {BenZvi}, {Beutler},
  {Bielby}, {Bolton}, {Brice{\~n}o}, {Buckley-Geer}, {Butler}, {Calamida},
  {Carlberg}, {Carter}, {Casas}, {Castander}, {Choi}, {Comparat},
  {Cukanovaite}, {Delubac}, {DeVries}, {Dey}, {Dhungana}, {Dickinson}, {Ding},
  {Donaldson}, {Duan}, {Duckworth}, {Eftekharzadeh}, {Eisenstein}, {Etourneau},
  {Fagrelius}, {Farihi}, {Fitzpatrick}, {Font-Ribera}, {Fulmer},
  {G{\"a}nsicke}, {Gaztanaga}, {George}, {Gerdes}, {Gontcho}, {Gorgoni},
  {Green}, {Guy}, {Harmer}, {Hernandez}, {Honscheid}, {Huang}, {James},
  {Jannuzi}, {Jiang}, {Joyce}, {Karcher}, {Karkar}, {Kehoe}, {Kneib},
  {Kueter-Young}, {Lan}, {Lauer}, {Le Guillou}, {Le Van Suu}, {Lee}, {Lesser},
  {Perreault Levasseur}, {Li}, {Mann}, {Marshall}, {Mart{\'\i}nez-V{\'a}zquez},
  {Martini}, {du Mas des Bourboux}, {McManus}, {Meier}, {M{\'e}nard},
  {Metcalfe}, {Mu{\~n}oz-Guti{\'e}rrez}, {Najita}, {Napier}, {Narayan},
  {Newman}, {Nie}, {Nord}, {Norman}, {Olsen}, {Paat}, {Palanque-Delabrouille},
  {Peng}, {Poppett}, {Poremba}, {Prakash}, {Rabinowitz}, {Raichoor}, {Rezaie},
  {Robertson}, {Roe}, {Ross}, {Ross}, {Rudnick}, {Safonova}, {Saha},
  {S{\'a}nchez}, {Savary}, {Schweiker}, {Scott}, {Seo}, {Shan}, {Silva},
  {Slepian}, {Soto}, {Sprayberry}, {Staten}, {Stillman}, {Stupak}, {Summers},
  {Sien Tie}, {Tirado}, {Vargas-Maga{\~n}a}, {Vivas}, {Wechsler}, {Williams},
  {Yang}, {Yang}, {Yapici}, {Zaritsky}, {Zenteno}, {Zhang}, {Zhang}, {Zhou}, \&
  {Zhou}}]{Dey+2019}
---. 2019{\natexlab{b}}, \aj, 157, 168, \dodoi{10.3847/1538-3881/ab089d}

\bibitem[{{Dichiara} {et~al.}(2021){Dichiara}, {Becerra}, {Chase}, {Troja},
  {Lee}, {Watson}, {Butler}, {O'Connor}, {Pereyra}, {L{\'o}pez}, {Lien},
  {Gottlieb}, \& {Kutyrev}}]{Dichiara+21}
{Dichiara}, S., {Becerra}, R.~L., {Chase}, E.~A., {et~al.} 2021, \apjl, 923,
  L32, \dodoi{10.3847/2041-8213/ac4259}

\bibitem[{{Dimitriadis} {et~al.}(2019){Dimitriadis}, {Jones}, {Siebert},
  {Brown}, {Arcavi}, {Bloom}, {Bostroem}, {Coulter}, {Drout}, {Ebeling},
  {Filippenko}, {Foley}, {Howell}, {Hung}, {Jha}, {Kasen}, {Kilpatrick},
  {Piro}, {Prochaska}, {Quataert}, {Ramirez-Ruiz}, {Riess}, {Rojas-Bravo},
  {Sand}, {Scolnic}, {Siellez}, {Valenti}, \& {Zheng}}]{GCN2019ebq_Keck2}
{Dimitriadis}, G., {Jones}, D.~O., {Siebert}, M.~R., {et~al.} 2019, GRB
  Coordinates Network, 24358, 1

\bibitem[{{Dobie} {et~al.}(2019){Dobie}, {Stewart}, {Murphy}, {Lenc}, {Wang},
  {Kaplan}, {Andreoni}, {Banfield}, {Brown}, {Corsi}, {De}, {Goldstein},
  {Hallinan}, {Hotan}, {Hotokezaka}, {Jaodand}, {Karambelkar}, {Kasliwal},
  {McConnell}, {Mooley}, {Moss}, {Newman}, {Perley}, {Prakash}, {Pritchard},
  {Sadler}, {Sharma}, {Ward}, {Whiting}, \& {Zhou}}]{Dobie+19}
{Dobie}, D., {Stewart}, A., {Murphy}, T., {et~al.} 2019, \apjl, 887, L13,
  \dodoi{10.3847/2041-8213/ab59db}

\bibitem[{{Dobie} {et~al.}(2021){Dobie}, {Stewart}, {Hotokezaka}, {Murphy},
  {Kaplan}, {Buckley}, {Cooke}, {Ho}, {Lenc}, {Leung}, {Gromadzki}, {O'Brien},
  {Pintaldi}, {Pritchard}, {Wang}, \& {Wang}}]{Dobie+21}
{Dobie}, D., {Stewart}, A., {Hotokezaka}, K., {et~al.} 2021, \mnras,
  \dodoi{10.1093/mnras/stab3628}

\bibitem[{{Doctor} {et~al.}(2017){Doctor}, {Kessler}, {Chen}, {Farr}, {Finley},
  {Foley}, {Goldstein}, {Holz}, {Kim}, {Morganson}, {Sako}, {Scolnic}, {Smith},
  {Soares-Santos}, {Spinka}, {Abbott}, {Abdalla}, {Allam}, {Annis}, {Bechtol},
  {Benoit-L{\'e}vy}, {Bertin}, {Brooks}, {Buckley-Geer}, {Burke}, {Carnero
  Rosell}, {Carrasco Kind}, {Carretero}, {Cunha}, {D'Andrea}, {da Costa},
  {DePoy}, {Desai}, {Diehl}, {Drlica-Wagner}, {Eifler}, {Frieman},
  {Garc{\'\i}a-Bellido}, {Gaztanaga}, {Gerdes}, {Gruendl}, {Gschwend},
  {Gutierrez}, {James}, {Krause}, {Kuehn}, {Kuropatkin}, {Lahav}, {Li}, {Lima},
  {Maia}, {March}, {Marshall}, {Menanteau}, {Miquel}, {Neilsen}, {Nichol},
  {Nord}, {Plazas}, {Romer}, {Sanchez}, {Scarpine}, {Schubnell},
  {Sevilla-Noarbe}, {Smith}, {Sobreira}, {Suchyta}, {Swanson}, {Tarle},
  {Walker}, {Wester}, \& {DES Collaboration}}]{Doctor+17}
{Doctor}, Z., {Kessler}, R., {Chen}, H.~Y., {et~al.} 2017, \apj, 837, 57,
  \dodoi{10.3847/1538-4357/aa5d09}

\bibitem[{{Drout} {et~al.}(2014){Drout}, {Chornock}, {Soderberg}, {Sanders},
  {McKinnon}, {Rest}, {Foley}, {Milisavljevic}, {Margutti}, {Berger},
  {Calkins}, {Fong}, {Gezari}, {Huber}, {Kankare}, {Kirshner}, {Leibler},
  {Lunnan}, {Mattila}, {Marion}, {Narayan}, {Riess}, {Roth}, {Scolnic},
  {Smartt}, {Tonry}, {Burgett}, {Chambers}, {Hodapp}, {Jedicke}, {Kaiser},
  {Magnier}, {Metcalfe}, {Morgan}, {Price}, \& {Waters}}]{Drout+14}
{Drout}, M.~R., {Chornock}, R., {Soderberg}, A.~M., {et~al.} 2014, \apj, 794,
  23, \dodoi{10.1088/0004-637X/794/1/23}

\bibitem[{{Drout} {et~al.}(2017){Drout}, {Piro}, {Shappee}, {Kilpatrick},
  {Simon}, {Contreras}, {Coulter}, {Foley}, {Siebert}, {Morrell}, {Boutsia},
  {Di Mille}, {Holoien}, {Kasen}, {Kollmeier}, {Madore}, {Monson},
  {Murguia-Berthier}, {Pan}, {Prochaska}, {Ramirez-Ruiz}, {Rest}, {Adams},
  {Alatalo}, {Ba{\~n}ados}, {Baughman}, {Beers}, {Bernstein}, {Bitsakis},
  {Campillay}, {Hansen}, {Higgs}, {Ji}, {Maravelias}, {Marshall}, {Moni Bidin},
  {Prieto}, {Rasmussen}, {Rojas-Bravo}, {Strom}, {Ulloa},
  {Vargas-Gonz{\'a}lez}, {Wan}, \& {Whitten}}]{Drout+17}
{Drout}, M.~R., {Piro}, A.~L., {Shappee}, B.~J., {et~al.} 2017, Science, 358,
  1570, \dodoi{10.1126/science.aaq0049}

\bibitem[{{Evans} {et~al.}(2017){Evans}, {Cenko}, {Kennea}, {Emery}, {Kuin},
  {Korobkin}, {Wollaeger}, {Fryer}, {Madsen}, {Harrison}, {Xu}, {Nakar},
  {Hotokezaka}, {Lien}, {Campana}, {Oates}, {Troja}, {Breeveld}, {Marshall},
  {Barthelmy}, {Beardmore}, {Burrows}, {Cusumano}, {D'A{\`\i}}, {D'Avanzo},
  {D'Elia}, {de Pasquale}, {Even}, {Fontes}, {Forster}, {Garcia}, {Giommi},
  {Grefenstette}, {Gronwall}, {Hartmann}, {Heida}, {Hungerford}, {Kasliwal},
  {Krimm}, {Levan}, {Malesani}, {Melandri}, {Miyasaka}, {Nousek}, {O'Brien},
  {Osborne}, {Pagani}, {Page}, {Palmer}, {Perri}, {Pike}, {Racusin}, {Rosswog},
  {Siegel}, {Sakamoto}, {Sbarufatti}, {Tagliaferri}, {Tanvir}, \&
  {Tohuvavohu}}]{Evans+17}
{Evans}, P.~A., {Cenko}, S.~B., {Kennea}, J.~A., {et~al.} 2017, Science, 358,
  1565, \dodoi{10.1126/science.aap9580}

\bibitem[{{Flesch}(2015)}]{Flesch15}
{Flesch}, E.~W. 2015, \pasa, 32, e010, \dodoi{10.1017/pasa.2015.10}

\bibitem[{{Flesch}(2021)}]{Flesch21}
---. 2021, VizieR Online Data Catalog, VII/290

\bibitem[{{Fong} \& {Berger}(2013)}]{FongBerger13}
{Fong}, W., \& {Berger}, E. 2013, \apj, 776, 18,
  \dodoi{10.1088/0004-637X/776/1/18}

\bibitem[{{Fong} {et~al.}(2015){Fong}, {Berger}, {Margutti}, \&
  {Zauderer}}]{fong+15}
{Fong}, W., {Berger}, E., {Margutti}, R., \& {Zauderer}, B.~A. 2015, ApJ, 815,
  102, \dodoi{10.1088/0004-637X/815/2/102}

\bibitem[{{Fong} {et~al.}(2021){Fong}, {Laskar}, {Rastinejad}, {Escorial},
  {Schroeder}, {Barnes}, {Kilpatrick}, {Paterson}, {Berger}, {Metzger}, {Dong},
  {Nugent}, {Strausbaugh}, {Blanchard}, {Goyal}, {Cucchiara}, {Terreran},
  {Alexander}, {Eftekhari}, {Fryer}, {Margalit}, {Margutti}, \&
  {Nicholl}}]{Fong+21}
{Fong}, W., {Laskar}, T., {Rastinejad}, J., {et~al.} 2021, \apj, 906, 127,
  \dodoi{10.3847/1538-4357/abc74a}

\bibitem[{{Foucart} {et~al.}(2013){Foucart}, {Deaton}, {Duez}, {Kidder},
  {MacDonald}, {Ott}, {Pfeiffer}, {Scheel}, {Szilagyi}, \&
  {Teukolsky}}]{Foucart2013}
{Foucart}, F., {Deaton}, M.~B., {Duez}, M.~D., {et~al.} 2013, \prd, 87, 084006,
  \dodoi{10.1103/PhysRevD.87.084006}

\bibitem[{{Fremling} {et~al.}(2019){Fremling}, {Dahiwale}, \&
  {Dugas}}]{TNS2019pzb}
{Fremling}, C., {Dahiwale}, A., \& {Dugas}, A. 2019, Transient Name Server
  Classification Report, 2019-1923, 1

\bibitem[{{Fryer} {et~al.}(2015){Fryer}, {Belczynski}, {Ramirez-Ruiz},
  {Rosswog}, {Shen}, \& {Steiner}}]{Fryer+15}
{Fryer}, C.~L., {Belczynski}, K., {Ramirez-Ruiz}, E., {et~al.} 2015, \apj, 812,
  24, \dodoi{10.1088/0004-637X/812/1/24}

\bibitem[{{Gaia Collaboration} {et~al.}(2016){Gaia Collaboration}, {Prusti},
  {de Bruijne}, {Brown}, {Vallenari}, {Babusiaux}, {Bailer-Jones}, {Bastian},
  {Biermann}, {Evans}, {Eyer}, {Jansen}, {Jordi}, {Klioner}, {Lammers},
  {Lindegren}, {Luri}, {Mignard}, {Milligan}, {Panem}, {Poinsignon},
  {Pourbaix}, {Randich}, {Sarri}, {Sartoretti}, {Siddiqui}, {Soubiran},
  {Valette}, {van Leeuwen}, {Walton}, {Aerts}, {Arenou}, {Cropper}, {Drimmel},
  {H{\o}g}, {Katz}, {Lattanzi}, {O'Mullane}, {Grebel}, {Holland}, {Huc},
  {Passot}, {Bramante}, {Cacciari}, {Casta{\~n}eda}, {Chaoul}, {Cheek}, {De
  Angeli}, {Fabricius}, {Guerra}, {Hern{\'a}ndez}, {Jean-Antoine-Piccolo},
  {Masana}, {Messineo}, {Mowlavi}, {Nienartowicz}, {Ord{\'o}{\~n}ez-Blanco},
  {Panuzzo}, {Portell}, {Richards}, {Riello}, {Seabroke}, \&
  {Tanga}}]{GaiaMission16}
{Gaia Collaboration}, {Prusti}, T., {de Bruijne}, J.~H.~J., {et~al.} 2016,
  \aap, 595, A1, \dodoi{10.1051/0004-6361/201629272}

\bibitem[{{Gaia Collaboration} {et~al.}(2021){Gaia Collaboration}, {Brown},
  {Vallenari}, {Prusti}, {de Bruijne}, {Babusiaux}, {Biermann}, {Creevey},
  {Evans}, {Eyer}, {Hutton}, {Jansen}, {Jordi}, {Klioner}, {Lammers},
  {Lindegren}, {Luri}, {Mignard}, {Panem}, {Pourbaix}, {Randich}, {Sartoretti},
  {Soubiran}, {Walton}, {Arenou}, {Bailer-Jones}, {Bastian}, {Cropper},
  {Drimmel}, {Katz}, {Lattanzi}, {van Leeuwen}, {Bakker}, {Cacciari},
  {Casta{\~n}eda}, {De Angeli}, {Ducourant}, {Fabricius}, {Fouesneau},
  {Fr{\'e}mat}, {Guerra}, {Guerrier}, {Guiraud}, {Jean-Antoine Piccolo},
  {Masana}, {Messineo}, {Mowlavi}, {Nicolas}, {Nienartowicz}, {Pailler},
  {Panuzzo}, {Riclet}, {Roux}, {Seabroke}, {Sordo}, {Tanga}, {Th{\'e}venin},
  {Gracia-Abril}, {Portell}, {Teyssier}, {Altmann}, {Andrae}, {Bellas-Velidis},
  {Benson}, {Berthier}, {Blomme}, {Brugaletta}, {Burgess}, {Busso}, {Carry},
  {Cellino}, {Cheek}, {Clementini}, {Damerdji}, {Davidson}, {Delchambre},
  {Dell'Oro}, {Fern{\'a}ndez-Hern{\'a}ndez}, {Galluccio}, {Garc{\'\i}a-Lario},
  {Garcia-Reinaldos}, {Gonz{\'a}lez-N{\'u}{\~n}ez}, {Gosset}, {Haigron},
  {Halbwachs}, {Hambly}, {Harrison}, {Hatzidimitriou}, {Heiter},
  {Hern{\'a}ndez}, {Hestroffer}, {Hodgkin}, {Holl}, {Jan{\ss}en}, {Jevardat de
  Fombelle}, {Jordan}, {Krone-Martins}, {Lanzafame}, {L{\"o}ffler}, {Lorca},
  {Manteiga}, {Marchal}, {Marrese}, {Moitinho}, {Mora}, {Muinonen}, {Osborne},
  {Pancino}, {Pauwels}, {Petit}, {Recio-Blanco}, {Richards}, {Riello},
  {Rimoldini}, {Robin}, {Roegiers}, {Rybizki}, {Sarro}, {Siopis}, {Smith},
  {Sozzetti}, {Ulla}, {Utrilla}, {van Leeuwen}, {van Reeven}, {Abbas}, {Abreu
  Aramburu}, {Accart}, {Aerts}, {Aguado}, {Ajaj}, {Altavilla}, {{\'A}lvarez},
  {{\'A}lvarez Cid-Fuentes}, {Alves}, {Anderson}, {Anglada Varela}, {Antoja},
  {Audard}, {Baines}, {Baker}, {Balaguer-N{\'u}{\~n}ez}, {Balbinot}, {Balog},
  {Barache}, {Barbato}, {Barros}, {Barstow}, {Bartolom{\'e}}, {Bassilana},
  {Bauchet}, {Baudesson-Stella}, {Becciani}, {Bellazzini}, {Bernet}, {Bertone},
  {Bianchi}, {Blanco-Cuaresma}, {Boch}, {Bombrun}, {Bossini}, {Bouquillon},
  {Bragaglia}, {Bramante}, {Breedt}, {Bressan}, {Brouillet}, {Bucciarelli},
  {Burlacu}, {Busonero}, {Butkevich}, {Buzzi}, {Caffau}, {Cancelliere},
  {C{\'a}novas}, {Cantat-Gaudin}, {Carballo}, {Carlucci}, {Carnerero},
  {Carrasco}, {Casamiquela}, {Castellani}, {Castro-Ginard}, {Castro Sampol},
  {Chaoul}, {Charlot}, {Chemin}, {Chiavassa}, {Cioni}, {Comoretto}, {Cooper},
  {Cornez}, {Cowell}, {Crifo}, {Crosta}, {Crowley}, {Dafonte}, {Dapergolas},
  {David}, {David}, {de Laverny}, {De Luise}, {De March}, {De Ridder}, {de
  Souza}, {de Teodoro}, {de Torres}, {del Peloso}, {del Pozo}, {Delbo},
  {Delgado}, {Delgado}, {Delisle}, {Di Matteo}, {Diakite}, {Diener},
  {Distefano}, {Dolding}, {Eappachen}, {Edvardsson}, {Enke}, {Esquej}, {Fabre},
  {Fabrizio}, {Faigler}, {Fedorets}, {Fernique}, {Fienga}, {Figueras},
  {Fouron}, {Fragkoudi}, {Fraile}, {Franke}, {Gai}, {Garabato},
  {Garcia-Gutierrez}, {Garc{\'\i}a-Torres}, {Garofalo}, \&
  {Gavras}}]{Gaia_eDR3}
{Gaia Collaboration}, {Brown}, A.~G.~A., {Vallenari}, A., {et~al.} 2021, \aap,
  649, A1, \dodoi{10.1051/0004-6361/202039657}

\bibitem[{{Garcia} {et~al.}(2020){Garcia}, {Morgan}, {Herner}, {Palmese},
  {Soares-Santos}, {Annis}, {Brout}, {Vivas}, {Drlica-Wagner}, {Santana-Silva},
  {Tucker}, {Allam}, {Wiesner}, {Garc{\'\i}a-Bellido}, {Gill}, {Sako},
  {Kessler}, {Davis}, {Scolnic}, {Casares}, {Chen}, {Conselice}, {Cooke},
  {Doctor}, {Foley}, {Horvath}, {Howell}, {Kilpatrick}, {Lidman}, {Olivares
  E.}, {Paz-Chinch{\'o}n}, {Pineda-G.}, {Quirola-V{\'a}squez}, {Rest},
  {Sherman}, {Abbott}, {Aguena}, {Avila}, {Bertin}, {Bhargava}, {Brooks},
  {Burke}, {Carnero Rosell}, {Carrasco Kind}, {Carretero}, {Costanzi}, {da
  Costa}, {Desai}, {Diehl}, {Dietrich}, {Doel}, {Everett}, {Flaugher},
  {Fosalba}, {Friedel}, {Frieman}, {Gaztanaga}, {Gerdes}, {Gruen}, {Gruendl},
  {Gschwend}, {Gutierrez}, {Hinton}, {Hollowood}, {Honscheid}, {James},
  {Kuehn}, {Kuropatkin}, {Lahav}, {Lima}, {Maia}, {March}, {Marshall},
  {Menanteau}, {Miquel}, {Ogando}, {Plazas}, {Romer}, {Roodman}, {Sanchez},
  {Scarpine}, {Schubnell}, {Serrano}, {Sevilla-Noarbe}, {Smith}, {Suchyta},
  {Swanson}, {Tarle}, {Thomas}, {Varga}, {Walker}, {Weller}, \& {DES
  Collaboration}}]{Garcia+20}
{Garcia}, A., {Morgan}, R., {Herner}, K., {et~al.} 2020, \apj, 903, 75,
  \dodoi{10.3847/1538-4357/abb823}

\bibitem[{{Goldstein} {et~al.}(2019){Goldstein}, {Andreoni}, {Nugent},
  {Kasliwal}, {Coughlin}, {Anand}, {Bloom}, {Mart{\'\i}nez-Palomera}, {Zhang},
  {Ahumada}, {Bagdasaryan}, {Cooke}, {De}, {Duev}, {Fremling}, {Gatkine},
  {Graham}, {Ofek}, {Singer}, \& {Yan}}]{Goldstein+19}
{Goldstein}, D.~A., {Andreoni}, I., {Nugent}, P.~E., {et~al.} 2019, \apjl, 881,
  L7, \dodoi{10.3847/2041-8213/ab3046}

\bibitem[{{Gomez} {et~al.}(2019){Gomez}, {Hosseinzadeh}, {Cowperthwaite},
  {Villar}, {Berger}, {Gardner}, {Alexander}, {Blanchard}, {Chornock}, {Drout},
  {Eftekhari}, {Fong}, {Gill}, {Margutti}, {Nicholl}, {Paterson}, \&
  {Williams}}]{Gomez+19}
{Gomez}, S., {Hosseinzadeh}, G., {Cowperthwaite}, P.~S., {et~al.} 2019, \apjl,
  884, L55, \dodoi{10.3847/2041-8213/ab4ad5}

\bibitem[{{Gompertz} {et~al.}(2018){Gompertz}, {Levan}, {Tanvir}, {Hjorth},
  {Covino}, {Evans}, {Fruchter}, {Gonz{\'a}lez-Fern{\'a}ndez}, {Jin}, {Lyman},
  {Oates}, {O'Brien}, \& {Wiersema}}]{Gompertz+18}
{Gompertz}, B.~P., {Levan}, A.~J., {Tanvir}, N.~R., {et~al.} 2018, \apj, 860,
  62, \dodoi{10.3847/1538-4357/aac206}

\bibitem[{{Gompertz} {et~al.}(2020){Gompertz}, {Cutter}, {Steeghs}, {Galloway},
  {Lyman}, {Ulaczyk}, {Dyer}, {Ackley}, {Dhillon}, {O'Brien}, {Ramsay},
  {Poshyachinda}, {Kotak}, {Nuttall}, {Breton}, {Pall{\'e}}, {Pollacco},
  {Thrane}, {Aukkaravittayapun}, {Awiphan}, {Brown}, {Burhanudin}, {Chote},
  {Chrimes}, {Daw}, {Duffy}, {Eyles-Ferris}, {Heikkil{\"a}}, {Irawati},
  {Kennedy}, {Killestein}, {Levan}, {Littlefair}, {Makrygianni}, {Marsh}, {Mata
  S{\'a}nchez}, {Mattila}, {Maund}, {McCormac}, {Mkrtichian}, {Mong},
  {Mullaney}, {M{\"u}ller}, {Obradovic}, {Rol}, {Sawangwit}, {Stanway},
  {Starling}, {Str{\o}m}, {Tooke}, {West}, \& {Wiersema}}]{Gompertz+20b}
{Gompertz}, B.~P., {Cutter}, R., {Steeghs}, D., {et~al.} 2020, \mnras, 497,
  726, \dodoi{10.1093/mnras/staa1845}

\bibitem[{{Graham} {et~al.}(2020){Graham}, {Ford}, {McKernan}, {Ross}, {Stern},
  {Burdge}, {Coughlin}, {Djorgovski}, {Drake}, {Duev}, {Kasliwal}, {Mahabal},
  {van Velzen}, {Belecki}, {Bellm}, {Burruss}, {Cenko}, {Cunningham}, {Helou},
  {Kulkarni}, {Masci}, {Prince}, {Reiley}, {Rodriguez}, {Rusholme}, {Smith}, \&
  {Soumagnac}}]{Graham+20}
{Graham}, M.~J., {Ford}, K.~E.~S., {McKernan}, B., {et~al.} 2020, \prl, 124,
  251102, \dodoi{10.1103/PhysRevLett.124.251102}

\bibitem[{{Gromadzki}(2019)}]{TNS2019qem}
{Gromadzki}, M. 2019, Transient Name Server Classification Report, 2019-2193, 1

\bibitem[{{Harris} {et~al.}(2020){Harris}, {Millman}, {van der Walt},
  {Gommers}, {Virtanen}, {Cournapeau}, {Wieser}, {Taylor}, {Berg}, {Smith},
  {Kern}, {Picus}, {Hoyer}, {van Kerkwijk}, {Brett}, {Haldane}, {del R{\'\i}o},
  {Wiebe}, {Peterson}, {G{\'e}rard-Marchant}, {Sheppard}, {Reddy}, {Weckesser},
  {Abbasi}, {Gohlke}, \& {Oliphant}}]{numpy}
{Harris}, C.~R., {Millman}, K.~J., {van der Walt}, S.~J., {et~al.} 2020, \nat,
  585, 357, \dodoi{10.1038/s41586-020-2649-2}

\bibitem[{{Hartley} {et~al.}(2021){Hartley}, {Choi}, {Amon}, {Gruendl},
  {Sheldon}, {Harrison}, {Bernstein}, {Sevilla-Noarbe}, {Yanny}, {Eckert},
  {Diehl}, {Alarcon}, {Banerji}, {Bechtol}, {Buchs}, {Cantu}, {Conselice},
  {Cordero}, {Davis}, {Davis}, {Dodelson}, {Drlica-Wagner}, {Everett},
  {Fert{\'e}}, {Gruen}, {Honscheid}, {Jarvis}, {Johnson}, {Kokron}, {MacCrann},
  {Myles}, {Pace}, {Palmese}, {Paz-Chinch{\'o}n}, {Pereira}, {Plazas}, {Prat},
  {Rodriguez-Monroy}, {Rykoff}, {Samuroff}, {S{\'a}nchez}, {Secco},
  {Tarsitano}, {Tong}, {Troxel}, {Vasquez}, {Wang}, {Zhou}, {Abbott}, {Aguena},
  {Allam}, {Annis}, {Bacon}, {Bertin}, {Bhargava}, {Brooks}, {Burke}, {Carnero
  Rosell}, {Carrasco Kind}, {Carretero}, {Castander}, {Costanzi}, {Crocce}, {da
  Costa}, {De Vicente}, {DeRose}, {Desai}, {Dietrich}, {Eifler}, {Elvin-Poole},
  {Ferrero}, {Flaugher}, {Fosalba}, {Garc{\'\i}a-Bellido}, {Gaztanaga},
  {Gerdes}, {Gschwend}, {Gutierrez}, {Hinton}, {Hollowood}, {Huterer}, {James},
  {Kent}, {Krause}, {Kuehn}, {Kuropatkin}, {Lahav}, {Lin}, {Maia}, {March},
  {Marshall}, {Martini}, {Melchior}, {Menanteau}, {Miquel}, {Mohr}, {Morgan},
  {Neilsen}, {Ogando}, {Pandey}, {Romer}, {Roodman}, {Sako}, {Sanchez},
  {Scarpine}, {Serrano}, {Smith}, {Soares-Santos}, {Suchyta}, {Swanson},
  {Tarle}, {Thomas}, {To}, {Varga}, {Walker}, {Wester}, {Wilkinson}, \&
  {Zuntz}}]{Hartley+21}
{Hartley}, W.~G., {Choi}, A., {Amon}, A., {et~al.} 2021, \mnras,
  \dodoi{10.1093/mnras/stab3055}

\bibitem[{{Hosseinzadeh} {et~al.}(2019){Hosseinzadeh}, {Cowperthwaite},
  {Gomez}, {Villar}, {Nicholl}, {Margutti}, {Berger}, {Chornock}, {Paterson},
  {Fong}, {Savchenko}, {Short}, {Alexander}, {Blanchard}, {Braga}, {Calkins},
  {Cartier}, {Coppejans}, {Eftekhari}, {Laskar}, {Ly}, {Patton}, {Pelisoli},
  {Reichart}, {Terreran}, \& {Williams}}]{Hosseinzadeh+19}
{Hosseinzadeh}, G., {Cowperthwaite}, P.~S., {Gomez}, S., {et~al.} 2019, \apjl,
  880, L4, \dodoi{10.3847/2041-8213/ab271c}

\bibitem[{{Hu} {et~al.}(2017){Hu}, {Wu}, {Andreoni}, {Ashley}, {Cooke}, {Cui},
  {Du}, {Dai}, {Gu}, {Hu}, {Lu}, {Li}, {Li}, {Liang}, {Liu}, {Ma}, {Shang},
  {Sun}, {Suntzeff}, {Tao}, {Udden}, {Wang}, {Wang}, {Wen}, {Xiao}, {Su},
  {Yang}, {Yang}, {Yuan}, {Zhou}, {Zhang}, {Zhou}, \& {Zhu}}]{Hu+17}
{Hu}, L., {Wu}, X., {Andreoni}, I., {et~al.} 2017, Science Bulletin, 62, 1433,
  \dodoi{10.1016/j.scib.2017.10.006}

\bibitem[{{Jayasinghe} {et~al.}(2019){Jayasinghe}, {Stanek}, {Kochanek},
  {Shappee}, {Holoien}, {Thompson}, {Prieto}, {Dong}, {Pawlak}, {Pejcha},
  {Shields}, {Pojmanski}, {Otero}, {Hurst}, {Britt}, \& {Will}}]{Jayasinghe+19}
{Jayasinghe}, T., {Stanek}, K.~Z., {Kochanek}, C.~S., {et~al.} 2019, \mnras,
  485, 961, \dodoi{10.1093/mnras/stz444}

\bibitem[{{Jencson} {et~al.}(2019){Jencson}, {de}, {Anand}, {Kasliwal},
  {Andreoni}, {Ahumada}, \& {Perley}}]{GCN2019ebq_Keck}
{Jencson}, J., {de}, K., {Anand}, S., {et~al.} 2019, GRB Coordinates Network,
  24233, 1

\bibitem[{{Kalogera} \& {Baym}(1996)}]{KalogeraBaym96}
{Kalogera}, V., \& {Baym}, G. 1996, \apjl, 470, L61, \dodoi{10.1086/310296}

\bibitem[{{Kasen} {et~al.}(2015){Kasen}, {Fern{\'a}ndez}, \&
  {Metzger}}]{kasen+15}
{Kasen}, D., {Fern{\'a}ndez}, R., \& {Metzger}, B.~D. 2015, MNRAS, 450, 1777,
  \dodoi{10.1093/mnras/stv721}

\bibitem[{{Kasen} {et~al.}(2017){Kasen}, {Metzger}, {Barnes}, {Quataert}, \&
  {Ramirez-Ruiz}}]{kasen+17}
{Kasen}, D., {Metzger}, B., {Barnes}, J., {Quataert}, E., \& {Ramirez-Ruiz}, E.
  2017, Nature, 551, 80, \dodoi{10.1038/nature24453}

\bibitem[{{Kasliwal}(2012)}]{Kasliwal12}
{Kasliwal}, M.~M. 2012, \pasa, 29, 482, \dodoi{10.1071/AS11061}

\bibitem[{{Kasliwal} {et~al.}(2017){Kasliwal}, {Nakar}, {Singer}, {Kaplan},
  {Cook}, {Van Sistine}, {Lau}, {Fremling}, {Gottlieb}, {Jencson}, {Adams},
  {Feindt}, {Hotokezaka}, {Ghosh}, {Perley}, {Yu}, {Piran}, {Allison},
  {Anupama}, {Balasubramanian}, {Bannister}, {Bally}, {Barnes}, {Barway},
  {Bellm}, {Bhalerao}, {Bhattacharya}, {Blagorodnova}, {Bloom}, {Brady},
  {Cannella}, {Chatterjee}, {Cenko}, {Cobb}, {Copperwheat}, {Corsi}, {De},
  {Dobie}, {Emery}, {Evans}, {Fox}, {Frail}, {Frohmaier}, {Goobar}, {Hallinan},
  {Harrison}, {Helou}, {Hinderer}, {Ho}, {Horesh}, {Ip}, {Itoh}, {Kasen},
  {Kim}, {Kuin}, {Kupfer}, {Lynch}, {Madsen}, {Mazzali}, {Miller}, {Mooley},
  {Murphy}, {Ngeow}, {Nichols}, {Nissanke}, {Nugent}, {Ofek}, {Qi}, {Quimby},
  {Rosswog}, {Rusu}, {Sadler}, {Schmidt}, {Sollerman}, {Steele}, {Williamson},
  {Xu}, {Yan}, {Yatsu}, {Zhang}, \& {Zhao}}]{Kasliwal+17}
{Kasliwal}, M.~M., {Nakar}, E., {Singer}, L.~P., {et~al.} 2017, Science, 358,
  1559, \dodoi{10.1126/science.aap9455}

\bibitem[{{Kasliwal} {et~al.}(2020){Kasliwal}, {Anand}, {Ahumada}, {Stein},
  {Carracedo}, {Andreoni}, {Coughlin}, {Singer}, {Kool}, {De}, {Kumar},
  {AlMualla}, {Yao}, {Bulla}, {Dobie}, {Reusch}, {Perley}, {Cenko}, {Bhalerao},
  {Kaplan}, {Sollerman}, {Goobar}, {Copperwheat}, {Bellm}, {Anupama}, {Corsi},
  {Nissanke}, {Agudo}, {Bagdasaryan}, {Barway}, {Belicki}, {Bloom}, {Bolin},
  {Buckley}, {Burdge}, {Burruss}, {Caballero-Garc{\'\i}a}, {Cannella},
  {Castro-Tirado}, {Cook}, {Cooke}, {Cunningham}, {Dahiwale}, {Deshmukh},
  {Dichiara}, {Duev}, {Dutta}, {Feeney}, {Franckowiak}, {Frederick},
  {Fremling}, {Gal-Yam}, {Gatkine}, {Ghosh}, {Goldstein}, {Golkhou}, {Graham},
  {Graham}, {Hankins}, {Helou}, {Hu}, {Ip}, {Jaodand}, {Karambelkar}, {Kong},
  {Kowalski}, {Khandagale}, {Kulkarni}, {Kumar}, {Laher}, {Li}, {Mahabal},
  {Masci}, {Miller}, {Mogotsi}, {Mohite}, {Mooley}, {Mroz}, {Newman}, {Ngeow},
  {Oates}, {Patil}, {Pandey}, {Pavana}, {Pian}, {Riddle},
  {S{\'a}nchez-Ram{\'\i}rez}, {Sharma}, {Singh}, {Smith}, {Soumagnac},
  {Taggart}, {Tan}, {Tzanidakis}, {Troja}, {Valeev}, {Walters}, {Waratkar},
  {Webb}, {Yu}, {Zhang}, {Zhou}, \& {Zolkower}}]{Kasliwal+20}
{Kasliwal}, M.~M., {Anand}, S., {Ahumada}, T., {et~al.} 2020, \apj, 905, 145,
  \dodoi{10.3847/1538-4357/abc335}

\bibitem[{{Kawaguchi} {et~al.}(2016){Kawaguchi}, {Kyutoku}, {Shibata}, \&
  {Tanaka}}]{kawaguchi+16}
{Kawaguchi}, K., {Kyutoku}, K., {Shibata}, M., \& {Tanaka}, M. 2016, ApJ, 825,
  52, \dodoi{10.3847/0004-637X/825/1/52}

\bibitem[{{Kawaguchi} {et~al.}(2020{\natexlab{a}}){Kawaguchi}, {Shibata}, \&
  {Tanaka}}]{Kawaguchi+20b}
{Kawaguchi}, K., {Shibata}, M., \& {Tanaka}, M. 2020{\natexlab{a}}, \apj, 889,
  171, \dodoi{10.3847/1538-4357/ab61f6}

\bibitem[{{Kawaguchi} {et~al.}(2020{\natexlab{b}}){Kawaguchi}, {Shibata}, \&
  {Tanaka}}]{Kawaguchi+20}
---. 2020{\natexlab{b}}, \apj, 893, 153, \dodoi{10.3847/1538-4357/ab8309}

\bibitem[{{Kilpatrick} {et~al.}(2021){Kilpatrick}, {Coulter}, {Arcavi},
  {Brink}, {Dimitriadis}, {Filippenko}, {Foley}, {Howell}, {Jones}, {Kasen},
  {Makler}, {Piro}, {Rojas-Bravo}, {Sand}, {Swift}, {Tucker}, {Zheng}, {Allam},
  {Annis}, {Antilen}, {Bachmann}, {Bloom}, {Bom}, {Bostroem}, {Brout}, {Burke},
  {Butler}, {Butner}, {Campillay}, {Clever}, {Conselice}, {Cooke}, {Dage}, {de
  Carvalho}, {de Jaeger}, {Desai}, {Garcia}, {Garcia-Bellido}, {Gill},
  {Girish}, {Hallakoun}, {Herner}, {Hiramatsu}, {Holz}, {Huber}, {Kawash},
  {McCully}, {Medallon}, {Metzger}, {Modak}, {Morgan}, {Mu{\~n}oz},
  {Mu{\~n}oz-Elgueta}, {Murakami}, {Felipe Olivares}, {Palmese}, {Patra},
  {Pereira}, {Pessi}, {Pineda-Garcia}, {Quirola-V{\'a}squez}, {Ramirez-Ruiz},
  {Rembold}, {Rest}, {Rodr{\'\i}guez}, {Santana-Silva}, {Sherman}, {Siebert},
  {Smith}, {Smith}, {Soares-Santos}, {Stacey}, {Stahl}, {Strader},
  {Strasburger}, {Sunseri}, {Tinyanont}, {Tucker}, {Ulloa}, {Valenti},
  {Vasylyev}, {Wiesner}, \& {Zhang}}]{Kilpatrick+21}
{Kilpatrick}, C.~D., {Coulter}, D.~A., {Arcavi}, I., {et~al.} 2021, \apj, 923,
  258, \dodoi{10.3847/1538-4357/ac23c6}

\bibitem[{{Korobkin} {et~al.}(2021){Korobkin}, {Wollaeger}, {Fryer},
  {Hungerford}, {Rosswog}, {Fontes}, {Mumpower}, {Chase}, {Even}, {Miller},
  {Misch}, \& {Lippuner}}]{Korobkin+21}
{Korobkin}, O., {Wollaeger}, R.~T., {Fryer}, C.~L., {et~al.} 2021, \apj, 910,
  116, \dodoi{10.3847/1538-4357/abe1b5}

\bibitem[{{Landsman}(1993)}]{IDLforever}
{Landsman}, W.~B. 1993, in Astronomical Society of the Pacific Conference
  Series, Vol.~52, Astronomical Data Analysis Software and Systems II, ed.
  R.~J. {Hanisch}, R.~J.~V. {Brissenden}, \& J.~{Barnes}, 246

\bibitem[{{Li} \& {Paczy{\'n}ski}(1998)}]{lipaczynski98}
{Li}, L.-X., \& {Paczy{\'n}ski}, B. 1998, \apjl, 507, L59,
  \dodoi{10.1086/311680}

\bibitem[{{Li} {et~al.}(2011){Li}, {Chornock}, {Leaman}, {Filippenko},
  {Poznanski}, {Wang}, {Ganeshalingam}, \& {Mannucci}}]{Li+11}
{Li}, W., {Chornock}, R., {Leaman}, J., {et~al.} 2011, \mnras, 412, 1473,
  \dodoi{10.1111/j.1365-2966.2011.18162.x}

\bibitem[{{LIGO Scientific Collaboration} \& {Virgo
  Collaboration}(2019{\natexlab{a}})}]{GCN25324}
{LIGO Scientific Collaboration}, \& {Virgo Collaboration}. 2019{\natexlab{a}},
  GRB Coordinates Network, 25324, 1

\bibitem[{{LIGO Scientific Collaboration} \& {Virgo
  Collaboration}(2019{\natexlab{b}})}]{GCN25333}
---. 2019{\natexlab{b}}, GRB Coordinates Network, 25333, 1

\bibitem[{{Lippuner} {et~al.}(2017){Lippuner}, {Fern{\'a}ndez}, {Roberts},
  {Foucart}, {Kasen}, {Metzger}, \& {Ott}}]{Lippuner+17}
{Lippuner}, J., {Fern{\'a}ndez}, R., {Roberts}, L.~F., {et~al.} 2017, \mnras,
  472, 904, \dodoi{10.1093/mnras/stx1987}

\bibitem[{{Lipunov} {et~al.}(2017){Lipunov}, {Gorbovskoy}, {Kornilov},
  {.~Tyurina}, {Balanutsa}, {Kuznetsov}, {Vlasenko}, {Kuvshinov}, {Gorbunov},
  {Buckley}, {Krylov}, {Podesta}, {Lopez}, {Podesta}, {Levato}, {Saffe},
  {Mallamachi}, {Potter}, {Budnev}, {Gress}, {Ishmuhametova}, {Vladimirov},
  {Zimnukhov}, {Yurkov}, {Sergienko}, {Gabovich}, {Rebolo}, {Serra-Ricart},
  {Israelyan}, {Chazov}, {Wang}, {Tlatov}, \& {Panchenko}}]{Lipunov+17}
{Lipunov}, V.~M., {Gorbovskoy}, E., {Kornilov}, V.~G., {et~al.} 2017, \apjl,
  850, L1, \dodoi{10.3847/2041-8213/aa92c0}

\bibitem[{{Lundquist} {et~al.}(2019){Lundquist}, {Paterson}, {Fong}, {Sand},
  {Andrews}, {Shivaei}, {Daly}, {Valenti}, {Yang}, {Christensen}, {Gibbs},
  {Shelly}, {Wyatt}, {Eskandari}, {Kuhn}, {Amaro}, {Arcavi}, {Behroozi},
  {Butler}, {Chomiuk}, {Corsi}, {Drout}, {Egami}, {Fan}, {Foley}, {Frye},
  {Gabor}, {Green}, {Grier}, {Guzman}, {Hamden}, {Howell}, {Jannuzi}, {Kelly},
  {Milne}, {Moe}, {Nugent}, {Olszewski}, {Palazzi}, {Paschalidis}, {Psaltis},
  {Reichart}, {Rest}, {Rossi}, {Schroeder}, {Smith}, {Smith}, {Spekkens},
  {Strader}, {Stark}, {Trilling}, {Veillet}, {Wagner}, {Weiner}, {Wheeler},
  {Williams}, \& {Zabludoff}}]{Lundquist+19}
{Lundquist}, M.~J., {Paterson}, K., {Fong}, W., {et~al.} 2019, \apjl, 881, L26,
  \dodoi{10.3847/2041-8213/ab32f2}

\bibitem[{{Lyke} {et~al.}(2020){Lyke}, {Higley}, {McLane}, {Schurhammer},
  {Myers}, {Ross}, {Dawson}, {Chabanier}, {Martini}, {Busca}, {Mas des
  Bourboux}, {Salvato}, {Streblyanska}, {Zarrouk}, {Burtin}, {Anderson},
  {Bautista}, {Bizyaev}, {Brandt}, {Brinkmann}, {Brownstein}, {Comparat},
  {Green}, {de la Macorra}, {Mu{\~n}oz Guti{\'e}rrez}, {Hou}, {Newman},
  {Palanque-Delabrouille}, {P{\^a}ris}, {Percival}, {Petitjean}, {Rich},
  {Rossi}, {Schneider}, {Smith}, {Vivek}, \& {Weaver}}]{Lyke+20}
{Lyke}, B.~W., {Higley}, A.~N., {McLane}, J.~N., {et~al.} 2020, \apjs, 250, 8,
  \dodoi{10.3847/1538-4365/aba623}

\bibitem[{{Margutti} {et~al.}(2019){Margutti}, {Metzger}, {Chornock}, {Vurm},
  {Roth}, {Grefenstette}, {Savchenko}, {Cartier}, {Steiner}, {Terreran},
  {Margalit}, {Migliori}, {Milisavljevic}, {Alexander}, {Bietenholz},
  {Blanchard}, {Bozzo}, {Brethauer}, {Chilingarian}, {Coppejans}, {Ducci},
  {Ferrigno}, {Fong}, {G{\"o}tz}, {Guidorzi}, {Hajela}, {Hurley}, {Kuulkers},
  {Laurent}, {Mereghetti}, {Nicholl}, {Patnaude}, {Ubertini}, {Banovetz},
  {Bartel}, {Berger}, {Coughlin}, {Eftekhari}, {Frederiks}, {Kozlova},
  {Laskar}, {Svinkin}, {Drout}, {MacFadyen}, \& {Paterson}}]{Margutti+19}
{Margutti}, R., {Metzger}, B.~D., {Chornock}, R., {et~al.} 2019, \apj, 872, 18,
  \dodoi{10.3847/1538-4357/aafa01}

\bibitem[{{Masci} {et~al.}(2019){Masci}, {Laher}, {Rusholme}, {Shupe}, {Groom},
  {Surace}, {Jackson}, {Monkewitz}, {Beck}, {Flynn}, {Terek}, {Landry},
  {Hacopians}, {Desai}, {Howell}, {Brooke}, {Imel}, {Wachter}, {Ye}, {Lin},
  {Cenko}, {Cunningham}, {Rebbapragada}, {Bue}, {Miller}, {Mahabal}, {Bellm},
  {Patterson}, {Juri{\'c}}, {Golkhou}, {Ofek}, {Walters}, {Graham}, {Kasliwal},
  {Dekany}, {Kupfer}, {Burdge}, {Cannella}, {Barlow}, {Van Sistine}, {Giomi},
  {Fremling}, {Blagorodnova}, {Levitan}, {Riddle}, {Smith}, {Helou}, {Prince},
  \& {Kulkarni}}]{ZTFproducts19}
{Masci}, F.~J., {Laher}, R.~R., {Rusholme}, B., {et~al.} 2019, \pasp, 131,
  018003, \dodoi{10.1088/1538-3873/aae8ac}

\bibitem[{{McCully} {et~al.}(2019){McCully}, {Hiramatsu}, {Hiramatsu},
  {Howell}, {Arcavi}, {Drout}, {Burke}, {Peligrino}, {de Carvalho}, {Forster},
  {Foley}, {Coulter}, {Kilpatrick}, {Sand}, {Valenti}, {Soares-Santos},
  {Rembold}, {Resti}, {Kasen}, {Metzger}, {Piro}, {Quataert}, {Ramirez-Ruiz},
  {Wheeler}, {Bauer}, {Brink}, {Cooke}, {Clocchiatti}, {Filippenko},
  {Freedman}, {Garnavich}, {Horvath}, {Jha}, {Kirshner}, {Krisciunas}, {Lin},
  {Madore}, {Makler}, {Prochaska}, {Riess}, {Sturani}, {Suntzeff}, {Tanaka},
  {Tucker}, {Vinko}, {Wang}, {Brown}, {Contrerasi}, {D'Andrea}, {Dimitriadis},
  {Jones}, {Lundquist}, {Narayan}, {Olivares}, {Palmese}, {Pan}, {Scolnic},
  {Zheng}, {Bernardo}, {Bostroem}, {Berthier}, {Rodriguez}, {Rojas-Bravo},
  {Siebert}, \& {Souza}}]{GCN2019ebq_Gemini}
{McCully}, C., {Hiramatsu}, D., {Hiramatsu}, D., {et~al.} 2019, GRB Coordinates
  Network, 24295, 1

\bibitem[{{Metcalfe} {et~al.}(2013){Metcalfe}, {Farrow}, {Cole}, {Draper},
  {Norberg}, {Burgett}, {Chambers}, {Denneau}, {Flewelling}, {Kaiser},
  {Kudritzki}, {Magnier}, {Morgan}, {Price}, {Sweeney}, {Tonry}, {Wainscoat},
  \& {Waters}}]{Metcalfe+13}
{Metcalfe}, N., {Farrow}, D.~J., {Cole}, S., {et~al.} 2013, \mnras, 435, 1825,
  \dodoi{10.1093/mnras/stt1343}

\bibitem[{{Metzger}(2019)}]{metzger19}
{Metzger}, B.~D. 2019, Living Reviews in Relativity, 23, 1,
  \dodoi{10.1007/s41114-019-0024-0}

\bibitem[{{Metzger} \& {Fern{\'a}ndez}(2014)}]{metzgerfernandez14}
{Metzger}, B.~D., \& {Fern{\'a}ndez}, R. 2014, \mnras, 441, 3444,
  \dodoi{10.1093/mnras/stu802}

\bibitem[{{Metzger} {et~al.}(2010){Metzger}, {Mart{\'{\i}}nez-Pinedo},
  {Darbha}, {Quataert}, {Arcones}, {Kasen}, {Thomas}, {Nugent}, {Panov}, \&
  {Zinner}}]{metzger+10}
{Metzger}, B.~D., {Mart{\'{\i}}nez-Pinedo}, G., {Darbha}, S., {et~al.} 2010,
  \mnras, 406, 2650, \dodoi{10.1111/j.1365-2966.2010.16864.x}

\bibitem[{{Morgan} {et~al.}(2020){Morgan}, {Soares-Santos}, {Annis}, {Herner},
  {Garcia}, {Palmese}, {Drlica-Wagner}, {Kessler}, {Garc{\'\i}a-Bellido},
  {Bachmann}, {Sherman}, {Allam}, {Bechtol}, {Bom}, {Brout}, {Butler},
  {Butner}, {Cartier}, {Chen}, {Conselice}, {Cook}, {Davis}, {Doctor}, {Farr},
  {Figueiredo}, {Finley}, {Foley}, {Galarza}, {Gill}, {Gruendl}, {Holz},
  {Kuropatkin}, {Lidman}, {Lin}, {Malik}, {Mann}, {Marriner}, {Marshall},
  {Mart{\'\i}nez-V{\'a}zquez}, {Meza}, {Neilsen}, {Nicolaou}, {Olivares E.},
  {Paz-Chinch{\'o}n}, {Points}, {Quirola-V{\'a}squez}, {Rodriguez}, {Sako},
  {Scolnic}, {Smith}, {Sobreira}, {Tucker}, {Vivas}, {Wiesner}, {Wood},
  {Yanny}, {Zenteno}, {Abbott}, {Aguena}, {Avila}, {Bertin}, {Bhargava},
  {Brooks}, {Burke}, {Carnero Rosell}, {Carrasco Kind}, {Carretero}, {da
  Costa}, {Costanzi}, {De Vicente}, {Desai}, {Diehl}, {Doel}, {Eifler},
  {Everett}, {Flaugher}, {Frieman}, {Gaztanaga}, {Gerdes}, {Gruen}, {Gschwend},
  {Gutierrez}, {Hartley}, {Hinton}, {Hollowood}, {Honscheid}, {James}, {Kuehn},
  {Lahav}, {Lima}, {Maia}, {March}, {Miquel}, {Ogando}, {Plazas}, {Roodman},
  {Sanchez}, {Scarpine}, {Schubnell}, {Serrano}, {Sevilla-Noarbe}, {Suchyta},
  \& {Tarle}}]{Morgan+20}
{Morgan}, R., {Soares-Santos}, M., {Annis}, J., {et~al.} 2020, \apj, 901, 83,
  \dodoi{10.3847/1538-4357/abafaa}

\bibitem[{{Morokuma} {et~al.}(2019){Morokuma}, {Ohta}, {Yoshida}, {Aoki},
  {Tanaka}, {Sasada}, {Nakaoka}, {Akitaya}, {Kawabata}, {Itoh}, \&
  {Utsumi}}]{GCN2019ebq_Subaru}
{Morokuma}, T., {Ohta}, K., {Yoshida}, M., {et~al.} 2019, GRB Coordinates
  Network, 24230, 1

\bibitem[{{Nicholl} {et~al.}(2021){Nicholl}, {Margalit}, {Schmidt}, {Smith},
  {Ridley}, \& {Nuttall}}]{Nicholl+21}
{Nicholl}, M., {Margalit}, B., {Schmidt}, P., {et~al.} 2021, \mnras, 505, 3016,
  \dodoi{10.1093/mnras/stab1523}

\bibitem[{{Oates} {et~al.}(2021){Oates}, {Marshall}, {Breeveld}, {Kuin},
  {Brown}, {De Pasquale}, {Evans}, {Fenney}, {Gronwall}, {Kennea}, {Klingler},
  {Page}, {Siegel}, {Tohuvavohu}, {Ambrosi}, {Barthelmy}, {Beardmore},
  {Bernardini}, {Campana}, {Caputo}, {Cenko}, {Cusumano}, {D'A{\`\i}},
  {D'Avanzo}, {D'Elia}, {Giommi}, {Hartmann}, {Krimm}, {Laha}, {Malesani},
  {Melandri}, {Nousek}, {O'Brien}, {Osborne}, {Pagani}, {Page}, {Palmer},
  {Perri}, {Racusin}, {Sakamoto}, {Sbarufatti}, {Schlieder}, {Tagliaferri}, \&
  {Troja}}]{Oates+21}
{Oates}, S.~R., {Marshall}, F.~E., {Breeveld}, A.~A., {et~al.} 2021, \mnras,
  507, 1296, \dodoi{10.1093/mnras/stab2189}

\bibitem[{{Ohgami} {et~al.}(2021){Ohgami}, {Tominaga}, {Utsumi}, {Niino},
  {Tanaka}, {Banerjee}, {Hamasaki}, {Yoshida}, {Terai}, {Takagi}, {Morokuma},
  {Sasada}, {Akitaya}, {Yasuda}, {Yanagisawa}, \& {Ohsawa}}]{Ohgami+21}
{Ohgami}, T., {Tominaga}, N., {Utsumi}, Y., {et~al.} 2021, \pasj, 73, 350,
  \dodoi{10.1093/pasj/psab002}

\bibitem[{{Paterson} {et~al.}(2021){Paterson}, {Lundquist}, {Rastinejad},
  {Fong}, {Sand}, {Andrews}, {Amaro}, {Eskandari}, {Wyatt}, {Daly}, {Bradley},
  {Zhou-Wright}, {Valenti}, {Yang}, {Christensen}, {Gibbs}, {Shelly},
  {Bilinski}, {Chomiuk}, {Corsi}, {Drout}, {Foley}, {Gabor}, {Garnavich},
  {Grier}, {Hamden}, {Krantz}, {Olszewski}, {Paschalidis}, {Reichart}, {Rest},
  {Smith}, {Strader}, {Trilling}, {Veillet}, {Wagner}, {Weiner}, \&
  {Zabludoff}}]{Paterson+21}
{Paterson}, K., {Lundquist}, M.~J., {Rastinejad}, J.~C., {et~al.} 2021, \apj,
  912, 128, \dodoi{10.3847/1538-4357/abeb71}

\bibitem[{{Perna} {et~al.}(2018){Perna}, {Chruslinska}, {Corsi}, \&
  {Belczynski}}]{Perna+18}
{Perna}, R., {Chruslinska}, M., {Corsi}, A., \& {Belczynski}, K. 2018, \mnras,
  477, 4228, \dodoi{10.1093/mnras/sty814}

\bibitem[{{Pian} {et~al.}(2017){Pian}, {D'Avanzo}, {Benetti}, {Branchesi},
  {Brocato}, {Campana}, {Cappellaro}, {Covino}, {D'Elia}, {Fynbo}, {Getman},
  {Ghirlanda}, {Ghisellini}, {Grado}, {Greco}, {Hjorth}, {Kouveliotou},
  {Levan}, {Limatola}, {Malesani}, {Mazzali}, {Melandri}, {M{\o}ller},
  {Nicastro}, {Palazzi}, {Piranomonte}, {Rossi}, {Salafia}, {Selsing},
  {Stratta}, {Tanaka}, {Tanvir}, {Tomasella}, {Watson}, {Yang}, {Amati},
  {Antonelli}, {Ascenzi}, {Bernardini}, {Bo{\"e}r}, {Bufano}, {Bulgarelli},
  {Capaccioli}, {Casella}, {Castro-Tirado}, {Chassande-Mottin}, {Ciolfi},
  {Copperwheat}, {Dadina}, {De Cesare}, {di Paola}, {Fan}, {Gendre},
  {Giuffrida}, {Giunta}, {Hunt}, {Israel}, {Jin}, {Kasliwal}, {Klose}, {Lisi},
  {Longo}, {Maiorano}, {Mapelli}, {Masetti}, {Nava}, {Patricelli}, {Perley},
  {Pescalli}, {Piran}, {Possenti}, {Pulone}, {Razzano}, {Salvaterra},
  {Schipani}, {Spera}, {Stamerra}, {Stella}, {Tagliaferri}, {Testa}, {Troja},
  {Turatto}, {Vergani}, \& {Vergani}}]{Pian+17}
{Pian}, E., {D'Avanzo}, P., {Benetti}, S., {et~al.} 2017, \nat, 551, 67,
  \dodoi{10.1038/nature24298}

\bibitem[{{Pozanenko} {et~al.}(2020){Pozanenko}, {Minaev}, {Grebenev}, \&
  {Chelovekov}}]{Pozanenko+20}
{Pozanenko}, A.~S., {Minaev}, P.~Y., {Grebenev}, S.~A., \& {Chelovekov}, I.~V.
  2020, Astronomy Letters, 45, 710, \dodoi{10.1134/S1063773719110057}

\bibitem[{{Rastinejad} {et~al.}(2021){Rastinejad}, {Fong}, {Kilpatrick},
  {Paterson}, {Tanvir}, {Levan}, {Metzger}, {Berger}, {Chornock}, {Cobb},
  {Laskar}, {Milne}, {Nugent}, \& {Smith}}]{Rastinejad+21}
{Rastinejad}, J.~C., {Fong}, W., {Kilpatrick}, C.~D., {et~al.} 2021, \apj, 916,
  89, \dodoi{10.3847/1538-4357/ac04b4}

\bibitem[{{Rhoades} \& {Ruffini}(1974)}]{RhoadesRuffini74}
{Rhoades}, C.~E., \& {Ruffini}, R. 1974, \prl, 32, 324,
  \dodoi{10.1103/PhysRevLett.32.324}

\bibitem[{{Rossi} {et~al.}(2020){Rossi}, {Stratta}, {Maiorano}, {Spighi},
  {Masetti}, {Palazzi}, {Gardini}, {Melandri}, {Nicastro}, {Pian}, {Branchesi},
  {Dadina}, {Testa}, {Brocato}, {Benetti}, {Ciolfi}, {Covino}, {D'Elia},
  {Grado}, {Izzo}, {Perego}, {Piranomonte}, {Salvaterra}, {Selsing},
  {Tomasella}, {Yang}, {Vergani}, {Amati}, \& {Stephen}}]{Rossi+20}
{Rossi}, A., {Stratta}, G., {Maiorano}, E., {et~al.} 2020, \mnras, 493, 3379,
  \dodoi{10.1093/mnras/staa479}

\bibitem[{{Schlafly} \& {Finkbeiner}(2011)}]{SchlaflyFinkbeiner11}
{Schlafly}, E.~F., \& {Finkbeiner}, D.~P. 2011, \apj, 737, 103,
  \dodoi{10.1088/0004-637X/737/2/103}

\bibitem[{{Shappee} {et~al.}(2017){Shappee}, {Simon}, {Drout}, {Piro},
  {Morrell}, {Prieto}, {Kasen}, {Holoien}, {Kollmeier}, {Kelson}, {Coulter},
  {Foley}, {Kilpatrick}, {Siebert}, {Madore}, {Murguia-Berthier}, {Pan},
  {Prochaska}, {Ramirez-Ruiz}, {Rest}, {Adams}, {Alatalo}, {Ba{\~n}ados},
  {Baughman}, {Bernstein}, {Bitsakis}, {Boutsia}, {Bravo}, {Di Mille}, {Higgs},
  {Ji}, {Maravelias}, {Marshall}, {Placco}, {Prieto}, \& {Wan}}]{Shappee+17}
{Shappee}, B.~J., {Simon}, J.~D., {Drout}, M.~R., {et~al.} 2017, Science, 358,
  1574, \dodoi{10.1126/science.aaq0186}

\bibitem[{{Shen} {et~al.}(2010){Shen}, {Kasen}, {Weinberg}, {Bildsten}, \&
  {Scannapieco}}]{Shen+10}
{Shen}, K.~J., {Kasen}, D., {Weinberg}, N.~N., {Bildsten}, L., \&
  {Scannapieco}, E. 2010, \apj, 715, 767, \dodoi{10.1088/0004-637X/715/2/767}

\bibitem[{{Shibata} \& {Hotokezaka}(2019)}]{Shibata+19}
{Shibata}, M., \& {Hotokezaka}, K. 2019, Annual Review of Nuclear and Particle
  Science, 69, 41, \dodoi{10.1146/annurev-nucl-101918-023625}

\bibitem[{{Silva} {et~al.}(2016){Silva}, {Blum}, {Allen}, {Dey}, {Schlegel},
  {Lang}, {Moustakas}, {Meisner}, {Valdes}, {Patej}, {Myers}, {Sprayberry},
  {Saha}, {Olsen}, {Safonova}, {Yang}, {Burleigh}, \& {MzLS Team}}]{Silva+2016}
{Silva}, D.~R., {Blum}, R.~D., {Allen}, L., {et~al.} 2016, in American
  Astronomical Society Meeting Abstracts, Vol. 228, American Astronomical
  Society Meeting Abstracts \#228, 317.02

\bibitem[{{Singer} {et~al.}(2016){Singer}, {Chen}, {Holz}, {Farr}, {Price},
  {Raymond}, {Cenko}, {Gehrels}, {Cannizzo}, {Kasliwal}, {Nissanke},
  {Coughlin}, {Farr}, {Urban}, {Vitale}, {Veitch}, {Graff}, {Berry},
  {Mohapatra}, \& {Mandel}}]{Singer+16_Supp}
{Singer}, L.~P., {Chen}, H.-Y., {Holz}, D.~E., {et~al.} 2016, \apjs, 226, 10,
  \dodoi{10.3847/0067-0049/226/1/10}

\bibitem[{{Smartt} {et~al.}(2017){Smartt}, {Chen}, {Jerkstrand}, {Coughlin},
  {Kankare}, {Sim}, {Fraser}, {Inserra}, {Maguire}, {Chambers}, {Huber},
  {Kr{\"u}hler}, {Leloudas}, {Magee}, {Shingles}, {Smith}, {Young}, {Tonry},
  {Kotak}, {Gal-Yam}, {Lyman}, {Homan}, {Agliozzo}, {Anderson}, {Angus},
  {Ashall}, {Barbarino}, {Bauer}, {Berton}, {Botticella}, {Bulla}, {Bulger},
  {Cannizzaro}, {Cano}, {Cartier}, {Cikota}, {Clark}, {De Cia}, {Della Valle},
  {Denneau}, {Dennefeld}, {Dessart}, {Dimitriadis}, {Elias-Rosa}, {Firth},
  {Flewelling}, {Fl{\"o}rs}, {Franckowiak}, {Frohmaier}, {Galbany},
  {Gonz{\'a}lez-Gait{\'a}n}, {Greiner}, {Gromadzki}, {Guelbenzu},
  {Guti{\'e}rrez}, {Hamanowicz}, {Hanlon}, {Harmanen}, {Heintz}, {Heinze},
  {Hernandez}, {Hodgkin}, {Hook}, {Izzo}, {James}, {Jonker}, {Kerzendorf},
  {Klose}, {Kostrzewa-Rutkowska}, {Kowalski}, {Kromer}, {Kuncarayakti},
  {Lawrence}, {Lowe}, {Magnier}, {Manulis}, {Martin-Carrillo}, {Mattila},
  {McBrien}, {M{\"u}ller}, {Nordin}, {O'Neill}, {Onori}, {Palmerio},
  {Pastorello}, {Patat}, {Pignata}, {Podsiadlowski}, {Pumo}, {Prentice}, {Rau},
  {Razza}, {Rest}, {Reynolds}, {Roy}, {Ruiter}, {Rybicki}, {Salmon}, {Schady},
  {Schultz}, {Schweyer}, {Seitenzahl}, {Smith}, {Sollerman}, {Stalder},
  {Stubbs}, {Sullivan}, {Szegedi}, {Taddia}, {Taubenberger}, {Terreran}, {van
  Soelen}, {Vos}, {Wainscoat}, {Walton}, {Waters}, {Weiland}, {Willman},
  {Wiseman}, {Wright}, {Wyrzykowski}, \& {Yaron}}]{Smartt+17}
{Smartt}, S.~J., {Chen}, T.~W., {Jerkstrand}, A., {et~al.} 2017, \nat, 551, 75,
  \dodoi{10.1038/nature24303}

\bibitem[{{Smith} {et~al.}(2020){Smith}, {Smartt}, {Young}, {Tonry}, {Denneau},
  {Flewelling}, {Heinze}, {Weiland}, {Stalder}, {Rest}, {Stubbs}, {Anderson},
  {Chen}, {Clark}, {Do}, {F{\"o}rster}, {Fulton}, {Gillanders}, {McBrien},
  {O'Neill}, {Srivastav}, \& {Wright}}]{ATLAS_smith+20}
{Smith}, K.~W., {Smartt}, S.~J., {Young}, D.~R., {et~al.} 2020, \pasp, 132,
  085002, \dodoi{10.1088/1538-3873/ab936e}

\bibitem[{{Soares-Santos} {et~al.}(2017){Soares-Santos}, {Holz}, {Annis},
  {Chornock}, {Herner}, {Berger}, {Brout}, {Chen}, {Kessler}, {Sako}, {Allam},
  {Tucker}, {Butler}, {Palmese}, {Doctor}, {Diehl}, {Frieman}, {Yanny}, {Lin},
  {Scolnic}, {Cowperthwaite}, {Neilsen}, {Marriner}, {Kuropatkin}, {Hartley},
  {Paz-Chinch{\'o}n}, {Alexander}, {Balbinot}, {Blanchard}, {Brown}, {Carlin},
  {Conselice}, {Cook}, {Drlica-Wagner}, {Drout}, {Durret}, {Eftekhari}, {Farr},
  {Finley}, {Foley}, {Fong}, {Fryer}, {Garc{\'\i}a-Bellido}, {Gill}, {Gruendl},
  {Hanna}, {Kasen}, {Li}, {Lopes}, {Louren{\c{c}}o}, {Margutti}, {Marshall},
  {Matheson}, {Medina}, {Metzger}, {Mu{\~n}oz}, {Muir}, {Nicholl}, {Quataert},
  {Rest}, {Sauseda}, {Schlegel}, {Secco}, {Sobreira}, {Stebbins}, {Villar},
  {Vivas}, {Walker}, {Wester}, {Williams}, {Zenteno}, {Zhang}, {Abbott},
  {Abdalla}, {Banerji}, {Bechtol}, {Benoit-L{\'e}vy}, {Bertin}, {Brooks},
  {Buckley-Geer}, {Burke}, {Carnero Rosell}, {Carrasco Kind}, {Carretero},
  {Castander}, {Crocce}, {Cunha}, {D'Andrea}, {da Costa}, {Davis}, {Desai},
  {Dietrich}, {Doel}, {Eifler}, {Fernand ez}, {Flaugher}, {Fosalba},
  {Gaztanaga}, {Gerdes}, {Giannantonio}, {Goldstein}, {Gruen}, {Gschwend},
  {Gutierrez}, {Honscheid}, {Jain}, {James}, {Jeltema}, {Johnson}, {Johnson},
  {Kent}, {Krause}, {Kron}, {Kuehn}, {Kuhlmann}, {Lahav}, {Lima}, {Maia},
  {March}, {McMahon}, {Menanteau}, {Miquel}, {Mohr}, {Nichol}, {Nord}, {Ogand
  o}, {Petravick}, {Plazas}, {Romer}, {Roodman}, {Rykoff}, {Sanchez},
  {Scarpine}, {Schubnell}, {Sevilla-Noarbe}, {Smith}, {Smith}, {Suchyta},
  {Swanson}, {Tarle}, {Thomas}, {Thomas}, {Troxel}, {Vikram}, {Wechsler},
  {Weller}, {Dark Energy Survey}, \& {Dark Energy Camera GW-EM
  Collaboration}}]{Soares-Santos+17}
{Soares-Santos}, M., {Holz}, D.~E., {Annis}, J., {et~al.} 2017, \apjl, 848,
  L16, \dodoi{10.3847/2041-8213/aa9059}

\bibitem[{{Strader}(2019)}]{TNS2019rwq}
{Strader}, J. 2019, Transient Name Server Classification Report, 2019-2125, 1

\bibitem[{{Tachibana} \& {Miller}(2018)}]{TachibanaMiller18}
{Tachibana}, Y., \& {Miller}, A.~A. 2018, \pasp, 130, 128001,
  \dodoi{10.1088/1538-3873/aae3d9}

\bibitem[{{Tak} {et~al.}(2021){Tak}, {Gibb}, {McGlynn}, {Smale}, {Jaffe},
  {Barthelmy}, {Burns}, {Cenko}, {Gonzales-Leon}, {Lorek}, {Martinez},
  {Perkins}, {Racusin}, {Sheets}, {Singer}, \& {TACH Team}}]{TACH}
{Tak}, D., {Gibb}, M., {McGlynn}, T., {et~al.} 2021, GRB Coordinates Network,
  31036, 1

\bibitem[{{Takada} {et~al.}(2014){Takada}, {Ellis}, {Chiba}, {Greene},
  {Aihara}, {Arimoto}, {Bundy}, {Cohen}, {Dor{\'e}}, {Graves}, {Gunn},
  {Heckman}, {Hirata}, {Ho}, {Kneib}, {Le F{\`e}vre}, {Lin}, {More},
  {Murayama}, {Nagao}, {Ouchi}, {Seiffert}, {Silverman}, {Sodr{\'e}},
  {Spergel}, {Strauss}, {Sugai}, {Suto}, {Takami}, \& {Wyse}}]{Takada+14}
{Takada}, M., {Ellis}, R.~S., {Chiba}, M., {et~al.} 2014, \pasj, 66, R1,
  \dodoi{10.1093/pasj/pst019}

\bibitem[{{Tanaka} \& {Hotokezaka}(2013)}]{tanaka+13}
{Tanaka}, M., \& {Hotokezaka}, K. 2013, ApJ, 775, 113,
  \dodoi{10.1088/0004-637X/775/2/113}

\bibitem[{{Tanvir} {et~al.}(2017){Tanvir}, {Levan},
  {Gonz{\'a}lez-Fern{\'a}ndez}, {Korobkin}, {Mandel}, {Rosswog}, {Hjorth},
  {D'Avanzo}, {Fruchter}, {Fryer}, {Kangas}, {Milvang-Jensen}, {Rosetti},
  {Steeghs}, {Wollaeger}, {Cano}, {Copperwheat}, {Covino}, {D'Elia}, {de Ugarte
  Postigo}, {Evans}, {Even}, {Fairhurst}, {Figuera Jaimes}, {Fontes}, {Fujii},
  {Fynbo}, {Gompertz}, {Greiner}, {Hodosan}, {Irwin}, {Jakobsson},
  {J{\o}rgensen}, {Kann}, {Lyman}, {Malesani}, {McMahon}, {Melandri},
  {O'Brien}, {Osborne}, {Palazzi}, {Perley}, {Pian}, {Piranomonte}, {Rabus},
  {Rol}, {Rowlinson}, {Schulze}, {Sutton}, {Th{\"o}ne}, {Ulaczyk}, {Watson},
  {Wiersema}, \& {Wijers}}]{Tanvir+17}
{Tanvir}, N.~R., {Levan}, A.~J., {Gonz{\'a}lez-Fern{\'a}ndez}, C., {et~al.}
  2017, \apjl, 848, L27, \dodoi{10.3847/2041-8213/aa90b6}

\bibitem[{{Thakur} {et~al.}(2020){Thakur}, {Dichiara}, {Troja}, {Chase},
  {S{\'a}nchez-Ram{\'\i}rez}, {Piro}, {Fryer}, {Butler}, {Watson}, {Wollaeger},
  {Ambrosi}, {Becerra Gonz{\'a}lez}, {Becerra}, {Bruni}, {Cenko}, {Cusumano},
  {D'A{\`\i}}, {Durbak}, {Fontes}, {Gatkine}, {Hungerford}, {Korobkin},
  {Kutyrev}, {Lee}, {Lotti}, {Minervini}, {Novara}, {La Parola}, {Pereyra},
  {Ricci}, {Tiengo}, \& {Veilleux}}]{Thakur+20}
{Thakur}, A.~L., {Dichiara}, S., {Troja}, E., {et~al.} 2020, \mnras, 499, 3868,
  \dodoi{10.1093/mnras/staa2798}

\bibitem[{{The LIGO Scientific Collaboration} {et~al.}(2021){The LIGO
  Scientific Collaboration}, {the Virgo Collaboration}, {the KAGRA
  Collaboration}, {Abbott}, {Abbott}, {Acernese}, {Ackley}, {Adams},
  {Adhikari}, {Adhikari}, {Adya}, {Affeldt}, {Agarwal}, {Agathos}, {Agatsuma},
  {Aggarwal}, {Aguiar}, {Aiello}, {Ain}, {Ajith}, {Akcay}, {Akutsu},
  {Albanesi}, {Allocca}, {Altin}, {Amato}, {Anand}, {Anand}, {Ananyeva},
  {Anderson}, {Anderson}, {Ando}, {Andrade}, {Andres}, {Andri{\'c}}, \&
  {Angelova}}]{GWTC-3}
{The LIGO Scientific Collaboration}, {the Virgo Collaboration}, {the KAGRA
  Collaboration}, {et~al.} 2021, arXiv e-prints, arXiv:2111.03606.
\newblock \doarXiv{2111.03606}

\bibitem[{{Tody}(1986)}]{Tody86}
{Tody}, D. 1986, in Society of Photo-Optical Instrumentation Engineers (SPIE)
  Conference Series, Vol. 627, Instrumentation in astronomy VI, ed. D.~L.
  {Crawford}, 733, \dodoi{10.1117/12.968154}

\bibitem[{{Tody}(1993)}]{Tody93}
{Tody}, D. 1993, in Astronomical Society of the Pacific Conference Series,
  Vol.~52, Astronomical Data Analysis Software and Systems II, ed. R.~J.
  {Hanisch}, R.~J.~V. {Brissenden}, \& J.~{Barnes}, 173

\bibitem[{{Tonry} {et~al.}(2018{\natexlab{a}}){Tonry}, {Denneau}, {Heinze},
  {Stalder}, {Smith}, {Smartt}, {Stubbs}, {Weiland }, \& {Rest}}]{Tonry+18}
{Tonry}, J.~L., {Denneau}, L., {Heinze}, A.~N., {et~al.} 2018{\natexlab{a}},
  \pasp, 130, 064505, \dodoi{10.1088/1538-3873/aabadf}

\bibitem[{{Tonry} {et~al.}(2018{\natexlab{b}}){Tonry}, {Denneau}, {Heinze},
  {Stalder}, {Smith}, {Smartt}, {Stubbs}, {Weiland}, \&
  {Rest}}]{ATLAS_tonry+18}
---. 2018{\natexlab{b}}, \pasp, 130, 064505, \dodoi{10.1088/1538-3873/aabadf}

\bibitem[{{Tucker} {et~al.}(2021){Tucker}, {Wiesner}, {Allam}, {Soares-Santos},
  {de Bom}, {Butner}, {Garcia}, {Morgan}, {Olivares}, {Palmese},
  {Santana-Silva}, {Shrivastava}, {Annis}, {Garcia-Bellido}, {Gill}, {Herner},
  {Kilpatrick}, {Makler}, {Sherman}, {Amara}, {Lin}, {Smith}, {Swann},
  {Arcavi}, {Bachmann}, {Bechtol}, {Berlfein}, {Briceno}, {Brout}, {Butler},
  {Cartier}, {Casares}, {Chen}, {Conselice}, {Contreras}, {Cook}, {Cooke},
  {Dage}, {D'Andrea}, {Davis}, {de Carvalho}, {Diehl}, {Dietrich}, {Doctor},
  {Drlica-Wagner}, {Drout}, {Farr}, {Finley}, {Fishbach}, {Foley},
  {Foerster-Buron}, {Fosalba}, {Friedel}, {Frieman}, {Frohmaier}, {Gruendl},
  {Hartley}, {Hiramatsu}, {Holz}, {Howell}, {Kawash}, {Kessler}, {Kuropatkin},
  {Lahav}, {Lundgren}, {Lundquist}, {Malik}, {Mann}, {Marriner}, {Marshall},
  {Martinez-Vazquez}, {McCully}, {Menanteau}, {Meza}, {Narayan}, {Neilsen},
  {Nicolaou}, {Nichol}, {Paz-Chinchon}, {Pereira}, {Pineda}, {Points},
  {Quirola}, {Rembold}, {Rest}, {Rodriguez}, {Romer}, {Sako}, {Salim},
  {Scolnic}, {Smith}, {Strader}, {Sullivan}, {Swanson}, {Thomas}, {Valenti},
  {Varga}, {Walker}, {Weller}, {Wood}, {Yanny}, {Zenteno}, {Aguena},
  {Andrade-Oliveira}, {Bertin}, {Brooks}, {Burke}, {Carnero Rosell}, {Carrasco
  Kind}, {Carretero}, {Costanzi}, {da Costa}, {De Vicente}, {Desai}, {Everett},
  {Ferrero}, {Flaugher}, {Gaztanaga}, {Gerdes}, {Gruen}, {Gschwend},
  {Gutierrez}, {Hinton}, {Hollowood}, {Honscheid}, {James}, {Kuehn}, {Lima},
  {Maia}, {Miquel}, {Ogando}, {Pieres}, {Plazas Malagon}, {Rodriguez Monroy},
  {Sanchez}, {Scarpine}, {Schubnell}, {Serrano}, {Sevilla}, {Smith}, {Suchyta},
  {Tarle}, {To}, \& {Zhang}}]{Tucker+21}
{Tucker}, D., {Wiesner}, M., {Allam}, S., {et~al.} 2021, arXiv e-prints,
  arXiv:2109.13351.
\newblock \doarXiv{2109.13351}

\bibitem[{{Valenti} {et~al.}(2017){Valenti}, {Sand}, {Yang}, {Cappellaro},
  {Tartaglia}, {Corsi}, {Jha}, {Reichart}, {Haislip}, \&
  {Kouprianov}}]{Valenti+17}
{Valenti}, S., {Sand}, D.~J., {Yang}, S., {et~al.} 2017, \apjl, 848, L24,
  \dodoi{10.3847/2041-8213/aa8edf}

\bibitem[{{Vieira} {et~al.}(2020){Vieira}, {Ruan}, {Haggard}, {Drout}, {Nynka},
  {Boyce}, {Spekkens}, {Safi-Harb}, {Carlberg}, {Fern{\'a}ndez}, {Piro},
  {Afsariardchi}, \& {Moon}}]{Vieira+20}
{Vieira}, N., {Ruan}, J.~J., {Haggard}, D., {et~al.} 2020, \apj, 895, 96,
  \dodoi{10.3847/1538-4357/ab917d}

\bibitem[{{Villar} {et~al.}(2017){Villar}, {Guillochon}, {Berger}, {Metzger},
  {Cowperthwaite}, {Nicholl}, {Alexand er}, {Blanchard}, {Chornock},
  {Eftekhari}, {Fong}, {Margutti}, \& {Williams}}]{villar+17}
{Villar}, V.~A., {Guillochon}, J., {Berger}, E., {et~al.} 2017, ApJL, 851, L21,
  \dodoi{10.3847/2041-8213/aa9c84}

\bibitem[{{Watson} {et~al.}(2020){Watson}, {Butler}, {Lee}, {Becerra},
  {Pereyra}, {Angeles}, {Farah}, {Figueroa}, {G{\'o}nzalez-Buitrago},
  {Quir{\'o}s}, {Ru{\'\i}z-D{\'\i}az-Soto}, {Tejada de Vargas}, {Tinoco}, \&
  {Wolfram}}]{Watson+20}
{Watson}, A.~M., {Butler}, N.~R., {Lee}, W.~H., {et~al.} 2020, \mnras, 492,
  5916, \dodoi{10.1093/mnras/staa161}

\bibitem[{{Wright} {et~al.}(2010){Wright}, {Eisenhardt}, {Mainzer}, {Ressler},
  {Cutri}, {Jarrett}, {Kirkpatrick}, {Padgett}, {McMillan}, {Skrutskie},
  {Stanford}, {Cohen}, {Walker}, {Mather}, {Leisawitz}, {Gautier}, {McLean},
  {Benford}, {Lonsdale}, {Blain}, {Mendez}, {Irace}, {Duval}, {Liu}, {Royer},
  {Heinrichsen}, {Howard}, {Shannon}, {Kendall}, {Walsh}, {Larsen}, {Cardon},
  {Schick}, {Schwalm}, {Abid}, {Fabinsky}, {Naes}, \& {Tsai}}]{WISE_Wright+10}
{Wright}, E.~L., {Eisenhardt}, P. R.~M., {Mainzer}, A.~K., {et~al.} 2010, \aj,
  140, 1868, \dodoi{10.1088/0004-6256/140/6/1868}

\bibitem[{{Wyatt} {et~al.}(2020){Wyatt}, {Tohuvavohu}, {Arcavi}, {Lundquist},
  {Howell}, \& {Sand}}]{Wyatt+20}
{Wyatt}, S.~D., {Tohuvavohu}, A., {Arcavi}, I., {et~al.} 2020, \apj, 894, 127,
  \dodoi{10.3847/1538-4357/ab855e}

\bibitem[{{Yang} {et~al.}(2017){Yang}, {Valenti}, {Cappellaro}, {Sand},
  {Tartaglia}, {Corsi}, {Reichart}, {Haislip}, \& {Kouprianov}}]{Yang+17}
{Yang}, S., {Valenti}, S., {Cappellaro}, E., {et~al.} 2017, \apjl, 851, L48,
  \dodoi{10.3847/2041-8213/aaa07d}

\bibitem[{{Zhou} {et~al.}(2021){Zhou}, {Newman}, {Mao}, {Meisner}, {Moustakas},
  {Myers}, {Prakash}, {Zentner}, {Brooks}, {Duan}, {Landriau}, {Levi}, {Prada},
  \& {Tarle}}]{Zhou+21}
{Zhou}, R., {Newman}, J.~A., {Mao}, Y.-Y., {et~al.} 2021, \mnras, 501, 3309,
  \dodoi{10.1093/mnras/staa3764}

\bibitem[{{Zimmerman} {et~al.}(2020){Zimmerman}, {Irani}, {Schulze}, {Bruch},
  \& {Yaron}}]{TNS2020pm}
{Zimmerman}, E., {Irani}, I., {Schulze}, S., {Bruch}, R., \& {Yaron}, O. 2020,
  Transient Name Server Classification Report, 2020-168, 1

\bibitem[{{Zou} {et~al.}(2017){Zou}, {Zhou}, {Fan}, {Zhang}, {Zhou}, {Nie},
  {Peng}, {McGreer}, {Jiang}, {Dey}, {Fan}, {He}, {Jiang}, {Lang}, {Lesser},
  {Ma}, {Mao}, {Schlegel}, \& {Wang}}]{Zou+2017}
{Zou}, H., {Zhou}, X., {Fan}, X., {et~al.} 2017, \pasp, 129, 064101,
  \dodoi{10.1088/1538-3873/aa65ba}

\end{thebibliography}

\end{document}